\documentclass[sn-mathphys,Numbered]{sn-jnl}

\usepackage{graphicx}%
\usepackage{multirow}%
\usepackage{amsmath,amssymb,amsfonts}%
\usepackage{amsthm}%
\usepackage{mathrsfs}%
\usepackage[title]{appendix}%
\usepackage{xcolor}%
\usepackage{textcomp}%
\usepackage{manyfoot}%
\usepackage{booktabs}%
\usepackage{algorithm}%
\usepackage{algorithmicx}%
\usepackage{algpseudocode}%
\usepackage{listings}%
\usepackage[frozencache]{minted}
\usepackage{mathtools}
\usepackage[disable]{todonotes}
\usepackage{tikz}
\usetikzlibrary{arrows, arrows.meta, shapes, trees, positioning, calc, fit}
\newcommand\textnode[2][]{\tikz[overlay]\node[fill=gray!60,inner sep=2pt,
  anchor=text, rounded rectangle, minimum height=14, #1] {#2};\phantom{#2}}
\tikzset{
  font={\fontsize{10pt}{12}\selectfont}}

\begin{document}

\title[The Tactician's Web of Large-Scale Formal Knowledge]{The Tactician's Web of Large-Scale Formal Knowledge}


\author[1]{\fnm{Lasse} \sur{Blaauwbroek}}

\affil[1]{\orgname{Institut des Hautes Études Scientifiques}, \orgaddress{\country{France}}}


\abstract{ The Tactician's Web is a platform offering a large web of strongly
  interconnected, machine-checked, formal mathematical knowledge conveniently
  packaged for machine learning, analytics, and proof engineering. Built on top
  of the Coq proof assistant, the platform exports a dataset containing a wide
  variety of formal theories, presented as a web of definitions, theorems, proof
  terms, tactics, and proof states. Theories are encoded both as a semantic
  graph (rendered below) and as human-readable text, each with a unique set of
  advantages and disadvantages. Proving agents may interact with Coq through the
  same rich data representation and can be automatically benchmarked on a set of
  theorems. Tight integration with Coq provides the unique possibility to make
  agents available to proof engineers as practical tools.
  \begin{center}
    \includegraphics[width=\textwidth]{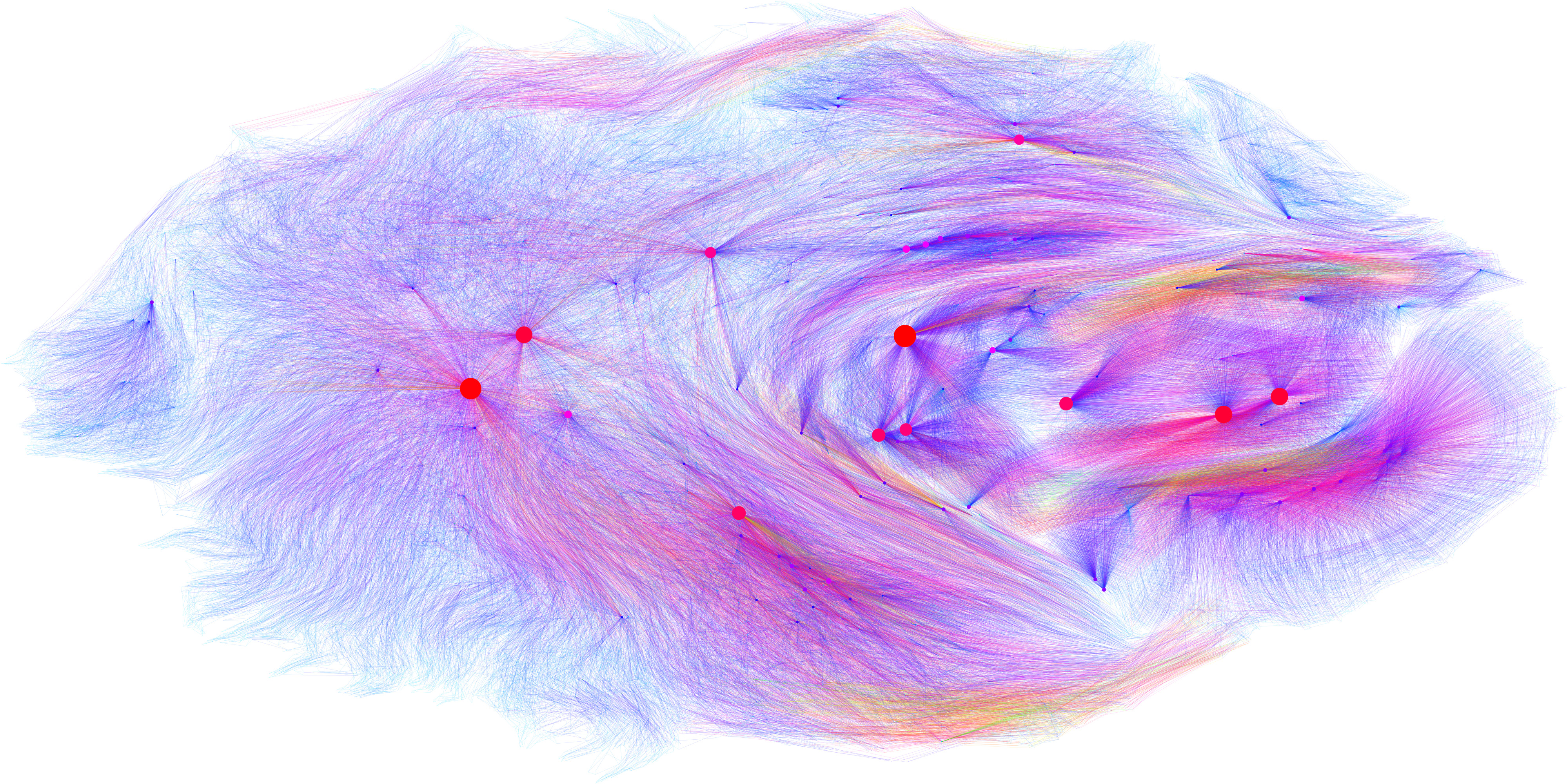}
  \end{center} 
}




\maketitle

\section{Introduction}

Proof assistants are systems that formally check the correctness of mathematical
theories. Users can input mathematical definitions and state theorems. Theorems
can then be proven interactively by the user, who inputs proof steps into the
assistant and receives feedback about the effect the step had on the state of the
proof. The proof assistant will mechanically verify the proof's correctness.

A large variety of mathematics has been formally verified in proof assistants
over the years. However, formally proving theorems is a painstaking and
time-consuming process. As such, there has been a long-running effort to
automate parts of this process. The ultimate goal for such automation is an
agent that is capable of proving complex theorems fully autonomously and can even
conjecture the existence of new theorems.

A practical problem that emerges while working on automation is how an agent
should interface with proof assistants. Most proof assistants, like
Coq~\cite{the_coq_development_team_2020_4021912}, are optimized to interact with
human agents. They input mathematics in a rich, text-based format that resembles
a blend between a traditional programming language and pen-and-paper
mathematics. A mathematical theory is built sentence by sentence. Each sentence,
called a vernacular command in Coq, builds on top of previous sentences.
However, the internal kernel knowledge maintained by proof assistants to ensure
correctness is quite different. Sentences undergo a complex transformation into
internal datastructures before they are added to the kernels knowledge base.
In complex developments, it may be difficult for the user to keep an accurate mental
model of the mathematical knowledge loaded into the kernel. Common
problems include (1) remembering what lemmas are available, (2) mapping
overloaded symbols and notations to the appropriate mathematical concept, and
(3) mental mismatches between the users' view of a mathematical concept versus
the details that exist in the kernel.

The interaction format between the proof assistant and a human is not
necessarily appropriate for interaction with a mechanical agent. Ideally, agents
should have access to a single structure that faithfully and conveniently
represents the formal knowledge that exists in the kernel. Such a structure
should be kept synchronized between the agent and proof assistant at all times.
Having explicit, shared knowledge saves an agent from having to
reconstruct an approximate model of the kernels knowledge.

In this work, we propose to encode such a shared knowledge base as a
large semantic graph. The graph encodes all available mathematical objects, including
definitions, theorems, and proofs, with explicit references between these
concepts. The structure of the graph is designed such that
information that usually remains implied in text-based
communication becomes explicit. One major source of implicit information is the
use of names, notations, and variables to reference objects. Consider the sentence
$\forall n,m:\mathbb{N},\ n+m=m+n$. The variable name $n$, refers to a
quantifier where the variable was bound. Determining which quantifier belongs to
$n$ requires a complex name-resolution procedure. This procedure is implicit
knowledge. In a graph, we can forego name resolution by replacing the name of a
variable with an explicit edge towards their binder. Such an encoding also
eliminates the need for an implicit algorithm to determine if two terms are
equal ($\alpha$-conversion). Two terms are equal if and only if their graph
encodings are equal (bisimilar).

The $+$ symbol in the example above refers to addition. However, the concept of
addition is highly overloaded. In this case, we can infer that $+$ means
addition of natural numbers, because $n$ and $m$ are
variables of type $\mathbb{N}$. Making such inferences requires a
combination of name resolution, type-inference, and a dictionary lookup of the
appropriate definition of addition based on the type. If we instead replace the
symbol $+$ with a direct edge to the appropriate definition, we can forego
such complexities. As a result of using explicit edges instead of names, a
mathematical theory now becomes a large mono-graph instead of a collection of loose
sentences connected only through implicit conventions. In fact, the entire known
universe of mathematical knowledge can be viewed as a single graph\footnote{A small piece
of the universe as formalized in Coq is rendered on the front page (see
Section~\ref{sec:datasets} for more explanation)}. We argue that this graph is
a much more direct representation of the semantic structure of a mathematical
theory than written text.

\subsection{Comparison of Math Representations for Agents}
Despite the philosophical arguments in favor of a graph-based representation
above, there are many other practical reasons for choosing a particular
representation. Here, we discuss the relative advantages and disadvantages of
representing mathematics as text, abstract syntax trees, or graphs in the
context of machine learning research for theorem proving.

\paragraph{Advantages of text}
Text is easy to work with. Because it is the format that mathematicians are used
to consume, it is easy to interpret and debug by humans. Due to the prevalence
of text-based machine learning research, text has become easy to process in
learning pipelines. One can take advantage of existing, battle-tested language
models such as transformers~\cite{DBLP:conf/nips/VaswaniSPUJGKP17} to quickly
prototype agents. Because formal mathematics tends resemble informal mathematics
(at least superficially), a language model can take advantage of a wealth of
background knowledge it has been pre-trained on. For example, when a model sees
the symbol $+$, it cannot know the exact definition of that symbol, but it may
have a fairly good sense of the intended meaning due to the prevalence of $+$ in
its background knowledge.

The existence of a built-in text-based printer in proof assistance is an
advantage. These printers support all possible syntactic constructs
out of the box. For graphs on the other hand, one needs to decide the best way
of encoding every syntactic feature of a formal language. Default printers also
support advanced features like notations, abbreviations, and even hiding of
unimportant information through implicit arguments and coercions. Because
these printers have been optimized for human consumption, one may assume that
they capture the important parts of a mathematical sentence while de-emphasizing
noise. For other representations, such optimizations have to be made separately.

\paragraph{Advantages of graphs}
Graphs are the natural datastructure to be processed by graph neural networks
(GNNs)~\cite{DBLP:journals/tnn/ScarselliGTHM09}.
As argued before, the flexibility of graphs allows one to
obtain the most direct representation of the semantic structure of a
mathematical theory, with nothing left to the imagination. When an agent does
not know the meaning of an object, it can simply follow an edge to look at its
definition. Graphs also lend themselves well to de-duplication. Whenever two
sub-graphs are identical (bisimilar), they can be shared.

Because of the possibility of faithfully encoding mathematics without using
identifiers (name-invariance), it becomes possible to create proving agents that
are oblivious to names. When an agent introduces a new object to the proof
assistant, such as a local hypothesis, it does not have to consider a name for
that object. If a name is required, any name will do, as the agent will not
observe it either now or in the future. This significantly reduces the size of
the action space of the agent.

\paragraph{Advantages of ASTs}
Abstract syntax trees are often used in the kernels of proof assistants and
other programming languages. They are trees, or sometimes directed acyclic
graphs, where each node represents a syntactic construct of the language. As
such, ASTs are a middle ground between text and fully-fledged graphs. They are
commonly preferred in kernels over graphs because tree structures are easier to
manipulate in practice, especially in functional programming languages. However,
in exchange for this convenience, ASTs leave more information implicit.

For Coq, it is possible to extract an s-expression based representation of the
AST's in the kernel through SerAPI~\cite{GallegoArias2016SerAPI}. This is used
by the CoqGym~\cite{DBLP:conf/icml/YangD19} machine learning environment. These
s-expressions are easy-to-maintain machine-readable datastructures that allow
external programs to observe the exact data in Coq's kernel. However, this also
means that they carry all of the technical warts and complexities that exist in the
kernel, even if those are irrelevant.

The individual s-expressions of definitions in SerAPI cannot reference each
other directly as a graph can. Instead, implicit named references and a lookup
table must be used. Similarly, bound variables are encoded using de Bruijn
indices, which are quite difficult to resolve correctly without access to the
name-resolution algorithms in the kernel.

\paragraph{Conclusion}
We argue that for machine learning and data analytics, a graph
representation is theoretically the most optimal representation. However, due to
practical considerations, a text-based approach may be easier and more effective
in the short term. To accommodate both approaches, we employ both encodings in
our platform. We hope this will enable direct experimental comparisons between
different approaches, and even to extract the best of both worlds.

\subsection{A Platform for Experimentation, Analytics and Practical Usage}
We present a platform where machine learning researchers and proof engineers can
come together.

For machine learning researchers we offer large-scale datasets of formal
knowledge encoded as graphs and text, debugging and
visualization tools, a massively parallel benchmarking system capable of running
on High-Performance Computing clusters, and protocols for interacting with the
Coq proof assistant. Because these interaction protocols are tightly integrated
into Coq, agents created by ML researchers can easily become available to proof engineers for
practical usage in their developments through a simple and convenient interface.
We hope that this will start a feedback loop
between proof engineers giving feedback on models and data scientists
incrementally refining their agents, ultimately creating powerful and
intelligent automation.

Additionally, we expect our datasets to be valuable to proof engineers who wish
to data-mine, visualize, and summarize the formal knowledge they have
created.

\subsection{Contributions}
We create a novel representation of Coq's Calculus of Inductive Constructions as
a semantic graph. Its design and implementation is described in
Section~\ref{sec:design-decisions}. The semantic graph also contains
machine-readable representations of tactic-based proofs
(Section~\ref{sec:tactic-representation}). Based on these representations, we
provide datasets~\cite{blaauwbroek_2023_10028721} and interaction
protocols~\cite{lasse_blaauwbroek_2023_10445966} for integration into Coq as
well as a benchmarking platform (Section~\ref{sec:modes-of-interaction}).
Libraries to help process the semantic graph by external agents are
provided~\cite{lasse_blaauwbroek_2023_10445964}. This also includes an extensive
data visualization tool\footnote{Visualizations of the dataset can be explored
  online: \url{http://grid01.ciirc.cvut.cz:8080}}.
Section~\ref{sec:evaluation} contains an experimental evaluation of some of the
aspects of our platform. A comprehensive evaluation of neural models
trained on the platform is available in a separate
publication~\cite{graph2tac}. Section~\ref{sec:threats-to-validity} contains a
discussion of potential threats to the validity of our data.

\section{Calculus of Inductive Constructions as a Graph}
\label{sec:cic-representation}

\subsection{Design Decisions}\label{sec:design-decisions}

Representing a $\lambda$-calculus term as a graph can potentially done in
many different ways. We aim for this translation to be as faithful, convenient,
and useful as possible. The graph structure that we target is an unordered
directed graph with both labeled nodes and labeled edges. Labels of nodes will
generally correspond to the labels of nodes in an abstract syntax tree. In an
abstract syntax tree, children of a node are typically ordered so that the
role of two subterms can be distinguished. As edges are not usually ordered in a
graph (as opposed to a tree), we instead use labeled edges to fulfill the purpose
of the ordering of children.

Having determined the target of our representation, we also need to discuss the
particular flavor of $\lambda$-terms that will be the target of the translation.
Coq is based on a flavor of the Calculus of Inductive Constructions~\cite{DBLP:conf/tlca/Paulin-Mohring93},
called Gallina. Like many programming languages, Gallina is the
user-facing language of Coq, which gets parsed, elaborated, globalized and simplified
before it moves into Coq's kernel. As such, Coq has a number of internal
representation ASTs for terms and the ability to translate freely back and
forth between these representations. This ability is needed due to the
interactive nature of Coq, where kernel terms, after having been processed, need to be
displayed in a palatable text-based format to users.

In principle, any of Coq's
intermediate internal representations can be used as a base to perform a
translation to graphs. However, some make more sense than others. Take, for
example, Coq's facility for syntax extensions and notations. These features
are explicitly geared towards text-based representations and may not make much
sense in a graph-based representation. More generally, we favor the more
structured, simple kernel-like language over a rich specification language. The
kernel language has few syntactic constructs and only contains identifiers that are
globalized into known entities (such as definitions). This fits well into the
goal of creating a large interconnected ``web'' of terms.

There are also some
choices in the design space that are less obvious, such as the inclusion of
implicit arguments. These are arguments of functions that, in normal
circumstances, can be automatically inferred by Coq from other arguments. Users
do not need to specify them manually, and will usually not encounter them.
Implicit arguments are an essential part of the CIC but are often large,
repetitive and somewhat ``boring'' (which is the reason they are usually not shown
to users). We choose to include implicit arguments in our graph representation
for the sake of consistency and regularity of the representation but at the
cost of larger terms. The wisdom of this decision will have to be experimentally
validated over time.

Having determined the input and output domain of our translation between terms
and graphs, we can discuss some more of its desired properties. We will
construct a function $f$ that takes a CIC-term parsed into Coq's kernel
representation and outputs a directed, labeled graph. To ensure that distinct
terms get mapped into distinct graphs, we require $f$ to be injective. On the
other hand, not all graphs have to be valid terms, so $f$ is not a bijection.
Furthermore, the notions of ``distinct terms'' and ``distinct graphs'' are up to
interpretation. As such, we will take the domain and image of $f$ modulo some
equivalence relations $\sigma$ and $\tau$.
\[ \text{CIC} \Big/ \sigma \quad\quad \xrightarrow[\text{ injective }]{f} \quad\quad G \Big/ \tau \]
Because the main goal for the graph representation is to manipulate Coq's
internal state through tactic commands (see Section~\ref{sec:tactic-representation}), the
leading factor in determining whether two terms are equal is if they can be
distinguished by a tactic (i.e. whether a tactic behaves differently on the two
terms)\footnote{Note that due to the complexity of Coq's tactic system, this is
  necessarily an approximation. In practice, there usually exists some
  exotic tactic that can distinguish even the most immaterial differences between
  terms.}. The equivalence relation $\sigma$ on the CIC, includes the following.
\begin{description}
\item[Name-invariance] The most obvious equivalence for $\lambda$-terms is
  $\alpha$-equivalence. In a graph representation, we are not usually interested
  in the name of any local binders. Although it can be argued that such names
  can carry semantically relevant information, in practice they almost never
  do. Binder names are usually single letters or a short abbreviation, and are
  often automatically generated by the proof assistant. The limited semantic
  value of local identifiers is confirmed by the Passport
  experiments~\cite{DBLP:journals/corr/abs-2204-10370}. As such, we expect
  graphs of terms to be equal modulo $\alpha$-equivalence. Furthermore, we
  extend this  name-invariance property to the local context of proof states.
  Similar to binders, local hypotheses are often short, generic names that can
  be generated by the proof assistant. A problem with making proof states name
  invariant is that tactics often have to reference hypotheses. In
  Section~\ref{sec:tactic-representation} we solve that by making tactics
  themselves name-invariant as well.

  Contrary to local names, we do not wish the graph representation to be
  invariant w.r.t. global names, such as definitions and inductives. Two
  identical definitions that differ only by name can still easily be
  distinguished by tactics such as \texttt{unfold}. Names of definitions are
  also more likely to carry useful semantic information. Furthermore, due to
  their injective nature, two isomorphic constructors of an inductive type can
  never be equal, making their name essential for distinguishing them.

\item[Universe collapse] Gallina terms are based on a stratified universe
  hierarchy starting with Set, Prop, and SProp, whose kinds are $\text{Type}_1$. Every
  universe $\text{Type}_i$ is then of kind $\text{Type}_{i+1}$. In the graph
  representation, we collapse this hierarchy into Set, Prop, SProp, and a single Type
  universe. Even though this makes the graph representation inconsistent, we
  consider this an acceptable simplification because, in many cases, a proper hierarchy
  can be automatically recovered. Furthermore, in the short term, we do not
  foresee any applications that might successfully utilize the information
  contained in the full universe hierarchy.

\item[Type-casting] A Gallina term can be type-casted from one type to another
  as long as those types are convertible. The cast of term $t$ to type $T$ is
  denoted as $t:T$. There are multiple possible reduction strategies to verify
  that a cast is valid, which can be chosen by the user. Since this is
  exclusively a performance optimization, we do not make any distinction between
  casting strategies.
\end{description}
In addition to these two equivalence classes, by the very nature of kernel
terms, many features like notations, implicit arguments, and type-class resolution
will also not be present in the graph representation. Some other potential
equivalence relations are intentionally omitted. For example, we do not consider
$\beta$-equivalence here, because reducing a term is easy to observe through a tactic.

The equivalence relation $\tau$ on the side of graphs is conceptually
simpler than the equivalence relation $\sigma$ on the side of the CIC. There are
three factors to keep in mind for $\tau$.
\begin{enumerate}
\item Because we define a directed graph, it makes sense to identify a term $t$ associated to a
  root node $n$ with exactly the sub-graph that is the forward closure from $n$. Any
  parts of the graph that are not in the forward closure are irrelevant for the
  interpretation of $t$.
\item Furthermore, we would like to take advantage of a graphs structural
  flexibility to share identical sub-terms. This allows us to
  substantially compress the graph, and add semantic information by easily
  identifying identical terms.
\item We want $\tau$ to respect $\alpha$-equivalence, just like $\sigma$ does.
  This means that two variable nodes can only be considered
  equivalent if the nodes that represent their binders are also equivalent.
\end{enumerate}
A direct consequence of points (1) and (3) is that the binding location of a
variable must be in its forward closure. As such, it is necessary to resolve a
variable using a back-edge to its binding site. Once we do this, we can take care
of point (2) by stipulating that graphs must be taken equal modulo
bisimulation.

As a reminder, we will restate
the definition of bisimilarity on a directed graph with labeled nodes and edges:
A relation $R$ between nodes is a bisimulation relation when for all nodes
$(p_1, p_2) \in R$ the labels of $p_1$ and $p_2$ are equal and
\begin{equation*}
  \begin{split}
    &\text{if}\ p_1\xrightarrow{a} q_1\ \text{then there exists}\ q_2\ \text{such that}\ p_2\xrightarrow{a}q_2\ \text{and}\ (q_1, q_2) \in R\\
    &\text{if}\ p_2\xrightarrow{a} q_2\ \text{then there exists}\ q_1\ \text{such that}\ p_1\xrightarrow{a}q_1\ \text{and}\ (q_1, q_2) \in R.
  \end{split}
\end{equation*}
Two nodes $p_1$ and $p_2$ in a graph are then considered bisimilar if there
exists a bisimulation relation such that $(p_1, p_2)\in R$.

The bisimulation relation captures exactly the notion of context-sensitive
$\alpha$-equivalence that we need. More information, justifications, proofs, and
algorithms related to this can be found in a separate
publication~\cite{sharing-paper}. Term sharing is further discussed in Section~\ref{sec:graph-sharing}.

Note that the requirement for back-edges from variables to binders is the main reason why,
within our design constraints, normal abstract syntax trees are not sufficient
to faithfully represent terms. We need to generalize from trees to DAGs in order
to allow term-sharing, and we need to generalize further to graphs to allow for
back-edges.

\subsection{Calculus of Inductive Constructions Gadgets}
In this section, we describe a concrete implementation of the graph translation
function $f$, that satisfies the design decisions of the previous section. The
translation is defined through a series of \textit{gadgets} that illustrate how
to translate a single piece of the syntax of the CIC, assuming that we already
know how to translate the subterms referred to in that syntactic construct. The
final translation is then constructed by recursively tying together all gadgets.

To illustrate this, we start with the simplest syntactic pieces of the CIC,
function application, type-casting, and type universes.
\begin{center}
\begin{tikzpicture}[
  lnode/.style = {fill=gray!60, circle, inner sep = 1.5, minimum height=14, },
  ]
  \node[lnode] (app) at (0, 0) {$@$};
  \node[above=0.1 of app.north] (lambda-text) {$T\cdot U$};
  \node[regular polygon, regular polygon sides=3, fill=gray!60, inner sep = 0.5,
  below left = 0.5 and 0.5 of app, anchor=north] (T) {$T_G$};
  \node[regular polygon, regular polygon sides=3, fill=gray!60, inner sep = 0.5,
  below right = 0.5 and 0.5 of app, anchor=north] (U) {$U_G$};
  \draw[->] (app) -- (T.north);
  \draw[->>] (app) -- (U.north);
\end{tikzpicture}
\qquad
\begin{tikzpicture}[
  lnode/.style = {fill=gray!60, circle, inner sep = 1.5, minimum height=14, },
  ]
  \node[lnode, rounded rectangle, inner sep = 3] (app) at (0, 0) {cast};
  \node[above=0.1 of app.north] (lambda-text) {$T : U$};
  \node[regular polygon, regular polygon sides=3, fill=gray!60, inner sep = 0.5,
  below left = 0.5 and 0.5 of app, anchor=north] (T) {$T_G$};
  \node[regular polygon, regular polygon sides=3, fill=gray!60, inner sep = 0.5,
  below right = 0.5 and 0.5 of app, anchor=north] (U) {$U_G$};
  \draw[->] (app) -- (T.north);
  \draw[->>] (app) -- (U.north);
\end{tikzpicture}
\qquad
\begin{tikzpicture}[
  lnode/.style = {fill=gray!60, circle, inner sep = 1.5, minimum height=14, },
  ]
  \node[lnode, rounded rectangle, inner sep = 3, label={$\text{Type}_i$}] (type) at (0, 0) {Type};
  \node[lnode, rounded rectangle, inner sep = 3, below = 1 of type, label=Set] (set) {Set};
  \node[lnode, rounded rectangle, inner sep = 3, right  = 1 of type, label=Prop] (prop) {Prop};
  \node[lnode, rounded rectangle, inner sep = 3, below = 1 of prop, label=SProp] (sprop) {SProp};
\end{tikzpicture}
\end{center}
Note how we have only four distinct nodes for universes. Every higher universe
$\text{Type}_i$ gets collapsed into a single node. Beyond that, we distinguish
between the universe of sets, the universe of propositions, and the universe of
proof-irrelevant propositions (SProp).

The function application $T\cdot U$ is translated as a single node with label ``@''. It has two
sub-trees corresponding to $T$ and $U$. For brevity, the translation $f(T)$ and
$f(U)$ is denoted as $T_G$ and $U_G$. Recall that the edges of the target graph $G$
are unordered. To distinguish the function and argument of the application, each
edge has a label. For brevity, we do not actually show this label. Instead, we
observe that each node has a small, fixed number of outgoing edge types. As such, the
label of an edge should be interpreted as a combination of the label of the source node of
the edge and the shape of the arrow of the edge. For example, the label of the
edge pointing towards $T_G$ is $@\!\!\to$ while the edge pointing towards $U_G$
has label $@\!\!\twoheadrightarrow$.

Apart from applications and casts, the other basic building blocks of the CIC are
binders. Those are more difficult to describe. In order to translate a term of
the shape $\lambda x:T.\ U$, we first need to know how to translate $U$. Because
$U$ is not closed, we need the ability to translate free variables. A free variable
$x$ is translated simply to a node with label $v_x$. Free
variables will never occur in the final translation of a term, because all terms
to be translated are closed. They are only needed as an
intermediate step during construction. As such, we show free variable nodes with a
dashed border. The graph resulting from translating the open term $U$ contains a
number of copies of $v_x$. To create the $\lambda$-abstraction, we
replace all instances of $v_x$ with a variable node with label $\uparrow$ that
contains back-edge to the binder. Similar translations are made for
let-abstraction and dependent products. We represent these translations as the
following gadgets.
\begin{center}
  \raisebox{-0.5\height}{
    \begin{tikzpicture}[
      lnode/.style = {fill=gray!60, circle, inner sep = 1.5, minimum height=14, },
      ]
      \node[regular polygon, regular polygon sides=3,
      fill=gray!60, inner sep = 7.8] (U) at (0,0) {$U'_G$};
      \node[above=0.1 of U.north] (text) {$U$};
      \node[lnode, draw, dashed,
      below= -0.1 of U.south, anchor=north] (v1) {$v_x$};
      \node[lnode, draw, dashed,
      left= 1.05 of v1.north, anchor=north] (v2) {$v_x$};
      \node[lnode, draw, dashed,
      right= 1.05 of v1.north, anchor=north] (v3) {$v_x$};
      \draw[dotted, shorten >= 5pt, shorten <= 5pt] (v1.east) -- (v3.west);
      \draw[dotted, shorten >= 5pt, shorten <= 5pt] (v1.west) -- (v2.east);
    \end{tikzpicture}}
  \kern-1.5em
  {\huge $\implies$}
  \raisebox{-0.5\height}{
\begin{tikzpicture}[
  lnode/.style = {fill=gray!60, circle, inner sep = 1.5, minimum height=14, },
  ]
  \node[lnode, rounded rectangle] (lambda) at (0, 0) {$\lambda$/$\forall$};
  \node[above=0.1 of lambda.north, align=center] (lambda-text) {$\lambda x:T.\ U$\\$\forall x:T.\ U$};
  \node[regular polygon, regular polygon sides=3, fill=gray!60, inner sep = 5.5,
  below right = 1.2 and 0.2 of lambda.south] (U) {$U'_G$};
  \node[regular polygon, regular polygon sides=3, fill=gray!60, inner sep = 0.5,
  left = 1.3 of U.north, anchor=north] (T) {$T_G$};
  \draw[->>] (lambda) -- (U.north);
  \draw[->] (lambda) -- (T.north);
  \node[lnode, below= 0.15 of U.south, anchor=center] (v1) {$\uparrow$};
  \node[lnode, left= 0.32 of v1] (v2) {$\uparrow$};
  \node[lnode, right= 0.32 of v1] (v3) {$\uparrow$};
  \draw[dotted, shorten >= 3pt, shorten <= 3pt] (v1) -- (v3);
  \draw[dotted, shorten >= 3pt, shorten <= 3pt] (v1) -- (v2);
  \draw[-latex] (v2.north west) to[out=150, in=-90] ($(T.south)-(0.7,0)$) to[out=90, in=180] (lambda);
  \draw[-latex] (v3.north east) to[bend right=45] (lambda);
  \draw[-latex] (v1.south east) to[out=-30, in=-125] ($(v3.south)+(0.3,0)$) to[out=55, in=-10] (lambda);
\end{tikzpicture}
\kern-1.3em
\begin{tikzpicture}[
  lnode/.style = {fill=gray!60, circle, inner sep = 1.5, minimum height=14, },
  ]
  \node[lnode, rounded rectangle, inner sep = 3] (lambda) at (0, 0) {let};
  \node[above=0.1 of lambda.north] (lambda-text) {$\text{let}\ x:T\coloneq V\ \text{in}\ U$};
  \node[regular polygon, regular polygon sides=3, fill=gray!60, inner sep = 5.5,
  below right = 1.2 and 0.85 of lambda.south] (U) {$U'_G$};
  \node[regular polygon, regular polygon sides=3, fill=gray!60, inner sep = 0.5,
  below left = -0.1 and 1.3 of U.north, anchor=north] (V) {$V_G$};
  \node[regular polygon, regular polygon sides=3, fill=gray!60, inner sep = 0.5,
  left = 2.2 of U.south, anchor=south] (T) {$T_G$};
  \draw[->] (lambda) -- (T.north);
  \draw[->>] (lambda) -- (V.north);
  \draw[->>>] (lambda) -- (U.north);
  \node[lnode, below= 0.15 of U.south, anchor=center] (v1) {$\uparrow$};
  \node[lnode, left= 0.32 of v1] (v2) {$\uparrow$};
  \node[lnode, right= 0.32 of v1] (v3) {$\uparrow$};
  \draw[dotted, shorten >= 3pt, shorten <= 3pt] (v1) -- (v3);
  \draw[dotted, shorten >= 3pt, shorten <= 3pt] (v1) -- (v2);
  \draw[-latex] (v2.west) to[out=180, in=-90] ($(T.south)-(0.7,-0.3)$) to[out=90, in=180] (lambda);
  \draw[-latex] (v3.north east) to[bend right=50] (lambda);
  \draw[-latex] (v1.south east) to[out=-30, in=-125] ($(v3.south)+(0.3,0)$) to[out=55, in=0] (lambda);
\end{tikzpicture}}
\end{center}

These gadgets have multiple variable nodes that point to the same binder.
Conceptually, each $\uparrow$ node in a graph corresponds to exactly one variable in
the original term. However, it should be noted that due to the bisimulation
equivalence on graphs, we can collapse all such nodes into a single node. For
example, the following two graphs that correspond to the term $\lambda x:T,\ C\ x
\ x$ are bisimilar.
\begin{center}
\begin{tikzpicture}[
  lnode/.style = {fill=gray!60, circle, inner sep = 1.5, minimum height=14, },
  ]
  \node[lnode] (lambda) at (0, 0) {$\lambda$};
  \node[lnode, below right = 0.5 and 0.5 of lambda] (app1) {$@$};
  \node[regular polygon, regular polygon sides=3, fill=gray!60, inner sep = 0.5,
  below left = 0.5 and 0.5 of lambda, anchor=north] (T) {$T_G$};
  \draw[->>] (lambda) -- (app1.north);
  \draw[->] (lambda) -- (T.north);
  \node[lnode, below left = 0.3 and 0.1 of app1] (app2) {$@$};
  \node[lnode, below right = 0.3 and 0.1 of app1] (v1) {$\uparrow$};
  \node[lnode, below right = 0.3 and 0.1 of app2] (v2) {$\uparrow$};
  \node[regular polygon, regular polygon sides=3, fill=gray!60, inner sep = 0.5,
  below left = 0.3 and 0.1 of app2, anchor=north] (C) {$C_G$};
  \draw[-latex] (v2.east) to[out=0, in=-90] ($(v1.east)+(0.3,0)$) to[out=90, in=0] (lambda);
  \draw[-latex] (v1.north east) to[out=55, in=-10] (lambda);
  \draw[->>] (app1) -- (v1);
  \draw[->] (app1) -- (app2);
  \draw[->>] (app2) -- (v2);
  \draw[->] (app2) -- (C.north);
\end{tikzpicture}
\qquad
\begin{tikzpicture}[
  lnode/.style = {fill=gray!60, circle, inner sep = 1.5, minimum height=14, },
  ]
  \node[lnode] (lambda) at (0, 0) {$\lambda$};
  \node[lnode, below right = 0.5 and 0.5 of lambda] (app1) {$@$};
  \node[regular polygon, regular polygon sides=3, fill=gray!60, inner sep = 0.5,
  below left = 0.5 and 0.5 of lambda, anchor=north] (T) {$T_G$};
  \draw[->>] (lambda) -- (app1.north);
  \draw[->] (lambda) -- (T.north);
  \node[lnode, below left = 0.3 and 0.1 of app1] (app2) {$@$};
  \node[lnode, below right = 0.3 and 0.5 of app2] (v2) {$\uparrow$};
  \node[regular polygon, regular polygon sides=3, fill=gray!60, inner sep = 0.5,
  below left = 0.3 and 0.1 of app2, anchor=north] (C) {$C_G$};
  \draw[-latex] (v2.north east) to[out=55, in=0] (lambda);
  \draw[->>] (app1) -- (v2);
  \draw[->] (app1) -- (app2);
  \draw[->>] (app2) -- (v2);
  \draw[->] (app2) -- (C.north);
\end{tikzpicture}
\end{center}
This observation justifies that going forward, gadgets that include binders and
variables will only contain a single $\uparrow$ node for each binder $x$, that
simultaneously represents all variables that refer to $x$.

\subsection{Definitions}\label{sec:definitions}
The gadgets above give a comprehensive translation of the traditional Calculus
of Constructions. The essential ingredient to transition into the Calculus of
Inductive Constructions is the notion of a \textit{definition}. We take this
notion to be rather broad, including function definitions, inductive
definitions, constructors, projections, axioms, section variables, and theorems.

Definitions are somewhat special, in that they form a directed acyclic graph.
Even though our graph translation contains cycles on the local level, when viewing the
graph from a bird's eye view, those cycles vanish. Definitions may reference
each other through their definitional body but the graph of dependencies between
definitions must form a DAG. An exception exists for inductive definitions and
constructors which are usually constructed mutually recursively. As such, it is
more accurate to say that the dependencies between \textit{clusters} of strongly
related definitions form a DAG.

The root node of a translated definition is marked with a label containing a
large amount of information. Besides the fully-qualified name of the definition,
additional knowledge is encoded around the different kinds of definitions.
\begin{itemize}
\item Inductives, constructors, and projections contain information about the
  cluster to which they belong. The cluster is the set of definitions that are
  simultaneously defined when an inductive structure is declared in Coq.
\item Theorems have a datastructure associated to them that encodes how the
  theorem was constructed through a \textit{tactical proof}. Tactical proofs
  are further described in Section~\ref{sec:tactic-representation}.
\item Every definition contains information about how and where it occurs in
  the global context, sections, and modules. We call this information the
  \textit{meta-graph}, which is further discussed in
  Section~\ref{sec:meta-graph}.
\item For every definition, the label includes a textual representation of
  both its type and its body (when applicable). This textual representation is
  produced by Coq's term printing functionality and corresponds directly to
  what is seen by an end-user when interacting with Coq. This information is
  meant to be consumed by text-based machine learning techniques.
\end{itemize}
Due to the wealth of information encoded in the label of a definition, it needs
to be treated somewhat special from the standpoint of the bisimulation relation.
Not all information in the label is always seen as ``relevant''. See
Section~\ref{sec:graph-sharing} for more information. In the following gadgets,
we will simply summarize the node of a definition by its unqualified name.

Non-inductive definitions can either be computationally relevant (traditional definitions,
where the body of the definition can be accessed during term conversion),
computationally irrelevant (usually theorems whose proof is not needed during conversion), or
axioms. Computationally relevant definitions are declared through one of these vernaculars.
\begin{minted}{coq}
Definition fun_x : T := U.    Definition fun_x : T. Proof. ... Defined.
\end{minted}
Here, the dots represent a tactical proof of $T$. Computationally irrelevant
definitions and axioms are declared as follows.
\begin{minted}{coq}
Definition thm_y : T. Proof. ... Qed.      Axiom axiom_z : T.
\end{minted}
These three vernaculars are translated according to the following gadgets.
\begin{center}
\begin{tikzpicture}[
  lnode/.style = {fill=gray!60, circle, inner sep = 1.5, minimum height=14, },
  ]
  \node[lnode, rounded rectangle, inner sep = 3] (funx) at (0, 0) {\texttt{fun\_x}};
  \node[regular polygon, regular polygon sides=3, fill=gray!60, inner sep = 0.5,
  below left = 0.5 and 0.5 of funx, anchor=north] (T) {$T_G$};
  \node[regular polygon, regular polygon sides=3, fill=gray!60, inner sep = 0.5,
  below right = 0.5 and 0.5 of funx, anchor=north] (U) {$U_G$};
  \draw[->] (funx) -- (T.north);
  \draw[->>] (funx) -- (U.north);
\end{tikzpicture}
\qquad
\begin{tikzpicture}[
  lnode/.style = {fill=gray!60, circle, inner sep = 1.5, minimum height=14, },
  ]
  \node[lnode, rounded rectangle, inner sep = 3] (funx) at (0, 0) {\texttt{thm\_y}};
  \node[regular polygon, regular polygon sides=3, fill=gray!60, inner sep = 0.5,
  below left = 0.5 and 0.5 of funx, anchor=north] (T) {$T_G$};
  \node[regular polygon, regular polygon sides=3, fill=gray!60, inner sep = 0.5,
  below right = 0.5 and 0.5 of funx, anchor=north] (U) {$U_G$};
  \draw[->] (funx) -- (T.north);
  \draw[->>>] (funx) -- (U.north);
\end{tikzpicture}
\qquad
\begin{tikzpicture}[
  lnode/.style = {fill=gray!60, circle, inner sep = 1.5, minimum height=14, },
  ]
  \node[lnode, rounded rectangle, inner sep = 3] (funx) at (0, 0) {\texttt{axiom\_z}};
  \node[regular polygon, regular polygon sides=3, fill=gray!60, inner sep = 0.5,
  below left = 0.5 and 0.5 of funx, anchor=north] (T) {$T_G$};
  \node[lnode, rounded rectangle, inner sep = 3,
  below right = 0.5 and 0.5 of funx, anchor=north] (U) {empty};
  \draw[->] (funx) -- (T.north);
  \draw[->>>>] (funx) -- (U.north);
\end{tikzpicture}
\end{center}
The only distinguishing factor between the three forms is the label of the edge
pointing to the body of the definition. The edge to the body of computationally
irrelevant definitions is special because it is optional. This allows the
sometimes rather large and expensive-to-generate body of a computationally
irrelevant definition to be omitted from the graph depending on the mode of
interaction with Coq (see Section~\ref{sec:modes-of-interaction}). In
offline datasets, such opaque proofs are always included (because generating and
storing them is a one-time cost), while during interactive operations they are
excluded by default. A consequence of this optionality is that as far as the
bisimulation relation is concerned, computationally irrelevant bodies are always
ignored.

Compared to ``regular'' definitions, inductive definitions are quite a bit more
intricate to define. As stated above, a single vernacular simultaneously defines a
cluster of inductives, constructors, and projections. (In addition, Coq might
automatically derive induction principles and other auxiliary definitions, but
those are not considered part of the cluster.) The basic blueprint of such a
mutually inductive vernacular is as follows, with the simple example of the
booleans and a more complex example of the mutually inductive propositions
\texttt{even} and \texttt{odd}.
\begin{minted}{coq}
Inductive I_A : T_A :=      Inductive bool : Set :=
| C_A1 : T_A1               | true  : bool
| ...                       | false : bool
| C_An : T_An
with ...                    Inductive even : nat -> Prop :=
with I_Z : T_Z :=           | ez : even 0
| C_Z : T_Z1                | eo : forall n, odd n -> even (S n)
| ...                       with odd : nat -> Prop :=
| C_Zn : T_Zn.              | oe : forall n, even n -> odd (S n).
\end{minted}
Note that (as with other syntactic elements of Gallina), parts of this
blueprint may sometimes be omitted by a user as it can be automatically inferred by Coq.
However, our translation always operates on the fully elaborated kernel terms.
Furthermore, this blueprint does not reflect some constraints that need to be
imposed on inductive definitions in order to make them valid and consistent,
such as the positivity condition and the need for the conclusion of all types
$T_X$ to be a universe. Fortunately, we can assume that these conditions
have already been checked by the kernel before the translation to graphs occurs,
allowing us to indeed get away with such a bare-bones blueprint.

The recursive nature of inductives is hidden within the terms $T_{Xi}$, which
represent the bodies of the constructors. As such, these terms can contain free
variables that reference the inductive types in the cluster being defined. The
translation of these terms is assumed to take the following shape.
\begin{center}
\begin{tikzpicture}[
  lnode/.style = {fill=gray!60, circle, inner sep = 1.5, minimum height=14, },
  ]
  \node[regular polygon, regular polygon sides=3,
  fill=gray!60, inner sep = -1.4] (U) at (0,0) {$T_{XiG}'$};
  \node[above=0.1 of U.north] (text) {$T_{Xi}$};
  \node[lnode, fill=gray!60, draw, dashed,
  below right= -0.1 and 0.6 of U.south, anchor=north] (v1) {$v_{I_Z}$};
  \node[lnode, fill=gray!60, draw, dashed,
  below left= -0.1 and 0.6 of U.south, anchor=north] (v2) {$v_{I_A}$};
  \draw[dotted, shorten >= 5pt, shorten <= 5pt] (v1.west) -- (v2.east);
\end{tikzpicture}
\end{center}
We then take the free variables nodes and ``tie the knot'' such that only bound
variables remain. The final cluster of definitions consists of the inductive
types $I_X$ and their constructors $C_{Xi}$.
\begin{center}
\begin{tikzpicture}[
  lnode/.style = {fill=gray!60, circle, inner sep = 1.5, minimum height=14, },
  ]
  \node[lnode, rounded rectangle] (IA) at (0, 0) {$I_A$};
  \node[lnode, rounded rectangle, below = 0.5 of IA] (CA1) {$C_{A1}$};
  \node[regular polygon, regular polygon sides=3, fill=gray!60, inner sep = -1.5,
  below left = 0.5 and 0 of CA1.south, anchor=north] (TA1) {$T'_{A1G}$};
  \node[lnode, below right = 0.15 and 0.5 of TA1.south, anchor=center] (v11) {$\uparrow$};
  \node[lnode, below left = 0.15 and 0.5 of TA1.south, anchor=center] (v12) {$\uparrow$};

  \node[lnode, rounded rectangle, below right = 0.5 and 1.5 of IA] (CAn) {$C_{An}$};
  \node[regular polygon, regular polygon sides=3, fill=gray!60, inner sep = -1.5,
  below left = 0.5 and 0 of CAn.south, anchor=north] (TAn) {$T'_{AnG}$};
  \node[lnode, below right = 0.15 and 0.5 of TAn.south, anchor=center] (vn1) {$\uparrow$};
  \node[lnode, below left = 0.15 and 0.5 of TAn.south, anchor=center] (vn2) {$\uparrow$};

  \node[regular polygon, regular polygon sides=3, fill=gray!60, inner sep = -1.5,
  below left = 0.5 and 1 of IA, anchor=north] (TA) {$T_{AG}$};

  \draw[->] (IA) -- (TA.north);
  \draw[->>] (IA) -- (CA1);
  \draw[->>] (IA) -- (CAn);
  \draw[->] (CA1) -- (TA1);
  \draw[->] (CAn) -- (TAn);

  \draw[dotted, shorten >= 10pt, shorten <= 10pt] (CA1) -- (CAn);
  \draw[dotted, shorten >= 10pt, shorten <= 10pt] (TA1.north east) -- (TAn.north west);
  \draw[dotted, shorten >= 5pt, shorten <= 5pt] (v11) -- (v12);
  \draw[dotted, shorten >= 5pt, shorten <= 5pt] (vn1) -- (vn2);

  \node[lnode, rounded rectangle, right = 6.5 of IA] (IZ) {$I_Z$};
  \node[lnode, rounded rectangle, below = 0.5 of IZ] (CZ1) {$C_{Zn}$};
  \node[regular polygon, regular polygon sides=3, fill=gray!60, inner sep = -1.5,
  below right = 0.5 and 0 of CZ1.south, anchor=north] (TZ1) {$T'_{ZnG}$};
  \node[lnode, below right = 0.15 and 0.5 of TZ1.south, anchor=center] (vz11) {$\uparrow$};
  \node[lnode, below left = 0.15 and 0.5 of TZ1.south, anchor=center] (vz12) {$\uparrow$};

  \node[lnode, rounded rectangle, below left = 0.5 and 1.5 of IZ] (CZn) {$C_{Z1}$};
  \node[regular polygon, regular polygon sides=3, fill=gray!60, inner sep = -1.5,
  below left = 0.5 and 0 of CZn.south, anchor=north] (TZn) {$T'_{Z1G}$};
  \node[lnode, below right = 0.15 and 0.5 of TZn.south, anchor=center] (vzn1) {$\uparrow$};
  \node[lnode, below left = 0.15 and 0.5 of TZn.south, anchor=center] (vzn2) {$\uparrow$};

  \node[regular polygon, regular polygon sides=3, fill=gray!60, inner sep = -1.5,
  below right = 0.5 and 1 of IZ, anchor=north] (TZ) {$T_{ZG}$};

  \draw[->] (IZ) -- (TZ.north);
  \draw[->>] (IZ) -- (CZ1);
  \draw[->>] (IZ) -- (CZn);
  \draw[->] (CZ1) -- (TZ1);
  \draw[->] (CZn) -- (TZn);

  \draw[dotted, shorten >= 10pt, shorten <= 10pt] (CZ1) -- (CZn);
  \draw[dotted, shorten >= 10pt, shorten <= 10pt] (TZ1.north west) -- (TZn.north east);
  \draw[dotted, shorten >= 5pt, shorten <= 5pt] (vz11) -- (vz12);
  \draw[dotted, shorten >= 5pt, shorten <= 5pt] (vzn1) -- (vzn2);

  \draw[-latex] (v12.west) to[out=180, in=-90] ($(TA.south)+(-1,0)$) to[out=90, in=180] (IA);
  \draw[-latex] (vn2.south west) to[out=225, in=-90] ($(TA.south)+(-1,0)$) to[out=90, in=180] (IA);

  \draw[-latex] (vz11.east) to[out=0, in=-90] ($(TZ.south)+(1,0)$) to[out=90, in=0] (IZ);
  \draw[-latex] (vzn1.south east) to[out=-45, in=-90] ($(TZ.south)+(1,0)$) to[out=90, in=0] (IZ);

  \coordinate (crosspoint) at ($(TAn.north)!0.5!(TZn.north)$);
  \draw[-latex] (v11.south east) to[out=-45, in=180] ($(vn1.south)+(.1,-.2)$) to[out=0, in=-135] (crosspoint) to[out=45, in=180] (IZ);
  \draw[-latex] (vn1.east) to[out=0, in=-135] (crosspoint) to[out=45, in=180] (IZ);

  \draw[-latex] (vz12.south west) to[out=-135, in=0] ($(vzn2.south)+(-.1,-.2)$) to[out=180, in=-45] (crosspoint) to[out=135, in=0] (IA);
  \draw[-latex] (vzn2.west) to[out=180, in=-45] (crosspoint) to[out=135, in=0] (IA);
\end{tikzpicture}
\end{center}
As an example, we will discuss the translation of booleans, pictured below.
\begin{center}
\begin{tikzpicture}[
  lnode/.style = {fill=gray!60, circle, inner sep = 1.5, minimum height=14, },
  ]
  \node[lnode, rounded rectangle] (bool) at (0, 0) {bool};
  \node[lnode, rounded rectangle, below left = 0.6 and 0.6 of bool] (set) {Set};
  \node[lnode, rounded rectangle, below = 0.6 of bool] (true) {true};
  \node[lnode, rounded rectangle, below right = 0.6 and 0.7 of bool] (false) {false};
  \node[lnode, below = 0.6 of $(true.east)!0.5!(false.west)$] (var) {$\uparrow$};

  \draw[->] (bool) -- (set);
  \draw[->>] (bool) -- (true);
  \draw[->>] (bool) -- (false);
  \draw[->] (true) -- (var);
  \draw[->] (false) -- (var);
  \draw[-latex] (var) to[out=0, in=0, looseness=2.5] (bool);
\end{tikzpicture}
\end{center}
Note that we once again take advantage of the bisimulation equivalence to merge
the variable nodes into a single node. The simplicity of the booleans illustrates
an important property of inductives: The graph structure of the constructors
$\texttt{true}$ and $\texttt{false}$ are identical. Their only distinguishing
factor is their node label. Without their label, the constructors would be
considered equal under the bisimulation relation, which would be incorrect due
to the disjointness property of constructors. The actual semantics of
$\texttt{true}$ and $\texttt{false}$ is not established in the definition of
$\texttt{bool}$ at all, but rather within the definitions of boolean operations
such as $\texttt{and}$ and $\texttt{or}$. Such symmetry breaking of otherwise identical
constructors happens quite frequently in practice. For the purpose of learning
the semantics of definitions from a graph representation, this leads to two
important lessons:
\begin{enumerate}
\item The name of a definition is important to take into account to
  distinguish two otherwise identical definition bodies. (The semantic
  information encoded in a name may also be used for learning purposes, but
  that is another matter.)
\item The semantic meaning of constructors and axioms is often encoded in
  secondary definitions. As such, in order to attain a full understanding of a
  mathematical theory, one cannot look at individual definitions in isolation.
\end{enumerate}

An example of translating a mutually inductive definition is $\texttt{even}$ and
$\texttt{odd}$ as defined above. In this translation, multiple parts of the
graph have been shared through the bisimulation equivalence. This is the case
for the sub-term $\mathbb{N} \to \text{Prop}$, which is the type of both
\texttt{even} and \texttt{odd}, has been shared. On the other hand, for presentational reasons, the
definitions $\mathbb{N}$ and $S$, have been duplicated in the graph (as is
allowed under bisimulation). Furthermore, these definitions are themselves also inductively
defined, but we omit the body of this inductive for the sake of brevity.
\begin{center}
\begin{tikzpicture}[
  lnode/.style = {fill=gray!60, circle, inner sep = 1.5, minimum height=14, },
  ]
  \node[lnode, rounded rectangle] (even) at (0, 0) {even};
  \node[lnode, rounded rectangle, right = 4 of even, anchor=center] (odd) {odd};

  \node[lnode, rounded rectangle, below = 0.8 of even.center, anchor=center] (eo) {eo};
  \node[lnode, below = 0.8 of eo.center, anchor=center] (eforall1) {$\forall$};
  \node[lnode, below left = 0.8 and 0.5 of eforall1.center, anchor=center] (enat) {$\mathbb{N}$};
  \node[lnode, below right = 0.8 and 0.5 of eforall1.center, anchor=center] (eforall2) {$\forall$};
  \node[lnode, below left = 0.8 and 0.5 of eforall2.center, anchor=center] (eapp1) {$@$};
  \node[lnode, below left = 0.9 and 0.5 of eapp1.center, anchor=center] (ev1) {$\uparrow$};
  \node[lnode, below right = 0.8 and 0.5 of eforall2.center, anchor=center] (eapp2) {$@$};
  \node[lnode, below left = 0.9 and 0.7 of eapp2.center, anchor=center] (ev2) {$\uparrow$};
  \node[lnode, below right = 0.9 and 0.7 of eapp2.center, anchor=center] (eapp3) {$@$};
  \node[lnode, below left = 0.9 and 0.5 of eapp3.center, anchor=center] (eS) {$S$};
  \node[lnode, below right = 0.9 and 0.5 of eapp3.center, anchor=center] (ev3) {$\uparrow$};

  \draw[->>] (eapp1) -- (ev3);
  \draw[->>] (eapp3) -- (ev3);
  \draw[->] (eapp3) -- (eS);
  \draw[->] (eapp2) -- (ev2);
  \draw[->>] (eapp2) -- (eapp3);
  \draw[->] (eapp1) -- (ev1);
  \draw[-latex] (ev3) to[out=90, in=-25] (eforall1);
  \draw[->] (eforall1) -- (enat);
  \draw[->>] (eforall1) -- (eforall2);
  \draw[->] (eforall2) -- (eapp1);
  \draw[->>] (eforall2) -- (eapp2);
  \draw[->] (eo) -- (eforall1);
  \draw[->>] (even) -- (eo);

  \node[lnode, rounded rectangle, below = 0.8 of odd.center, anchor=center] (oe) {oe};
  \node[lnode, below = 0.8 of oe.center, anchor=center] (oforall1) {$\forall$};
  \node[lnode, below left = 0.8 and 0.5 of oforall1.center, anchor=center] (onat) {$\mathbb{N}$};
  \node[lnode, below right = 0.8 and 0.5 of oforall1.center, anchor=center] (oforall2) {$\forall$};
  \node[lnode, below left = 0.8 and 0.5 of oforall2.center, anchor=center] (oapp1) {$@$};
  \node[lnode, below left = 0.9 and 0.5 of oapp1.center, anchor=center] (ov1) {$\uparrow$};
  \node[lnode, below right = 0.8 and 0.5 of oforall2.center, anchor=center] (oapp2) {$@$};
  \node[lnode, below left = 0.9 and 0.7 of oapp2.center, anchor=center] (ov2) {$\uparrow$};
  \node[lnode, below right = 0.9 and 0.7 of oapp2.center, anchor=center] (oapp3) {$@$};
  \node[lnode, below left = 0.9 and 0.5 of oapp3.center, anchor=center] (oS) {$S$};
  \node[lnode, below right = 0.9 and 0.5 of oapp3.center, anchor=center] (ov3) {$\uparrow$};

  \draw[->>] (oapp1) -- (ov3);
  \draw[->>] (oapp3) -- (ov3);
  \draw[->] (oapp3) -- (oS);
  \draw[->] (oapp2) -- (ov2);
  \draw[->>] (oapp2) -- (oapp3);
  \draw[->] (oapp1) -- (ov1);
  \draw[-latex] (ov3) to[out=90, in=-25] (oforall1);
  \draw[->] (oforall1) -- (onat);
  \draw[->>] (oforall1) -- (oforall2);
  \draw[->] (oforall2) -- (oapp1);
  \draw[->>] (oforall2) -- (oapp2);
  \draw[->] (oe) -- (oforall1);
  \draw[->>] (odd) -- (oe);

  \node[lnode, below right = 0.8 and 2.2 of even.center, anchor=center] (tforall) {$\forall$};
  \node[lnode, below left = 0.8 and 0.5 of tforall.center, anchor=center] (tnat) {$\mathbb{N}$};
  \node[lnode, rounded rectangle, below right = 0.8 and 0.5 of tforall.center, anchor=center] (prop) {Prop};

  \draw[->] (even) -- (tforall);
  \draw[->] (odd) -- (tforall);
  \draw[->] (tforall) -- (tnat);
  \draw[->>] (tforall) -- (prop);

  \draw[-latex] (ov1) to[out=145, in=-55] (even);
  \draw[-latex, looseness=1.2] (ov2) to[out=-135, in=-135] (odd);

  \draw[-latex, looseness=2] (ev1) to[out=105, in=165] (odd);
  \draw[-latex, looseness=1.2] (ev2) to[out=-135, in=-135] (even);

  \node[lnode, rounded rectangle, below left = 0.8 and 2.2 of even.center, anchor=center] (ez) {ez};
  \node[lnode, below = 0.8 of ez.center, anchor=center] (zapp) {$@$};
  \node[lnode, below left = 0.8 and 0.5 of zapp.center, anchor=center] (zv) {$\uparrow$};
  \node[lnode, below right = 0.8 and 0.5 of zapp.center, anchor=center] (zz) {$0$};

  \draw[->>] (even) -- (ez);
  \draw[->] (ez) -- (zapp);
  \draw[->] (zapp) -- (zv);
  \draw[->>] (zapp) -- (zz);
  \draw[-latex, looseness=1.5] (zv) to[out=100, in=180] (even);
\end{tikzpicture}
\end{center}

We must also discuss one particular special case of inductive types, namely
records with primitive projections. A record is a simple, non-recursive
inductive that can conveniently be described as follows:
\begin{minted}{coq}
Record I : T := C { P1 : T1; ... ; Pn : Tn }.
\end{minted}
A record is simply syntactic sugar for the following inductive with a single
constructor, together with a set of projections $P_i$ that extract data from the
record.
\begin{minted}{coq}
Inductive I : T := C : T1 -> ... -> Tn -> I.
\end{minted}
The projections of an inductive can specified to be \textit{primitive}, which
means that they are not ordinary projections defined using pattern
matching, but a new type of definition supported specially by the kernel for
efficiency purposes. The translation of such records proceeds identically to how
the inductive would normally be generated, with the addition of nodes corresponding to
these projections:
\begin{center}
\begin{tikzpicture}[
  lnode/.style = {fill=gray!60, circle, inner sep = 1.5, minimum height=14, },
  ]
  \node[lnode] (I) at (0, 0) {$I$};
  \node[lnode, below = 0.8 of I.center] (C) {$C$};

  \node[lnode, right = 1 of C] (at1) {$\forall$};
  \node[regular polygon, regular polygon sides=3, fill=gray!60, inner sep = -1.5,
  below left = 0.5 and 0 of at1.south, anchor=north] (Tp1) {$T_{1G}$};
  \node[lnode, above = 0.8 of at1.center, rounded rectangle] (p1) {$P_1$};

  \node[lnode, right = 1 of at1] (at2) {$\forall$};
  \node[regular polygon, regular polygon sides=3, fill=gray!60, inner sep = -1.5,
  below left = 0.5 and 0 of at2.south, anchor=north] (Tp2) {$T_{nG}$};
  \node[lnode, above = 0.8 of at2.center, rounded rectangle] (p2) {$P_n$};

  \node[lnode, right = 1 of at2] (v) {$\uparrow$};

  \node[regular polygon, regular polygon sides=3, fill=gray!60, inner sep = 0,
  below left = 0.8 and 1 of IA.center, anchor=north] (T) {$T_{G}$};

  \draw[->] (I) -- (T.north);
  \draw[->>] (I) -- (C);
  \draw[->] (at1) -- (Tp1.north);

  \draw[->] (C) -- (at1);
  \draw[->>, dotted] (at1) -- (at2);
  \draw[->] (at2) -- (Tp2.north);

  \draw[->] (p1) -- (at1);
  \draw[->] (p2) -- (at2);
  \draw[->>] (at2) -- (v);

  \draw[-latex] (v.north) to[out=90, in=-25] ($(p2)+(.4,.4)$) to[out=155, in=25] (I.north east);
\end{tikzpicture}
\end{center}

\subsection{Case Analysis and Fixpoints}
The final features to make the Calculus of Inductive Constructors complete are
case analysis and (co-)fixpoints. These constructs allow for the recursive
deconstruction of the data encoded using (co-)inductive datatypes. Case analysis
on a datatype is done using the following syntactic template in Coq.
\begin{minted}{coq}
match U as x in (I a1 .. an) return T with
| C1 y11 .. y1n => V1
| ...
| Cn yn1 .. ynn => Vn
end
\end{minted}
The return clause $T$ is an elimination clause that specifies the type of the
returned terms $V_i$. These types may be dependent on the variable $x$ and the
annotations $a_i$ of the inductive type $I$ that is being matched. Similarly,
the returned terms $V_i$ may depend on the variables $y_{ij}$ that are being
matched. In order to simplify the surface area of this syntactic construct, this
is translated to the following equivalent syntax.
\begin{minted}{coq}
match U return (fun x a1 .. an => T) with
| C1 => (fun y11 .. y1n => V1)
| ...
| Cn => (fun yn1 .. ynn => Vn)
end
\end{minted}
This allows the following much simpler blueprint for case analysis, which can be straightforwardly
translated.
\begin{minted}{coq}
match U return T with
| C_1 => V_1
| ...
| C_n => V_n
end
\end{minted}
\begin{center}
\begin{tikzpicture}[
  lnode/.style = {fill=gray!60, circle, inner sep = 1.5, minimum height=14, },
  ]
  \node[lnode, rounded rectangle] (case) {case};
  \node[regular polygon, regular polygon sides=3, fill=gray!60, inner sep = 0,
  below left = 0.8 and 3 of case.center, anchor=north] (U) {$U_{G}$};
  \node[regular polygon, regular polygon sides=3, fill=gray!60, inner sep = 0,
  below left = 0.8 and 1.5 of case.center, anchor=north] (T) {$T_{G}$};
  \node[lnode, rounded rectangle, below = 0.8 of case.center, anchor=north] (branch1) {branch};
  \node[lnode, rounded rectangle, below right = 0.8 and 2 of case.center, anchor=north] (branch2) {branch};
  \node[regular polygon, regular polygon sides=3, fill=gray!60, inner sep = 0,
  below right = 0.8 and 0.5 of branch1.center, anchor=north] (V1) {$V_{1G}$};
  \node[regular polygon, regular polygon sides=3, fill=gray!60, inner sep = 0,
  below right = 0.8 and 0.5 of branch2.center, anchor=north] (V2) {$V_{nG}$};
  \node[lnode, below left = 0.8 and 0.5 of branch1.center, anchor=north] (C1) {$C_1$};
  \node[lnode, below left = 0.8 and 0.5 of branch2.center, anchor=north] (C2) {$C_n$};
  \node[lnode, below right = 0.8 and 3.5 of case.center, anchor=north] (I) {$I$};

  \draw[->] (case) -- (U.north);
  \draw[->>] (case) -- (T.north);
  \draw[->>>] (case) -- (branch1);
  \draw[->>>] (case) -- (branch2);
  \draw[->] (branch1) -- (C1);
  \draw[->] (branch2) -- (C2);
  \draw[->>] (branch1) -- (V1.north);
  \draw[->>] (branch2) -- (V2.north);
  \draw[->>>>] (case) -- (I);
  \draw[dotted, shorten >= 5pt, shorten <= 5pt] (branch1.east) -- (branch2.west);
\end{tikzpicture}
\end{center}

The syntax of (co-)fixpoints is similarly complex. The expression of a fixpoint
can consist of a set of mutually recursive functions, with an arbitrary number
of parameters. Each function has a parameter that is marked as
\textit{structurally decreasing} to ensure the consistency of the calculus.
Finally, the $\texttt{for fi}$ clause selects one function $f_i$ to be the
representative for the expression.
\begin{minted}{coq}
fix f1 (p11 : P11) ... (p1n : P1n) {struct P1x} : T1 := U1
...
with fi (pi1 : Pi1) ... (pin : Pin) {struct Piy} : Ti := Ui
...
with fn (pn1 : Pn1) ... (pnn : Pnn) {struct Pnz} : Tn := Un
for fi
\end{minted}
To simplify this syntax, we move the parameters of the fixpoints into their body
using ordinary $\lambda$-abstraction. It should be noted that this translation
loses information on which parameter is structurally decreasing. However, in nearly all
cases, this can be reconstructed, unless there are multiple valid
choices. Different choices can lead to changes in the reduction behavior of the
fixpoint, and hence this translation loses a marginal amount of information.
\begin{minted}{coq}
fix f1 : P11 -> ... -> P1n -> T1 := fun p11 ... p1n => U1
...
with fi : Pi1 -> ... -> Pin -> Ti := fun pi1 ... pin => Ui
...
with fn : Pn1 -> ... -> Pnn -> Tn := fun pn1 ... pnn => Un
for fi
\end{minted}
The final, simplified blueprint for fixpoints takes the following shape.
\begin{minted}{coq}
fix f1 : T1 := U1
...
with fi : Ti := Ui
...
with fn : Tn := Un
for fi
\end{minted}
The bodies $U_i$ of the functions may reference themselves through free
variables. As such, each $U_i$ is assumed to be translated as follows.
\begin{center}
  \begin{tikzpicture}[
    lnode/.style = {fill=gray!60, circle, inner sep = 1.5, minimum height=14, },
    ]
    \node[regular polygon, regular polygon sides=3,
    fill=gray!60, inner sep = 8] (U) at (0,0) {$U_{iG}'$};
    \node[above=0.1 of U.north] (text) {$U_{i}$};
    \node[lnode, fill=gray!60, draw, dashed,
    below right= -0.1 and 1.15 of U.south, anchor=north] (v1) {$v_{f_n}$};
    \node[lnode, fill=gray!60, draw, dashed,
    below right= -0.1 and 0 of U.south, anchor=north] (vi) {$v_{f_i}$};
    \node[lnode, fill=gray!60, draw, dashed,
    below left= -0.1 and 1.15 of U.south, anchor=north] (v2) {$v_{f_1}$};
    \draw[dotted, shorten >= 5pt, shorten <= 5pt] (v1.west) -- (vi.east);
    \draw[dotted, shorten >= 5pt, shorten <= 5pt] (v2.east) -- (vi.west);
  \end{tikzpicture}
\end{center}
The final gadget for a fixpoint follows the familiar pattern of ``tying the
knot''. Co-fixpoints are translated analogously, such that only the labels of
the fixpoint nodes differ.
\begin{center}
\begin{tikzpicture}[
  lnode/.style = {fill=gray!60, circle, inner sep = 1.5, minimum height=14, },
  ]
  \node[lnode, rounded rectangle] (fix) {fix};
  \node[lnode, rounded rectangle, below left = 0.5 and 3.5 of fix] (fixfun1) {fixf};
  \node[lnode, rounded rectangle, below = 0.5 of fix] (fixfuni) {fixf};
  \node[lnode, rounded rectangle, below right = 0.5 and 3.5 of fix] (fixfunn) {fixf};
  \node[regular polygon, regular polygon sides=3, fill=gray!60, inner sep = -1.5,
  below left = 0.5 and 0.8 of fixfun1.south, anchor=north] (T1) {$T_{1G}$};
  \node[regular polygon, regular polygon sides=3, fill=gray!60, inner sep = -1.5,
  below left = 0.5 and 0.8 of fixfuni.south, anchor=north] (Ti) {$T_{iG}$};
  \node[regular polygon, regular polygon sides=3, fill=gray!60, inner sep = -1.5,
  below left = 0.5 and 0.8 of fixfunn.south, anchor=north] (Tn) {$T_{nG}$};

  \node[regular polygon, regular polygon sides=3,
  fill=gray!60, inner sep = 5, below right = 0.5 and 0.5 of fixfun1.south, anchor = north] (U1) {$U_{1G}'$};
  \node[lnode, fill=gray!60,
  below right= -0.1 and 1 of U1.south, anchor=north] (v11) {$\uparrow$};
  \node[lnode, fill=gray!60,
  below right= -0.1 and 0 of U1.south, anchor=north] (v1i) {$\uparrow$};
  \node[lnode, fill=gray!60,
  below left= -0.1 and 1 of U1.south, anchor=north] (v1n) {$\uparrow$};
  \draw[dotted, shorten >= 5pt, shorten <= 5pt] (v11.west) -- (v1i.east);
  \draw[dotted, shorten >= 5pt, shorten <= 5pt] (v1n.east) -- (v1i.west);

  \node[regular polygon, regular polygon sides=3,
  fill=gray!60, inner sep = 5.8, below right = 0.5 and 0.5 of fixfuni.south, anchor = north] (Ui) {$U_{iG}'$};
  \node[lnode, fill=gray!60,
  below right= -0.1 and 1 of Ui.south, anchor=north] (vi1) {$\uparrow$};
  \node[lnode, fill=gray!60,
  below right= -0.1 and 0 of Ui.south, anchor=north] (vii) {$\uparrow$};
  \node[lnode, fill=gray!60,
  below left= -0.1 and 1 of Ui.south, anchor=north] (vin) {$\uparrow$};
  \draw[dotted, shorten >= 5pt, shorten <= 5pt] (vi1.west) -- (vii.east);
  \draw[dotted, shorten >= 5pt, shorten <= 5pt] (vin.east) -- (vii.west);

  \node[regular polygon, regular polygon sides=3,
  fill=gray!60, inner sep = 5, below right = 0.5 and 0.5 of fixfunn.south, anchor = north] (Un) {$U_{nG}'$};
  \node[lnode, fill=gray!60,
  below right= -0.1 and 1 of Un.south, anchor=north] (vn1) {$\uparrow$};
  \node[lnode, fill=gray!60,
  below right= -0.1 and 0 of Un.south, anchor=north] (vni) {$\uparrow$};
  \node[lnode, fill=gray!60,
  below left= -0.1 and 1 of Un.south, anchor=north] (vnn) {$\uparrow$};
  \draw[dotted, shorten >= 5pt, shorten <= 5pt] (vn1.west) -- (vni.east);
  \draw[dotted, shorten >= 5pt, shorten <= 5pt] (vnn.east) -- (vni.west);

  \draw[->] (fix) -- (fixfun1);
  \draw[->>] (fix.south east) -- (fixfuni.north east);
  \draw[->] (fix.south west) -- (fixfuni.north west);
  \draw[->] (fix) -- (fixfunn);

  \draw[->] (fixfun1) -- (T1.north);
  \draw[->] (fixfuni) -- (Ti.north);
  \draw[->] (fixfunn) -- (Tn.north);

  \draw[->>] (fixfun1) -- (U1.north);
  \draw[->>] (fixfuni) -- (Ui.north);
  \draw[->>] (fixfunn) -- (Un.north);

  \coordinate (crosspoint1) at ($(U1.south west)+(-0.5,-0.7)$);
  \draw[-latex] (crosspoint1) to[out=180, in=180] (fixfun1);
  \draw[rounded corners] (v1n.south) -- (crosspoint1-|v1n.south) -- (crosspoint1);
  \draw[rounded corners] (vin.south) -- (crosspoint1-|vin.south) -- (crosspoint1);
  \draw[rounded corners] (vnn.south) -- (crosspoint1-|vnn.south) -- (crosspoint1);

  \coordinate (crosspoint2) at ($(Un.south east)+(0.4,-0.9)$);
  \draw[-latex] (crosspoint2) to[out=0, in=0] (fixfunn);
  \draw[rounded corners] (v11.south) -- (crosspoint2-|v11.south) -- (crosspoint2);
  \draw[rounded corners] (vi1.south) -- (crosspoint2-|vi1.south) -- (crosspoint2);
  \draw[rounded corners] (vn1.south) -- (crosspoint2-|vn1.south) -- (crosspoint2);

  \coordinate (crosspoint3) at ($(U1.south east)!0.4!(Ui.south west)+(0, -1.2)$);
  \draw[-latex] (crosspoint3) to[out=90, in=180] (fixfuni);
  \draw[rounded corners] (v1i.south) -- (crosspoint3-|v1i.south) -- (crosspoint3);
  \draw[rounded corners] (vii.south) -- (crosspoint3-|vii.south) -- (crosspoint3);
  \draw[rounded corners] (vni.south) -- (crosspoint3-|vni.south) -- (crosspoint3);
\end{tikzpicture}
\end{center}

\subsection{Existential Variables}\label{sec:existential-variables}

The final crucial piece of Gallina's syntax is the ability to leave
\textit{holes} inside of terms. The purpose of a hole is to allow users to
incrementally build a term by iteratively refining holes with new terms that may
in turn contain holes themselves. The term is completed once no hole remains.
This refinement process is often performed using a sequence of \textit{tactics}.
Details about tactic-based proofs can be found in
Section~\ref{sec:tactic-representation}. In this section, we are concerned with
the low-level machinery that underlies a tactical proof.

A term with a hole is usually inputted by the user using an underscore \_. For
example, the term $\lambda x : T.\ \_$ represents a function whose body has not been
filled in yet. Before such a term enters the kernel, the hole must be properly
typed. This is done by turning it into an existential variable $?e$. An
existential variable is associated with a typing $\Gamma\vdash U$. The contract
for such a typing is that $?e$ can later be substituted for any term $T$ such
that $\Gamma\vdash T:U$ type-checks. When an existential variable occurs in a
term, it is associated with a list of substitutions that resolve the context
$\Gamma$ associated with the variable. For example, once passed through the
kernel, $\lambda x : T,\ \_$ will take the shape $\lambda x : T,\ ?e\{h \coloneq
x\}$ where $e$ has the associated typing $h : T \vdash U$ and where $U$ is an
arbitrary term that may even be an existential variable itself.

The reason why holes receive names during type-checking is that a term may
contain the same variable several times, each time potentially with a different
substitution list. As a contrived example, consider the term
\[ \text{let}\ f : (\forall y : \mathbb{B},\ \_) \coloneq \_\ \text{in}\
  \text{@conj}\ \_\ \_\ (f\ \top)\ (f\ \bot).\]
This term posits the existence of a function $f$ that takes a boolean as input.
The output type of the function may be dependent on the input. Then, we
construct the conjunction of the output of $f$ when applied to both true and false.
Once type-checked and moved into the kernel, such a term would be represented
as follows. (The reader is encouraged to ask Coq to verify this elaboration.)
\[ \text{let}\ f : (\forall y : \mathbb{B},\ ?p\{x\coloneq y\}) \coloneq\ ?e\{\}\
  \text{in}\ \text{@conj}\ ?p\{x\coloneq\top\}\
  ?p\{x\coloneq\bot\}\ (f\ \top)\ (f\ \bot)\]
Here, the variable $?e$ receives the typing $\cdot\vdash\forall y:\mathbb{B},\
?p\{x\coloneq y\}$ and
$?p$ receives the typing $x:\mathbb{B}\vdash \text{Prop}$. Three of the four
holes have been converted into the same existential variable (with different
substitution lists) because otherwise the term would not type-check. This shows
that existential variables really refer to a named entity rather than just an
anonymous typing $\Gamma\vdash T$. We will write the named entity associated to
an existential variable $?e$ as $\Gamma\vdash_e T$. We usually refer to this as
a \textit{proof state}, because such a named typing represents the
current state of an interactive proof\footnote{This terminology is somewhat
  ambiguous because sometimes a proof state is also considered to be the set
of all named typings that are currently open.}. Every term has an (often implicit) ordered list
of proof states associated to it that may be referenced through existential
variables. Additionally, the typing associated to each proof state may in turn
contain existential variables that refer to proof states preceding it. In this
way, much like definitions, proof states form a directed acyclic graph.

To translate a proof state $\Gamma\vdash_e T$ into a graph, we need to know how
to translate the context $\Gamma$. The hypotheses within this context also form
a directed acyclic graph. Due to our wish for name-invariance, we cannot label
the nodes that represent a hypothesis with its name. On the other hand, having
fully anonymous hypotheses is also troublesome. Consider the typing $x :
\mathbb{N}, y : \mathbb{N} \vdash x = y$. Using a naive, anonymous translation
of the context, this would be bisimulation equivalent to $x :
\mathbb{N} \vdash x = x$ because both hypotheses are identical. However, only
the latter typing can be inhabited, which means that these typings should not be
equivalent.

A pragmatic solution to this issue is to find a middle ground between the
nameless and named representation of hypotheses. Instead of labeling hypothesis
nodes using their name, we label them by their position index within the context.\footnote{This
is still not fully satisfactory, however, because the ordering of hypotheses in
the context is often somewhat arbitrary. A more faithful graph-based
representation of proof states remains an open problem.} This semi-anonymous representation
leads to the following gadget for proof states.

\begin{center}
\begin{tikzpicture}[
  lnode/.style = {fill=gray!60, circle, inner sep = 1.5, minimum height=14, },
  ]
  \node[regular polygon, regular polygon sides=3,
  fill=gray!60, inner sep = -2.5] (T) at (0,0) {$T_{i+1_G}'$};
  \node[above=0.1 of T.north] (text) {$T_{i+1}$};
  \node[lnode, fill=gray!60, draw, dashed,
  below right= -0.1 and 0.6 of T.south, anchor=north] (vi) {$v_{h_i}$};
  \node[lnode, fill=gray!60, draw, dashed,
  below left= -0.1 and 0.6 of T.south, anchor=north] (v1) {$v_{h_1}$};
  \draw[dotted, shorten >= 5pt, shorten <= 5pt] (v1.east) -- (vi.west);

  \node[regular polygon, regular polygon sides=3,
  fill=gray!60, inner sep = 3, below = 1.5 of T.south] (U) {$U_{G}'$};
  \node[above=0.1 of U.north] (text2) {$U$};
  \node[lnode, fill=gray!60, draw, dashed,
  below right= -0.1 and 0.6 of U.south, anchor=north] (vi) {$v_{h_n}$};
  \node[lnode, fill=gray!60, draw, dashed,
  below left= -0.1 and 0.6 of U.south, anchor=north] (v1) {$v_{h_1}$};
  \draw[dotted, shorten >= 5pt, shorten <= 5pt] (v1.east) -- (vi.west);
\end{tikzpicture}
\qquad\qquad
\begin{tikzpicture}[
  lnode/.style = {fill=gray!60, circle, inner sep = 1.5, minimum height=14, },
  ]
  \node[lnode, rounded rectangle] (ps) at (0, 0) {proof state $e$};
  \node[above=0.1 of ps.north, align=center] (lambda-text) {$h_1 : T_1, \ldots, h_n : T_n \vdash_e U$};
  \node[regular polygon, regular polygon sides=3, fill=gray!60, inner sep = 0,
  below = 0.4 of ps] (U) {$U'_G$};
  \node[lnode, rounded rectangle, below left = 0.6 and 1.5 of U] (h1) {hyp 1};
  \node[regular polygon, regular polygon sides=3, fill=gray!60, inner sep = -1,
  below = 0.4 of h1] (T1) {$T_{1G}$};
  \node[lnode, rounded rectangle, below = 0.6 of U] (h2) {hyp 2};
  \node[regular polygon, regular polygon sides=3, fill=gray!60, inner sep = -1,
  below = 0.4 of h2] (T2) {$T'_{2G}$};
  \node[lnode, rounded rectangle, below right = 0.6 and 1.5 of U] (hn) {hyp n};
  \node[regular polygon, regular polygon sides=3, fill=gray!60, inner sep = -1,
  below = 0.4 of hn] (Tn) {$T'_{nG}$};

  \draw[->] (ps) -- (U);
  \draw[->] (h1) -- (T1);
  \draw[->] (h2) -- (T2);
  \draw[->] (hn) -- (Tn);

  \coordinate (crosspoint) at ($(T2.south west)+(0, -0.4)$);
  \draw[->, dashed] (crosspoint) to[out=180, in=-90] ($(T1.south east)!0.5!(T2.south west)$) to[out=90, in=-45] (h1);
  \draw[dashed, rounded corners=10] (T2.south) -- (crosspoint-|T2.south) -- (crosspoint);
  \draw[dashed, rounded corners=10] ($(Tn.south)+(-0.3, 0)$) -- ($(Tn.south)+(-0.3, -0.4)$) -- (crosspoint);
  \draw[dashed, rounded corners=10] (Tn.south) -- ($(Tn.south)+(0, -0.8)$) -- ($(Tn.south west)+(0, -0.8)$);
  \draw[->, dashed] ($(Tn.south west)+(0, -0.8)$) to[out=180, in=-90] ($(T2.south east)!0.5!(Tn.south west)$) to[out=90, in=-45] (h2);
  \draw[dashed] ($(Tn.south)+(0.3, 0)$) -- ($(Tn.south)+(0.3, -0.4)$);

  \draw[dotted, shorten >= 10pt, shorten <= 10pt] (h2.east) -- (hn.west);

  \draw[->, dashed] ($(U.south)+(-0.3, 0)$) to[out=-90, in=45] (h1);
  \draw[->, dashed] ($(U.south)+(0, 0)$) to[out=-90, in=90] (h2);
  \draw[->, dashed] ($(U.south)+(0.3, 0)$) to[out=-90, in=135] (hn);

  \draw[->>] (ps) to[out=-135, in=90] (h1);
  \draw[->>] (ps) to[out=-45, in=90] (hn);
  \draw[->>] (ps) to[out=-135, in=135, looseness=2] (h2);
\end{tikzpicture}
\end{center}

Every type $T_{i+1}$ of a hypothesis is assumed to contain up to $i$ free
variables. These variables are resolved to the node that represents the
appropriate hypothesis. The edges that point towards a hypothesis are drawn
dashed because the label on those edges depends on their source node, which is
not drawn in the gadget.

Now that we can translate proof states, translating existential variables
becomes a matter of dealing with the substitution list. In the gadget below, we
assume that the proof state $e$ has already been appropriately translated and
that the hypothesis nodes ``belong'' to the proof state $e$ as one would expect.

\begin{center}
\begin{tikzpicture}[
  lnode/.style = {fill=gray!60, circle, inner sep = 1.5, minimum height=14,},
  ]
  \node[lnode, rounded rectangle] (evar) at (0, 0) {evar};
  \node[above=0.1 of evar.north, align=center] (text) {$?e\{h_1 \coloneq T_1, \ldots, h_n \coloneq T_n\}$};
  \node[lnode, rounded rectangle, below left = 0.6 and 1.5 of evar, anchor=north] (ps) {proof state $e$};
  \node[lnode, below = 0.6 of evar, anchor=north] (s1) {$\coloneq$};
  \node[lnode, below right = 0.6 and 2 of evar, anchor=north] (sn) {$\coloneq$};
  \node[lnode, rounded rectangle, below left = 0.6 and 0.1 of s1] (h1) {hyp 1};
  \node[lnode, rounded rectangle, below left = 0.6 and 0.1 of sn] (hn) {hyp n};
  \node[regular polygon, regular polygon sides=3, fill=gray!60, inner sep = -1,
  below right = 0.6 and 0.3 of s1, anchor = north] (T1) {$T_{1G}$};
  \node[regular polygon, regular polygon sides=3, fill=gray!60, inner sep = -1,
  below right = 0.6 and 0.3 of sn, anchor = north] (Tn) {$T_{nG}$};

  \draw[->] (evar) -- (ps);
  \draw[->>] (evar) -- (s1);
  \draw[->>] (evar) -- (sn);
  \draw[->] (s1) -- (h1);
  \draw[->] (sn) -- (hn);
  \draw[->>] (s1) -- (T1.north);
  \draw[->>] (sn) -- (Tn.north);

  \node[regular polygon, regular polygon sides=3, fill=gray!60, inner sep = 4,
  below = 0 of ps, anchor = north] (Tps) {};
  \node[regular polygon, regular polygon sides=3, fill=gray!60, inner sep = 4,
  below = 0 of h1, anchor = north] (Tps) {};
  \node[regular polygon, regular polygon sides=3, fill=gray!60, inner sep = 4,
  below = 0 of hn, anchor = north] (Tps) {};
\end{tikzpicture}
\end{center}

To demonstrate the translation of proof states and existential variables, we
translate the example term from earlier in this section. The definition of
\textit{conj} is omitted from the translation from brevity. Note that all
subterms are fully shared under the bisimulation relation in the graph below.
\[ \text{let}\ f : (\forall y : \mathbb{B},\ ?p\{x\coloneq y\}) \coloneq\ ?e\{\}\
  \text{in}\ \text{@conj}\ ?p\{x\coloneq\top\}\
  ?p\{x\coloneq\bot\}\ (f\ \top)\ (f\ \bot)\]
\begin{center}
\begin{tikzpicture}[
  lnode/.style = {fill=gray!60, circle, inner sep = 1.5, minimum height=14,},
  ]
  \node[lnode, rounded rectangle] (let) at (0, 0) {let};
  \node[lnode, rounded rectangle, below left = 0.8 and 0 of let.center, anchor=center] (ev1) {evar};
  \node[lnode, rounded rectangle, below left = 0.8 and 0 of ev1.center, anchor=center] (pse) {proof state $e$};
  \node[lnode, below left = 0.8 and 0.8 of pse.center, anchor=north] (forall1) {$\forall$};
  \node[lnode, rounded rectangle, below right = 0.8 and 0.8 of forall1.center, anchor=center] (ev2) {evar};
  \node[lnode, rounded rectangle, below right = 1 and 0.8 of ev2.center, anchor=center] (psp) {proof state $p$};
  \node[lnode, rounded rectangle, below left = 1 and 0.8 of psp.center, anchor=center] (prop) {Prop};
  \node[lnode, below left = 1 and 0.8 of ev2.center, anchor=center] (s1) {$\coloneq$};
  \node[lnode, below left = 1 and 0.4 of s1.center, anchor=center] (v1) {$\uparrow$};
  \node[lnode, rounded rectangle, below right = 1 and 0.8 of psp.center, anchor=center] (h1) {hyp 1};

  \node[lnode, right = 0.5 of psp.east, anchor=center] (s2) {$\coloneq$};
  \node[lnode, rounded rectangle, above = 1 of s2.center, anchor=center] (ev3) {evar};
  \node[lnode, above = 0.8 of ev3.center, anchor=center] (app1) {$@$};
  \node[lnode, rounded rectangle, left = 1.2 of app1.center, anchor=center] (conj) {conj};
  \node[lnode, right = 1 of app1.center, anchor=center] (app2) {$@$};
  \node[lnode, rounded rectangle, below = 0.8 of app2.center, anchor=center] (ev4) {evar};
  \node[lnode, right = 1.3 of app2.center, anchor=center] (app3) {$@$};
  \node[lnode, right = 1 of app3.center, anchor=center] (app4) {$@$};
  \node[lnode, below = 1 of ev4.center, anchor=center] (s3) {$\coloneq$};
  \node[lnode, below = 0.8 of app3.center, anchor=center] (app5) {$@$};
  \node[lnode, below = 0.8 of app4.center, anchor=center] (app6) {$@$};
  \node[lnode, below left = 1 and 0.5 of app5.center, anchor=center] (v2) {$\uparrow$};

  \node[lnode, below = 0.8 of h1.center, anchor=center] (bool) {$\mathbb{B}$};
  \node[lnode, right = 1.4 of bool.center, anchor=center] (true) {$\top$};
  \node[lnode, right = 1.4 of true.center, anchor=center] (false) {$\bot$};
  \node[lnode, rounded rectangle, below = 1 of bool.center, anchor=center] (set) {Set};
  \node[lnode, below = 1 of true.center, anchor=center] (v3) {$\uparrow$};

  \node[regular polygon, regular polygon sides=3, fill=gray!60, inner sep = 4,
  below = 0 of conj, anchor = north] (conjt) {};

  \draw[->] (let) to[out=-145, in=135] (forall1);
  \draw[->>] (let) -- (ev1);
  \draw[->>] (ev1) -- (pse);
  \draw[->] (pse) -- (forall1);
  \draw[->>] (forall1) -- (ev2);
  \draw[->] (ev2) -- (psp);
  \draw[->>] (ev2) -- (s1);
  \draw[->] (psp) -- (prop);
  \draw[->>] (psp) -- (h1);
  \draw[->>] (s1) -- (v1);
  \draw[-latex] (v1) to[out=100, in=-105] (forall1);

  \draw[->] (s1) -- (h1);
  \draw[->] (s2) -- (h1);
  \draw[->] (s3) -- (h1);

  \draw[->] (app1) -- (conj);
  \draw[->] (app2) -- (app1);
  \draw[->] (app3) -- (app2);
  \draw[->] (app4) -- (app3);

  \draw[->>] (app1) -- (ev3);
  \draw[->>] (app2) -- (ev4);
  \draw[->>] (app3) -- (app5);
  \draw[->>] (app4) -- (app6);
  \draw[->] (ev3) -- (psp);
  \draw[->] (ev4) -- (psp);
  \draw[->>] (ev3) -- (s2);
  \draw[->>] (ev4) -- (s3);

  \draw[->] (app5) -- (v2);
  \draw[->] (app6) -- (v2);
  \draw[-latex, looseness=1.5] (v2) to[out=90, in=-20] (let);
  \draw[->>>] (let) to[out=0, in=90] (app4);

  \draw[->>] (s2) -- (true);
  \draw[->>] (app5) to[out=-45, in=45] (true);
  \draw[->>] (s3) -- (false);
  \draw[->>] (app6) to[out=-45, in=45] (false);
  \draw[->, looseness=2] (forall1) to[out=-125, in=180] (bool);

  \draw[->] (bool) -- (set);
  \draw[->>] (bool) -- (true);
  \draw[->>] (bool) to[out=-25, in=-155] (false);
  \draw[->] (true) -- (v3);
  \draw[->] (false) -- (v3);
  \draw[-latex] (v3) -- (bool);
  \draw[->] (h1) -- (bool);
\end{tikzpicture}
\end{center}

\subsection{The Meta-graph}\label{sec:meta-graph}
The preceding sections exhaustively describe every aspect of kernel-level
Gallina terms (modulo some of the stated limitations). However, outside of the kernel,
proof developments in the Coq Proof Assistant contain a large amount of
additional subtleties. As discussed in Section~\ref{sec:design-decisions}, we
do not consider most of this extra-kernel information relevant for our
translation, with two important exceptions. One is tactic-based proofs,
discussed in \ref{sec:tactic-representation} and the other is a treatment of the
global context we discuss in this section.

We consider the global context to be the collection of definitions (see
Section~\ref{sec:definitions}) that are loaded and available for use in Coq at
any given time. This collection is dependent on the state of the current Coq
document and which other documents are loaded (\texttt{Require}d) as it evolves
over time either through compilation or interactive exploration. In order to
synthesize new proofs, a learning agent needs a proper understanding of the
mathematical knowledge that is available. How one constructs a new proof may
depend heavily on the set of
lemmas may already be available for use in the proof. As such, when extracting a dataset from
a mathematical development (see Section~\ref{sec:datasets}), this needs to
contain a copy of the global context at each time-point during the compilation
of the development.

Including a completely separate copy of the global context at each time point
would be inefficient. Instead, each definition contains a pointer to the
definition that chronologically precedes it, such that the global context can be
recovered by following this sequence of pointers to the end. In addition,
definition $d_{i+1}$ in a file may contain a pointer to the
\textit{representative} definition of other files that have been
loaded after the declaration of definition $d_i$. A files
representative is the definition whose associated global context coincides
exactly with the global context that becomes available after loading that file.

These pointers that form the global context induce a directed acyclic graph on
definitions that we call the \textit{meta-graph}, as it encodes information that
is not part of Coq's core calculus but rather a representation of abstract time.
The formal contract specified on the global context of a definition $d$ induced
by the pointers described above is that any definition that is transitively
reachable through the body of $d$ must also be part of the global context of
$d$. (Note that if $d$ is part of a cluster of mutually inductive definitions,
that cluster is considered part of the global context.) In addition, there is the
expectation that definitions are ordered within the global context according to
their respective order within the source Coq document.

Although at first approximation the global context grows linearly as definitions
are added, things are complicated once Coqs section mechanism and module
system come into play. Both these features
allow for a form of abstraction by
parameterizing a theory over a fixed set of hypotheses or over another theory.
The act of opening and closing a section or instantiating a module functor
causes abstract time to ``branch''. The effects of these features on the
meta-graph are as follows.

\paragraph{Sections} Within an open section, one or more section hypotheses may be
declared. Coq treats such hypotheses in a dual matter. On one hand, it is
possible to see a hypothesis as an axiom that can be referred to within the
body of new definitions. On the other hand, when in tactic proof mode, section
hypotheses are treated similarly to how ordinary hypotheses in a proof are
treated.\footnote{This dual treatment of section hypotheses is known to cause
  many problems. See \url{https://github.com/coq/coq/issues/6254}.}
For the sake of translating to a single, uniform graph representation such a
dual representation is not acceptable. As such, we always translate section
hypotheses as an axiom (with an annotation that marks it as coming from a
section).

When a section is closed, the global context is backtracked up to the point
where the section was opened. All definitions occurring within the section are
then replayed, but now with the section hypotheses ``discharged'' into their
body. This causes the global context to branch into two. One branch corresponds to
the section while it is open. It is abandoned after the section is closed. The second branch
contains the updated definitions that include the discharged hypotheses.

It is important to note that this causes a kind of almost-duplication of data in
datasets in the sense that the definitions in the two branches can be
mechanically derived from each other. This can be rather dangerous from a
machine learning perspective because extra care has to be taken while generating
a training-testing-validation split of the data to avoid data
leakage.\footnote{The authors have had the misfortune of accidentally creating a data-split
  with leakage on multiple occasions, which was only discovered much later. In
  general, creating a fully randomized split of the data is highly
  discouraged. Instead, one should choose a split such that it is obvious that
  none of the training  data depends on the testing data.}
To highlight the relationship between the
two branches in the meta-graph, every derived definition contains a pointer to
the original definition it was derived from.

To illustrate all this, we given a simple example of a Coq document and the
meta-graph that is derived from such a document.
\begin{minted}{coq}
Definition X := ...
Require F.
Section S.
Hypothesis H1 : ... Hypothesis H2 : ...
Definition D1 : ... Definition D2 : ...
End S.
Definition Y := ...
\end{minted}

\begin{center}
\begin{tikzpicture}[
  lnode/.style = {fill=gray!60, circle, inner sep = 1.5, minimum height=14,},
  ]
  \node[lnode, rounded rectangle] (X) at (0, 0) {$X$};
  \node[lnode, rounded rectangle, below = 0.5 of X] (F1) {File $F$};
  \node[lnode, rounded rectangle, right = 1 of X] (H1) {$H_1$};
  \node[lnode, rounded rectangle, right = 1 of H1] (H2) {$H_2$};
  \node[lnode, rounded rectangle, right = 1 of H2] (SD1) {$S.D_1$};
  \node[lnode, rounded rectangle, right = 1 of SD1] (SD2) {$S.D_2$};
  \node[lnode, rounded rectangle, below = 0.5 of SD1] (D1) {$D_1$};
  \node[lnode, rounded rectangle, below = 0.5 of SD2] (D2) {$D_2$};
  \node[lnode, rounded rectangle, right = 1 of D2] (Y) {$Y$};

  \draw[<-] (X) -- (H1);
  \draw[<-] (F1) -- (H1);
  \draw[<-] (H1) -- (H2);
  \draw[<-] (H2) -- (SD1);
  \draw[<-] (SD1) -- (SD2);

  \draw[<-] (X) to[out=-25, in=180] (D1);
  \draw[<-] (D1) -- (D2);
  \draw[<-] (F1) -- (D1);
  \draw[<-] (D2) -- (Y);

  \draw[->>] (D1) -- (SD1);
  \draw[->>] (D2) -- (SD2);
\end{tikzpicture}
\end{center}
In this meta-graph, for example, the global context of $S.D_2$ consists of
$S.D_1$, $H_2$, $H_1$, $X$ and the global context of the representative of file
$F$. Definition $D_2$, which is derived from $S.D_2$, has a global context
consisting of $D_1$, $X$ and file $F$. The hypotheses $H_1$ and $H_2$ have been
discharged into $D_1$ and $D_2$.

\paragraph{Modules} A module $\mathcal{F}$ may depend on a module type $\mathcal{T}$ that specifies the
signature of a theory. In such a case, we refer to $\mathcal{F}$ as a module functor. If
one has a module $\mathcal{M}$ that implements the signature of $\mathcal{T}$, then one may
instantiate $\mathcal{F}$ using $\mathcal{M}$, written as
$\mathcal{F}(\mathcal{M})$. Both the creation of the module functor
$\mathcal{F}$ and its instantiation cause a branch in abstract time. While
$\mathcal{F}$ is being (interactively) created, the signatures specified by
$\mathcal{T}$ are placed in the global context as a collection of axioms.
Any definitions that are part of $\mathcal{F}$ may depend on these axioms. Once
$\mathcal{F}$ is finished and closed, all definitions related to both
$\mathcal{T}$ and $\mathcal{F}$ are backtracked, and their related branch is
abandoned. Then, when the instantiation $\mathcal{F}(\mathcal{M})$ is made, a
copy of all definitions in $\mathcal{F}$ is made, where all references to
signatures of $\mathcal{T}$ are substituted with the corresponding concrete
implementations $\mathcal{M}$.

Similar to the treatment of sections, to clearly indicate which definitions are
near-duplicates of each other, the meta-graph includes a pointer from any
definition belonging to an instantiated functor to the corresponding definition
in the original declaration of the functor. Note that the potential for
data duplication in modules is much larger than in sections, because a single
functor may be instantiated many times. Each time this happens, a complete copy
of the functor is created. Additionally, the instantiation of a functor may happen
far away from the creation of a functor, potentially even in a completely new
project. As such, extreme care must be taken when dealing with data coming from
modules.

An example of the meta-graph in the presence of modules is provided below. Three
different branches are created, for the module type $T$, the functor $F$, and the
modules $M$ and $R$. Cross-references are made between the derived parameter
$U.P$ of $T.P$ and the derived definition $R.D$ of $F.D$. Note that $M.P$ is not
considered derived from $T.P$, because its declaration does not mention $T$ in
any way.
\begin{minted}{coq}
Definition X := ...
Module Type T.   Parameter  P :  ... End T.
Module F(U : T). Definition D := ... End F.
Module M.        Definition P := ... End M.
Module R := F(M).
Definition Y := ...
\end{minted}
\begin{center}
\begin{tikzpicture}[
  lnode/.style = {fill=gray!60, circle, inner sep = 1.5, minimum height=14,},
  ]
  \node[lnode, rounded rectangle] (X) at (0, 0) {$X$};
  \node[lnode, rounded rectangle, right = 1 of X] (XP) {$U.P$};
  \node[lnode, rounded rectangle, above = 0.5 of XP] (TP) {$T.P$};
  \node[lnode, rounded rectangle, right = 1 of XP] (FD) {$F.D$};
  \node[lnode, rounded rectangle, below = 0.5 of XP] (MP) {$M.P$};
  \node[lnode, rounded rectangle, below = 0.5 of FD] (RD) {$R.D$};
  \node[lnode, rounded rectangle, right = 1 of RD] (Y) {$Y$};

  \draw[<-] (X) -- (TP);
  \draw[<-] (X) -- (XP);
  \draw[<-] (X) -- (MP);
  \draw[<-] (XP) -- (FD);
  \draw[<-] (MP) -- (RD);
  \draw[<-] (RD) -- (Y);

  \draw[->>] (XP) -- (TP);
  \draw[->>] (RD) -- (FD);
\end{tikzpicture}
\end{center}

\subsection{Graph Sharing}\label{sec:graph-sharing}
While providing the gadgets and examples for Gallina terms in the preceding
sections, we have taken advantage of the fact that graphs are taken modulo bisimulation to
de-duplicate or duplicate parts of a graph for presentational purposes. In
reality, we wish to share as many sub-terms as possible across the entire web of
definitions and proofs. By treating the entire universe of mathematics defined
within Coq as one unified graph, we can share subterms not only within a single
definition but across the entire universe. We do this by calculating a hash for
each node in the graph that represents its \textit{identity}. The identity is
designed such that two hashes are equal if and only if their corresponding nodes
are bisimilar (modulo hash collisions). We will discuss some potential use cases
for this identity information.
\begin{itemize}
\item By presenting a fully shared graph to a learning model, that model can
  take advantage of the additional semantic information that is provided by 
  the shared graph. For example, consider a proof state $h : T \vdash T$, where
  $T$ is a large term. A model tasked with learning how to complete such a proof
  state should clearly immediately invoke the assumption $h$. However, given
  that $T$ is large, such a decision may be non-trivial to make. On the other
  hand, if the proof state is presented as a graph such that the two occurrences
  of $T$ are shared, this becomes a trivial decision. In this case, the model does not even
  need to care about the particulars of $T$.
\item Sharing subterms across a large dataset allows for massive compression of
  the physical representation of the graph. For example, sharing subterms in
  a dataset extracted from Coq's standard library allows for an 88\% reduction in
  the number of nodes. As such, in practice, it pays very well to perform such
  de-duplication. Note that in order to ensure the modularity of a dataset, we
  only allow the sharing of graphs between files that already depend on each
  other.\footnote{However, the existence of the identity hash of nodes makes it
    easy to share terms even more aggressively as a post-processing step.}
\item Identity information can be directly used as a quick summary of a
  subgraph. For example, this can be useful for creating an oracle. If one has a
  large dataset of proof states and corresponding actions, the hashes of those
  proof states can be compared to a query proof state to quickly find whether
  this query has already been answered before.
\end{itemize}

The algorithm for calculating the identity hash of a node in $O(n\log n)$ time
is described in a separate publication~\cite{sharing-paper}. Here, we suffice with the
statement that the hash respects the bisimulation relation.

We calculate three variants of the identity hash, all with a subtly different purpose.
\begin{description}
  \item[Semantic] The semantic identity hash corresponds to the bisimulation
    relation on the graph induced by translating Gallina terms.
  \item[Meta] The meta-identity hash includes the meta-graph in addition to the
    graph obtained from Gallina terms. As such, two definitions that are
    semantically equal (and even have an equal name), but have a different
    associated global context still receive a different meta-identity.
  \item[Physical] The physical-identity hash of a definition node not only
    includes the meta-graph, but additionally includes the filename to which the
    definition belongs. This is the hash used to de-duplicate the graph in a
    dataset. Including the filename in the hash is important in the presence of
    diamond dependencies between files. Consider, for example, two unrelated
    files $X$ and $Y$ that both contain a definition $d$ with the same
    meta-identity. In a dataset, these definitions would not be de-duplicated
    because $X$ and $Y$ do not depend on one another. However, now consider a
    third file $Z$ that loads both $X$ and $Y$. If $Z$ references $d$ from file
    $X$, then the identity information would allow changing this reference to
    the identical definition in file $Y$. This way, two references that are
    intended to point to the same definition may end up pointing to separate
    definitions in different files, leading to subtle bugs. By including the
    filename in the hash, this problem is avoided.
\end{description}

\section{Tactic-based Proof Representations}\label{sec:tactic-representation}
In the Coq Proof Assistant, most proofs are not written by directly inputting
proof terms. Instead, the user writes tactic scripts. Tactics are meta-programs
that may analyze the current state of the open proof and advance the proof
towards qed. The actions take by tactics ranges from simple inference steps and
reduction strategies to decision procedures, domain-specific heuristics and
general purpose search procedures. In addition to built-in tactics, users may
write their own procedures either as an OCaml plugin or as a program written in
the one of various tactic programming languages such as
Ltac~\cite{DBLP:conf/lpar/Delahaye00}, Ltac2~\cite{pedrot2019ltac2},
Mtac~\cite{DBLP:journals/pacmpl/KaiserZKRD18} and Elpi~\cite{tassi2018elpi}.

Writing proofs through tactic scripts allows users to input potentially very
large proof terms using short scripts by exploiting the automation offered by
the various available tactics. Analyzing such scripts is interesting, for
example for a machine learning model. Just like end-users, ML models can take
advantage of the wealth of available automation such that they do not have to
generate low-level proof terms. As such, we are interested in representing tactic
scripts in a machine-readable format that is amenable to easy analysis and
machine learning purposes. We require the ability to analyze a tactic script and
the effect that individual tactics in the script have on the state of the proof.

Contrary to the highly structured calculus of constructions, tactic languages--in particular
Ltac--do not have very well-defined semantics. As a programming language, Ltac
is highly dynamic in nature and a precise operational semantics is nearly
impossible to define. Tactics may manipulate the state of the proof in
arbitrary ways. If desired, proof states may be added, refined, backtracked,
reordered, and arbitrarily modified with little regard to any kind of structured
proof editing or even the correctness of the proof term that is being
generated\footnote{Coq will still type-check the generated proof at qed-time.}.
As such, a fully faithful machine-readable representation of how a proof evolves
during the execution of a tactic script is close to unachievable. In
the next section, we will describe a simplified, approximate model of how
tactics are executed by the proof engine.

\subsection{A Simplified Model of the Tactic Engine}\label{sec:simplified-tactic-model}
The execution of a tactic script is determined by the semantics of the tactic
programming language and the proof representation that is being manipulated by
the tactics. Here, we will take a simplified view of both components.

\paragraph{Proof Representation} The entire state of the current proof is
represented as a set of open proof states, which are described in
Section~\ref{sec:existential-variables}.\footnote{Note the potential confusion
  in terminology: A single proof state is only one piece of the entire state,
  which may encompass many proof states.} Under the hood, a partial proof term
is maintained that may contain existential variables that reference these open
proof states. Tactics may refine proof states by providing a new term that will
be substituted for the existential variables pointing to that proof state. This
term may itself contain existential variables that reference new proof states, which
will be added to the set of open proof states. The proof is complete once the
set of open proof states is empty, and hence the proof term no longer contains
existential variables.

Proof states may themselves also contain existential variables, in which case
the proof state is dependent on another proof state. Refining a proof state
often also causes its dependency to be refined or solved entirely as a
side-effect through unification. For this reason, Coq splits the set of open
proof states into \textit{shelved} and \textit{unshelved} sets. The shelved
states are expected to be solved entirely as a side-effect through unification,
while the unshelved states are expected to be the target of explicit refinement
by tactics. This dichotomy is only a heuristic. In practice, proof states can be
moved freely onto the shelf and off the shelf. For example, shelved proof
states that were not solved by side-effect may need to be moved off the shelf to
be solved explicitly.

In our simplified model, we assume that the heuristic employed by the shelf is
indeed accurate:
\begin{itemize}
  \item Unshelved proof states are never a dependency of another proof state,
    and will be refined only through direct manipulation, never through a
    side-effect. This guarantees the existence of a tree
    between all the unshelved proof states that exist over time, where the edges
    are formed by tactics that refine a proof state into subsequent states.
    The violation of this assumption will cause the tree to be
    broken, causing ``dangling'' proof states that were never explicitly
    refined by a tactic.
    In practice, this assumption is rarely broken. In our extracted dataset
    (see section~\ref{sec:datasets}), we encountered only 76 violations out of 4.6
    million proof state transitions recorded.
  \item Shelved proof states are fully solved as a side-effect. When a proof
    state remains unsolved, it needs to be moved off the shelf and solved explicitly,
    causing it to ``spontaneously'' appear in the directed acyclic graph of
    unshelved states. Violations of this assumption happen more often, although
    they are still rare.\footnote{It is not possible to measure the number of
      violations in datasets, because the ``spontaneous'' appearance of unshelved proof
      states are indistinguishable from the set of initial proof states when a
      new proof is started.}
  \end{itemize}

\paragraph{Semantics of Tactics} A tactic is executed on a subset of the
unshelved proof states, called the \textit{focus}. The focus of a tactic is
determined through some form of structured proof editing, either through proof
bullets, because it was inherited from the previous tactic, or through manual
selection by the user. We assume that tactics only refine the focused proof
states, even though technically speaking any proof state can be manipulated.
(Shelved proof states may still be refined through side effects.) Furthermore,
we assume that the refinements made by tactics are valid and that every new
proof state is indeed created through a refinement of another proof state,
creating a clear parent-child relationship between states. Finally, we assume
that the actions performed by tactics on a state in the focus are independent from
the other states in the focus. In this way, the purpose of having multiple
states focused at once is to run a tactic in parallel that could be
sequentialized. Note that in the presence of side effects, the order of the
sequentialization is important.

The Ltac tactic engine is a fully-fledged programming language, including
binders, prolog-style backtracking, pattern matching, and proof matching, that
can be used to write complex proof scripts and even decision procedures that
automatically solve classes of proof states. When representing
tactical proofs in a dataset for machine learning purposes, we are generally not
interested in such advanced concepts. Rather, we view a tactic script as a
series of commands, where each command may require some parameters. We take this
view, because we are interested in exploiting already-existing decision
procedures, but not learning to write new procedures. Attempting to synthesize
complex tactic procedures would likely be more involved than directly
synthesizing the proof term that would be generated by such a procedure. This
has the following consequences:
\begin{enumerate}
\item Tactic scripts that contain syntactic sugar, or other shorthands that
  shorten the script should be expanded and decomposed. For example, the tactic
  \texttt{rewrite plus\_n\_O, app\_nil\_r}, that rewrites using two lemmas
  simultaneously, should rather be split into two separate rewrites
  \texttt{rewrite plus\_n\_O; rewrite app\_nil\_r}. In exchange for increasing
  the size of the proof script by a small constant factor, this ensures one
  receives early feedback from the kernel while synthesizing a proof.
  Additionally, it reduces
  the number of tokens that have to be correctly synthesized at once, and
  reduces families of tactics that differ only in their number of
  parameters to a single tactic. Tactic decomposition is described in
  Section~\ref{sec:static-tactic-decomposition}.
\item Some tactical proofs involve binders, pattern matching, and goal matching
  to act as ad-hoc heuristic automation that is only meant to prove a single
  (usually long and repetitive) theorem. Such scripts are often a main loop that
  dynamically determines the next tactic to execute based on the shape of the
  focused proof state. Because synthesizing such a loop dynamic loop is
  difficult, we prefer to record the trace of tactics executed by the loop.
  Decomposing such ad-hoc dynamic scripts is described in
  Section~\ref{sec:dynamic-tactic-decomposition}.
\item The need for nameless representations of CIC terms as described in
  Section~\ref{sec:design-decisions} also extends to tactics. Tactics often
  introduce hypotheses in the local context of a proof state, and those
  hypotheses can be named. For scripts written by humans, it is generally
  considered good practice to give hypotheses descriptive names. For example,
  while using the \texttt{intro} tactic, it is preferred to use the named
  variant \texttt{intro H} over the unnamed variant. However, because the
  graph-based representation of hypotheses does not have names, there is no
  advantage to synthesizing a tactic that invents a name. As such, tactics are
  post-processed to remove as many names as possible, as described in
  Section~\ref{sec:tactic-anonymization}.
\item Tactic scripts are executed in a logic
  monad~\cite{DBLP:conf/icfp/KiselyovSFS05} that allows for prolog-style
  backtracking. A single tactic may produce multiple results, and each of these
  results may be explored in combination with the results of other tactics in the
  script until a proof is found. For example, the \texttt{constructor} tactic
  will attempt to apply all constructors of an inductive datatype to a proof
  state and backtrack between them. When treating tactics as commands that can
  be executed sequentially during a proof search, such backtracking behavior may
  cause unexpected results. Therefore, any executed tactic expression
  containing an atomic tactic that employs backtracking is treated as a
  black box and will not be decomposed as described above.
\end{enumerate}

We will illustrate the model of tactical scripts outlined above with an example
in propositional logic from Coq's standard library.
\begin{minted}{coq}
Theorem and_iff_compat_l : forall A B C : Prop,
  (B <-> C) -> (A /\ B <-> A /\ C).
Proof.
  intros ? ? ? [Hl Hr]; split; intros [? ?];
  (split; [ assumption | ]); [apply Hl | apply Hr]; assumption.
Qed.
\end{minted}
The proof of this theorem consists of a single somewhat complex compound tactic
expression. Individual tactics in the expression are separated by semicolons,
which sequences two tactics together such that the goals generated by the first
tactic are in focus for the second tactic. Additionally, square brackets allow
running different tactics for each individual proof state in the focus (within
the brackets, the focus is temporarily restricted). This tactic
expression does not backtrack (it does not contain any backtracking primitives),
and hence is eligible to be decomposed. The only syntactic sugar present in the
expression are intro-patterns, which can be expanded into regular tactics:
\begin{minted}{coq}
intros ? ? ? [Hl Hr] --> intro; intro; intro; intro tmp;
                         destruct tmp as [Hl Hr]
intros [? ?]         --> intro tmp; destruct tmp
\end{minted}
Finally, as a post-processing step, we can anonymize the name-generating
tactics. This means converting \texttt{intro tmp} into \texttt{intro} and
\texttt{destruct tmp as [Hl Hr]} into \texttt{destruct tmp}. Note that this step
breaks the proof when viewing it from a text-based perspective because names
will now be mechanically generated by Coq. For example, the tactic \texttt{apply Hl}
will now fail, because \texttt{Hl} has been renamed. For this reason,
anonymization is a post-processing step that occurs after tactics have been
executed in Coq. This problem does not occur when viewing tactics as commands
that manipulate graph-based proof states, because the argument to \texttt{apply}
would simply be a pointer into the name-invariant graph.

After these transformations, the proof script can be graphically represented as
follows:

\begin{center}
  \begin{tikzpicture}[
    ps/.style = {fill=gray!60, circle, inner sep = 4},
    ]
    \node[ps] (ps1) at (0, 0) {};
    \node[ps, right = 0.4 of ps1.east] (ps2) {};
    \node[ps, right = 0.4 of ps2.east] (ps3) {};
    \node[ps, right = 0.4 of ps3.east] (ps4) {};
    \node[ps, right = 0.4 of ps4.east] (ps5) {};
    \node[ps, right = 0.4 of ps5.east] (ps6) {};
    \node[ps, below right = 0.25 and 0.5 of ps6.east] (ps7a) {};
    \node[ps, above right = 0.25 and 0.5 of ps6.east] (ps7b) {};
    \node[ps, right = 1.0 of ps7a.east] (ps8a) {};
    \node[ps, right = 1.0 of ps7b.east] (ps8b) {};
    \node[ps, right = 1.0 of ps8a.east] (ps9a) {};
    \node[ps, right = 1.0 of ps8b.east] (ps9b) {};
    \node[ps, right = 0.8 of ps9a.east] (ps10a) {};
    \node[ps, right = 0.8 of ps9b.east] (ps10b) {};
    \node[ps, above = 1.0 of ps10b.center] (ps10c) {};
    \node[ps, below = 1.0 of ps10a.center] (ps10d) {};
    \node[ps, right = 1.3 of ps10a.east] (ps11a) {};
    \node[ps, right = 1.3 of ps10b.east] (ps11b) {};

    \coordinate[right = 0.5 of ps10c.east] (qed1) {};
    \coordinate[right = 0.5 of ps10d.east] (qed2) {};
    \coordinate[right = 0.5 of ps11a.east] (qed3) {};
    \coordinate[right = 0.5 of ps11b.east] (qed4) {};

    \node[draw, rounded corners=9, fit=(ps1), label=above:\text{intro}] (t1) {};
    \node at (t1.center) {1};
    \node[draw, rounded corners=9, fit=(ps2), label=below:\text{intro}] (t2) {};
    \node at (t2.center) {2};
    \node[draw, rounded corners=9, fit=(ps3), label=above:\text{intro}] (t3) {};
    \node at (t3.center) {3};
    \node[draw, rounded corners=9, fit=(ps4), label=below:\text{intro}] (t4) {};
    \node at (t4.center) {4};
    \node[draw, rounded corners=9, fit=(ps5), label=above:\text{destruct H}] (t5) {};
    \node at (t5.center) {5};
    \node[draw, rounded corners=9, fit=(ps6), label=below:\text{split}] (t6) {};
    \node at (t6.center) {6};
    \node[draw, rounded corners=9, fit=(ps7a) (ps7b), label=\text{intro}] (t7) {};
    \node at (t7.center) {7};
    \node[draw, rounded corners=9, fit=(ps8a), label=below:\text{destruct H1}] (t8) {};
    \node at (t8.center) {8};
    \node[draw, rounded corners=9, fit=(ps8b), label=above:\text{destruct H1}] (t9) {};
    \node at (t9.center) {9};
    \node[draw, rounded corners=9, fit=(ps9a), label=below:\text{split}] (t10) {};
    \node at (t10.center) {10};
    \node[draw, rounded corners=9, fit=(ps9b), label=above:\text{split}] (t11) {};
    \node at (t11.center) {12};
    \node[draw, rounded corners=9, fit=(ps10c), label=above:\text{assumption}] (t12) {};
    \node at (t12.center) {13};
    \node[draw, rounded corners=9, fit=(ps10d), label=below:\text{assumption}] (t13) {};
    \node at (t13.center) {11};
    \node[draw, rounded corners=9, fit=(ps10a), label=below:\text{apply Hl}] (t14) {};
    \node at (t14.center) {14};
    \node[draw, rounded corners=9, fit=(ps10b), label=above:\text{apply Hr}] (t15) {};
    \node at (t15.center) {15};
    \node[draw, rounded corners=9, fit=(ps11a) (ps11b), label=\text{assumption}] (t16) {};
    \node at (t16.center) {16};

    \draw[-latex] (ps1) -- (ps2);
    \draw[-latex] (ps2) -- (ps3);
    \draw[-latex] (ps3) -- (ps4);
    \draw[-latex] (ps4) -- (ps5);
    \draw[-latex] (ps5) -- (ps6);
    \draw[-latex] (ps6) -- (ps7a);
    \draw[-latex] (ps6) -- (ps7b);
    \draw[-latex] (ps7a) -- (ps8a);
    \draw[-latex] (ps7b) -- (ps8b);
    \draw[-latex] (ps8a) -- (ps9a);
    \draw[-latex] (ps8b) -- (ps9b);
    \draw[-latex] (ps9a) -- (ps10a);
    \draw[-latex] (ps9b) -- (ps10b);
    \draw[-latex] (ps10a) -- (ps11a);
    \draw[-latex] (ps10b) -- (ps11b);
    \draw[-latex] (ps9a) to[out=-45, in=155] (ps10d);
    \draw[-latex] (ps9b) to[out=45, in=-155] (ps10c);

    \draw[-|] (ps10c.east) -- (qed1);
    \draw[-|] (ps10d.east) -- (qed2);
    \draw[-|] (ps11a.east) -- (qed3);
    \draw[-|] (ps11b.east) -- (qed4);
  \end{tikzpicture}
\end{center}
We represent a tactical proof as a sequence of \textit{proof steps}. Each proof
step corresponds to the execution of a single tactic on a set of focused proof
states. In the example, a proof state is represented as a grey dot. The
execution of a tactic is represented by encircling the focus on which the tactic
it executed. In each proof step, a
tactic may refine the proof states in its focus into other proof states by
generating a \textit{proof term}. We call such a refinement an \textit{outcome} of running
the tactic. An outcome can be represented by a 4-tuple $(\Delta\vdash_x B,
\texttt{tac}, U, \langle\Gamma_i\vdash_{y_i} A_i\rangle_{i\in[1\ldots n]})$, where \texttt{tac} refines proof state
$x$ into $n$ proof states $y_i$ by generating proof term $U$. Note that when $n
= 0$, that tactic has generated a proof term without holes, finishing the
current subtree of the proof.

The proof states and proof term are
represented in the data both in textual and graph format. Furthermore, textual
metadata about hypotheses is present, such as the name and textual
representation of a hypothesis. To represent a tactic in a way such that it can be
processed in the context of graphical representations, we split it into a
\textit{base tactic} and arguments. This happens even when the tactic is a
non-decomposable compound expression. For example, for \texttt{apply Hl}, the
base tactic is \texttt{apply \_}, which comes with a single argument \texttt{Hl}.
The base tactic is then treated as an opaque tactic identifier, while arguments
are interpreted as kernel terms that can be converted into graphs. These terms may reference
the hypotheses of the proof state on which it is executed. This also means
that the arguments of a tactic that is executed on multiple proof states
simultaneously may have different graph-based representations for each proof
state, because they may be interpreted under a different local context.
Arguments that are not terms, such as primitive integers, primitive strings, and
tactics (supplied to higher-order tacticals) are not currently represented as
arguments and remain part of the base tactic. In addition to a graph-friendly
representation, datasets also contain textual representations of the base tactic
and the full tactic, both anonymized and non-anonymized.

The sequence of proof steps is in the order in which tactics were originally
executed by the engine. Re-executing the tactics in that order will reproduce the same
proof. However, the refinement arrows also induce a more relaxed, partial order
on the proof steps. It is possible to execute the steps in any sequence that
respects the partial order and still obtain a proof, as long as no shelved proof
states are refined as a side-effect. Such cases can significantly ease the synthesis
of a proof, because one can then solve separate branches completely
independently. Ideally, one would also like to represent proofs with dependent (shelved)
proof states faithfully using a partial order between proof steps. Such a
representation remains future work.

Below, we show the graphical representation of the outcome tuple corresponding
to proof step 10 in the example. The starting proof state $x$ of this step is
textually represented as follows.
\begin{minted}{coq}
A, B, C : Prop
Hl : B -> C, Hr : C -> B
H1 : A, H2 : C
________________________
A /\ B
\end{minted}
This proof state is \texttt{split} into two states $y_1$ and $y_2$, with
conclusions $A$ and $B$ and the same hypotheses. This is done by generating the
proof term $\text{@conj}\ A\ B\ ?y_1\ ?y_2$. All this information can be
compactly represented as the following graph.

\begin{center}
\begin{tikzpicture}[
  lnode/.style = {fill=gray!60, circle, inner sep = 1.5, minimum height=14,},
  ]
  \node[lnode, rounded rectangle] (prop) at (0, 0) {Prop};
  \node[lnode, rounded rectangle, above = 0.8 of prop.center] (hypb) {hyp $B$};
  \node[lnode, rounded rectangle, left = 0.8 of hypb.center] (hypa) {hyp $A$};
  \node[lnode, rounded rectangle, right = 4 of hypb.center] (hypc) {hyp $C$};
  \node[lnode, above right = 0.8 and 0.9 of hypb.east, anchor=south] (forall1) {$\forall$};
  \node[lnode, above = 0.8 of hypc.center, anchor=south] (forall2) {$\forall$};
  \node[lnode, rounded rectangle, above = 0.8 of forall1.center] (hyphl) {hyp $H_l$};
  \node[lnode, rounded rectangle, above = 0.8 of forall2.center] (hyphr) {hyp $H_r$};
  \node[lnode, above = 0.8 of hypb.center] (app1) {$@$};
  \node[lnode, above = 0.8 of hypa.center] (app2) {$@$};
  \node[lnode, left = 1 of app2.center] (and) {$\wedge$};
  \node[lnode, rounded rectangle, above = 1.8 of app1.center] (psb) {proof state $x$};
  \node[lnode, rounded rectangle, left = 1.3 of psb.center] (psa1) {proof state $y_1$};
  \node[lnode, rounded rectangle, right = 1.3 of psb.center] (psa2) {proof state $y_2$};

  \node[lnode, rounded rectangle, above = 0.8 of psa2.center] (ev1) {evar};
  \node[lnode, rounded rectangle, above = 0.8 of psa1.center] (ev2) {evar};
  \node[lnode, above = 0.8 of ev1.center] (app3) {$@$};
  \node[lnode, above = 0.8 of ev2.center] (app4) {$@$};
  \node[lnode, left = 0.6 of app4.center] (app5) {$@$};
  \node[lnode, left = 0.6 of app5.center] (app6) {$@$};
  \node[lnode, rounded rectangle, left = 0.6 of app6.center] (conj) {conj};

  \node[lnode, rounded rectangle, left = 0.6 of and.center] (hyph1) {hyp $H_1$};
  \node[lnode, rounded rectangle, right = 0.6 of forall2.center] (hyph2) {hyp $H_2$};

  \node[lnode, left = 0.9 of hypa.center] (subst1) {$\coloneq$};
  \node[lnode, right = 0.9 of hypb.center] (subst2) {$\coloneq$};
  \node[lnode, left = 0.9 of hypc.center] (subst3) {$\coloneq$};
  \node[lnode, right = 0.9 of hyphl.center] (subst4) {$\coloneq$};
  \node[lnode, left = 0.9 of hyphr.center] (subst5) {$\coloneq$};
  \node[lnode, above = 0.8 of hyph1.center] (subst6) {$\coloneq$};
  \node[lnode, above = 0.8 of hyph2.center] (subst7) {$\coloneq$};

  \node[above = 0.6 of app3.center, label=above:\text{split}] (start) {$\text{@conj}\ A\ B\ ?y_1\ ?y_2$};

  \node[above = 0.6 of psb.center] (note1) {towards\ \ \textnode{hyp}\ \ nodes};
  \node[above = 0.4 of note1.center] (note2) {towards \textnode{$\coloneq$} nodes};

  \draw[->] (forall1) to[in=60, out=-160] (hypb);
  \draw[->>] (forall1) to[in=140, out=-20, looseness=0.9] (hypc);
  \draw[->>] (forall2) to[in=40, out=-155, looseness=0.9] (hypb);
  \draw[->] (forall2) -- (hypc);
  \draw[->] (hyphl) -- (forall1);
  \draw[->] (hyphr) -- (forall2);
  \draw[->] (hypa) -- (prop);
  \draw[->] (hypc) to[out=-90, in=10] (prop);
  \draw[->] (hypb) -- (prop);

  \draw[->] (psa1) to[out=-90, in=120] (hypa);
  \draw[->] (psa2) to[out=-165, in=60, looseness=1.3] (hypb);
  \draw[->] (psb) -- (app1);
  \draw[->] (app1) -- (app2);
  \draw[->] (app2) -- (and);
  \draw[->>] (app1) -- (hypb);
  \draw[->>] (app2) -- (hypa);

  \draw[->] (app3) -- (app4);
  \draw[->] (app4) -- (app5);
  \draw[->] (app5) -- (app6);
  \draw[->] (app6) -- (conj);
  \draw[->>] (app3) -- (ev1);
  \draw[->>] (app4) -- (ev2);
  \draw[->>] (app5) to[out=-110, in=160, looseness=1.9] ($(app2)+(0, 1)$) to[out=-20, in=130] (hypb);
  \draw[->>] (app6) to[out=-110, in=-150, looseness=1.5] (hypa);
  \draw[->] (ev1) -- (psa2);
  \draw[->] (ev2) -- (psa1);

  \draw[->] (hyph1) -- (hypa);
  \draw[->] (hyph2) -- (hypc);

  \draw[->] (subst1) to[out=25, in=170] (hypa);
  \draw[->>] (subst1) to[out=-25, in=-170] (hypa);
  \draw[->] (subst2) to[out=155, in=10] (hypb);
  \draw[->>] (subst2) to[out=-155, in=-10] (hypb);
  \draw[->] (subst3) to[out=25, in=170] (hypc);
  \draw[->>] (subst3) to[out=-25, in=-170] (hypc);
  \draw[->] (subst4) to[out=155, in=10] (hyphl);
  \draw[->>] (subst4) to[out=-155, in=-10] (hyphl);
  \draw[->] (subst5) to[out=25, in=170] (hyphr);
  \draw[->>] (subst5) to[out=-25, in=-170] (hyphr);
  \draw[->] (subst6) to[out=-65, in=65] (hyph1);
  \draw[->>] (subst6) to[out=-115, in=115] (hyph1);
  \draw[->] (subst7) to[out=-65, in=65] (hyph2);
  \draw[->>] (subst7) to[out=-115, in=115] (hyph2);

  \draw[->>] (psa1) -- (note1.south west);
  \draw[->>] (psa2) -- (note1.south east);
  \draw[->>] (psb) -- (note1);
  \draw[->>] (ev1) -- (note2.east);
  \draw[->>] (ev2) -- (note2.west);

  \draw[->] (start) -- (app3);

\end{tikzpicture}
\end{center}

In this graph, to avoid clutter, we draw the arrows from the proof states to the hypotheses they
consist of implicitly. Because the hypotheses remain constant
among the proof states, they can all be efficiently shared between them. Only
the conclusions differ. Note that each evar node is connected to a substitution node
($\coloneq$) for each hypothesis. These
nodes represent the ``paper trail'' that allows for tracing the origin of a
hypothesis. In this example, this paper trail is trivial, because the hypotheses
remain constant. For example, the substitution node encodes that hypothesis
$H_1$ of proof state $y_1$ can be traced back to hypothesis $H_1$ of proof state
$x$, which happens to be physically the same node.

To illustrate the purpose of the paper trail left by the substitution nodes, we
give a rather contrived example of a proof of $\text{False}\to\text{True}$. We
prove this theorem, by first running \texttt{intro}, converting the proof state
into $H : \text{False}\vdash \text{True}$. Then, the \texttt{contradiction}
tactic recognizes that we can derive any term from $H$, including one of type
$\text{True}$. This entire proof sequence can be represented as the graph below.

\begin{center}
\hspace*{-25pt}
\begin{tikzpicture}[
  lnode/.style = {fill=gray!60, circle, inner sep = 1.5, minimum height=14,},
  ]
  \node[lnode, draw=red] (lambda) at (0, 0) {$\lambda$};
  \node[lnode, draw=red, rounded rectangle, below right = 0.8 and 0.5 of lambda.center] (ev) {evar};
  \node[lnode, draw=red, rounded rectangle, below left = 0.8 and 0.5 of ev.center] (subst) {$\coloneq$};
  \node[lnode, rounded rectangle, right = 0.7 of subst.center] (psa) {proof state $y$};
  \node[lnode, rounded rectangle, left = 0.7 of subst.center] (psb) {proof state $x$};
  \node[lnode, below = 0.8 of psb.center] (forall) {$\forall$};
  \node[lnode, below = 0.8 of forall.center] (false) {$\bot$};
  \node[lnode, rounded rectangle, below right = 0.8 and 0.5 of subst.center] (hyp) {hyp $H$};
  \node[lnode, draw=red, rounded rectangle, below left = 0.8 and 0.5 of subst.center] (v) {$\uparrow$};
  \node[lnode, below = 0.8 of hyp.center] (true) {$\top$};

  \node[lnode, draw=blue, rounded rectangle, right = 2.0 of lambda.center] (false-ind) {$\bot$-ind};
  \node[lnode, draw=blue, right = 0.8 of false-ind.center] (app1) {$@$};
  \node[lnode, draw=blue, right = 0.5 of app1.center] (app2) {$@$};

  \draw[->, red] (lambda) to[out=-170, in=135, looseness=1.6] (false);
  \draw[->>, red] (lambda) -- (ev);
  \draw[->, red] (ev) -- (psa);
  \draw[->>, red] (ev) -- (subst);
  \draw[->] (psb) -- (forall);
  \draw[->] (forall) -- (false);
  \draw[->>] (forall) -- (true);
  \draw[->>] (psa) -- (hyp);
  \draw[->, red] (subst) -- (hyp);
  \draw[->>, red] (subst) -- (v);
  \draw[-latex, red] (v) to[out=90, in=-100] (lambda);
  \draw[->] (hyp) -- (false);
  \draw[->] (psa) to[out=-60, in=30] (true);

  \draw[->, blue] (app1) -- (false-ind);
  \draw[->, blue] (app2) -- (app1);
  \draw[->>, blue] (app1) to[out=-90,in=0] (true);
  \draw[->>, blue] (app2) to[out=-90,in=0] (hyp);

  \node[above = 0.5 of lambda.center, label=above:\texttt{intro}] (start1) {$\lambda f:\text{False},\ ?y\{H\coloneq f\}$};
  \draw[->] (start1) -- (lambda);

  \node[above = 0.5 of app2.center, label=above:\texttt{contradiction}] (start2) {$\text{False\_ind}\ \text{True}\ H$};
  \draw[->] (start2) -- (app2);

  \node[lnode, draw=blue, rounded rectangle, right = 4.0 of psa.center] (m-false-ind) {$\bot$-ind};
  \node[lnode, draw=blue, right = 0.8 of m-false-ind.center] (m-app1) {$@$};
  \node[lnode, draw=blue, right = 0.5 of m-app1.center] (m-app2) {$@$};
  \node[lnode, draw=red] (m-lambda) at (app2-|m-app1) {$\lambda$};
  \node[lnode, below left = 0.8 and 0.5 of m-lambda.center] (m-bot) {$\bot$};
  \node[lnode, below = 0.8 of m-app1.center] (m-top) {$\top$};
  \node[lnode, draw=red, below = 0.8 of m-app2.center] (m-v) {$\uparrow$};

  \draw[->, red] (m-lambda) -- (m-bot);
  \draw[->>, red] (m-lambda) -- (m-app2);
  \draw[->, blue] (m-app2) -- (m-app1);
  \draw[->, blue] (m-app1) -- (m-false-ind);
  \draw[->>, blue] (m-app1) -- (m-top);
  \draw[->>, blue] (m-app2) -- (m-v);
  \draw[-latex, red] (m-v) to[out=45, in=-45] (m-lambda);

  \draw[-latex, line width = 4, shorten >= 15, shorten <= 35] (psa) -- (m-false-ind);

  \node[above = 0.5 of m-lambda.center, label=above:\text{final}] (start3) {proof term};
  \draw[->] (start3) -- (m-lambda);

\end{tikzpicture}
\end{center}

The sequence starts with proof state $x$, corresponding to
$\cdot\vdash\text{False}\to\text{True}$. Then, \texttt{intro} generates the term
$\lambda f:\text{False},\ ?y\{H\coloneq f\}$ (displayed in red), which comes
with a new proof term $y$. Then, \texttt{contradiction} will generate the proof
term $\text{False\_ind}\ \text{True}\ H$ (displayed in blue), which references
the hypothesis $H$. At the end of the proof, Coq's kernel will compose these two
proof steps into a single term, displayed on the right.

Such a composition can
be performed rather easily in graph-based form with the help of substitution
nodes. One simply starts with the proof term generated by the first tactic,
and locates all evar nodes and their corresponding proof states, in this case
only $y$. Then, one locates the proof term that solves $y$, in this case
$\text{False\_ind}\ \text{True}\ H$. The root of this term is substituted into
the location of the existential variable. Finally, one needs to resolve the
hypothesis $H$ that is referenced. This is done by changing any pointer to the
hypothesis to the term indicated by the corresponding substitution node.

In this way, as long as no dependent existential variables exist, one can
mechanically obtain the proof term corresponding to a tactical proof sequence.

\subsection{Tactic Instrumentation\todo{sync with code}}

To record the proof step that is the result of running a tactic, namely a list
of 4-tuples $(\Delta\vdash_x B, \texttt{tac}, U, \langle\Gamma_i\vdash_{y_i}
A_i\rangle_{i\in[1\ldots n]})$, a special instrumentation tactical
\texttt{record tac} was created. This tactical takes \texttt{tac} as an argument
and behaves exactly as that tactic, with the side-effect that the proof states
before and after execution of \texttt{tac} are stored in a database, as well as
the generated proof term $U$.

As motivated in Section~\ref{sec:simplified-tactic-model}, we do not wish to
simply record the outcomes of each top-level tactic expression. That would be too imprecise
because tactic expressions often consist of many tactics glued together by
the composition operator, $\texttt{tac1 ; tac2}$, and other tacticals.
We recursively insert our
recording tactic into these composed expressions, as
illustrated by the ``instr'' transformation below.
\begin{align*}
  \text{instr}(\texttt{t1; t2}) &= \text{instr}(\texttt{t1})\texttt{; }\text{instr}(\texttt{t2}) \\
  \text{instr}(\texttt{t1; [t2 | t3 | ..]}) &= \text{instr}(\texttt{t1})\texttt{; [}\text{instr}(\texttt{t2})\texttt{ | }\text{instr}(\texttt{t3})\texttt{ | ..]} \\
  \text{instr}(\texttt{only n: t}) &= \texttt{only n: } \text{instr}(\texttt{t}) \\
  \text{instr}(\texttt{t1 + t2}) &= \text{instr}(\texttt{t1})\texttt{ + }\text{instr}(\texttt{t2}) \\
  \text{instr}(\texttt{first [t1 | t2 | ..]}) &= \texttt{first [}\text{instr}(\texttt{t1})\texttt{ | }\text{instr}(\texttt{t2})\texttt{ | ..]} \\
  \text{instr}(\texttt{t1 || t2}) &= \text{instr}(\texttt{t1})\texttt{ || }\text{instr}(\texttt{t2}) \\
  \text{instr}(\texttt{solve [t]}) &= \texttt{solve [} \text{instr}(\texttt{t})\texttt{]} \\
  \text{instr}(\texttt{tryif t1 then t2 else t3}) &= \texttt{tryif } \text{instr}(\texttt{t1})\texttt{ then } \text{instr}(\texttt{t2}) \\
                                &\phantom{=} \texttt{ else }\text{instr}(\texttt{t3})\\
  \text{instr}(\texttt{once t}) &= \texttt{once } \text{instr}(\texttt{t}) \\
  \text{instr}(\texttt{try t}) &= \texttt{try } \text{instr}(\texttt{t}) \\
  \text{instr}(\texttt{progress t}) &= \texttt{progress } \text{instr}(\texttt{t}) \\
  \text{instr}(\texttt{do n t}) &= \texttt{do n } \text{instr}(\texttt{t}) \\
  \text{instr}(\texttt{repeat t}) &= \texttt{repeat } \text{instr}(\texttt{t}) \\
  \text{instr}(\texttt{time t}) &= \texttt{time } \text{instr}(\texttt{t}) \\
  \text{instr}(\texttt{timeout n t}) &= \texttt{timeout n } \text{instr}(\texttt{t}) \\
  \text{instr}(\texttt{t}) &= \texttt{record t} \text{ (when \texttt{t} is atomic)}
\end{align*}

Although the tacticals listed above differ widely in behavior, all of those
behaviors can be strictly described in terms of the arguments they receive.
The tacticals may execute their arguments in a particular order, multiplicity,
with custom backtracking behavior, or with a modified focus, but otherwise do not
modify the state of the proof. As such, it is possible to instrument only the
execution of the arguments, and not the
tacticals themselves while still obtaining a valid proof tree. We illustrate
this with some examples:
\begin{itemize}
\item Restricting the focus of a tactic, for example through (\texttt{only 1: t}),
  which executes \texttt{t} on the first proof state in the focus, is an
  implicit operation in the proof tree. No semantic information is lost by
  omitting it.
\item The \texttt{try t} tactical executes \texttt{t} and absorbs any error that
  \texttt{t} might generate. When an error occurs, the expression behaves as a
  no-op. If \texttt{t} fails, no trace of its execution will be found within the
  proof tree. As such, some information about the nature of the proof script
  execution is lost. However, the proof tree still represents a correct proof
  that is more succinct. If an inappropriate attempt was made to execute a
  tactic, the proof tree does not need to record this.
\item The \texttt{first [t1 | t2]} tactic behaves similar to \texttt{try}, but
  will execute \texttt{t2} when \texttt{t1} fails. In such an event, only the
  execution of the successful \texttt{t2} tactic can be found in the proof tree.
  This holds for all backtracking primitives: A correct proof tree only needs to
  record the successful portion of a proof search.
\item Sequentially executing \texttt{t} in a loop until failure is done through
  \texttt{repeat t}. Hence, this can be translated as \texttt{t; t; t; ..}, with
  an appropriate amount of repetitions. Although such a translation indeed
  results in a valid proof tree, there is no bound on its size. A small
  tactic expression may result in a large proof tree. As such, it is not obvious
  whether this transformation is advantageous. When \texttt{t} is simple, it may
  indeed be possible for an AI to profitably utilize its repetition. On the
  other hand, when \texttt{t} is itself a compound expression, an AI has to
  perform full program synthesis. As argued in
  Section~\ref{sec:simplified-tactic-model}, in such a case it might be easier
  to immediately synthesize a kernel-level proof term.
\item The \texttt{once t} tactical restricts \texttt{t} to have a single
  success. This prevents \texttt{t} from participating in a backtracking search.
  Such a restriction has no negative effect on the proof tree. In fact, as discussed in
  Section~\ref{sec:simplified-tactic-model}, any atomic tactic of which more
  than one result was consumed cannot be properly represented in the proof tree
  because one would not know which of the successes of \texttt{t} was
  responsible for the final outcome.
  Hence, \texttt{once t} ensures that \texttt{t} can indeed be represented.
  Note that when it is detected for an atomic tactic \texttt{t} that multiple
  successes were consumed, the instrumentation of the entire surrounding
  expression is aborted. Instead, one simply records the outcome of the entire
  expression. This bailout strategy ensures a valid proof tree because the
  execution engine implicitly wraps any top-level tactic
  expression inside \texttt{once}.
\end{itemize}

\subsubsection{Static Tactic Decomposition}
\label{sec:static-tactic-decomposition}
Many of the syntactic constructs of the Ltac language are syntactic sugar, or
otherwise derived from simpler constructs. This type of compound expression
cannot be as easily instrumented as tacticals like \texttt{repeat} and
\texttt{try}, because their behavior cannot be described by some permutation of
their arguments. Instead, we use a series of rewrite rules that convert complex
tactic invocations into a series of simpler tactics that are combined using
tacticals that can be instrumented as normal. Note that the intent of the
rewrite rules is to expand syntactic sugar, but not to decompose complex tactic
procedures.

Due to the highly complex behavior of some tactics, creating rewrite rules that
fully preserve the semantics of the original tactic is next to impossible.
Instead, we create rules that preserve semantics in most situations. To
ensure that proof scripts remain functional, we execute both the original tactic and the
decomposed tactics and compare their result. If the result differs meaningfully,
the decomposition is aborted and the original tactic is recorded. Comparing the
result of two tactic executions entails checking that the generated proof states
cannot be distinguished by subsequent tactics. Such a check goes beyond normal
equality modulo $\alpha$-equivalence. In addition, the names and ordering of
hypotheses must be identical, as well as the names of the binders in the spine
of the proof states' conclusion.\footnote{Some tactics can still distinguish
  proof states that are equal according to this notion. However, in practice, this
  equality relation has proven to be the sweet spot between flexibility
  and strictness.}

Below, we include some of the more important rewrite rules. For presentational
reasons, these rules have been simplified. We omit some more general rewriting
forms that can be easily deduced from these simpler rules. The rules utilize
two helper tacticals \texttt{sf} and \texttt{sl}, which execute their argument
by selecting only the first respectively the last proof state in the focus.
\begin{align*}
  \texttt{apply l1, l2,..,ln} &\to \texttt{apply l1; sl apply l2; ..; sl apply ln}\\
  \texttt{apply l1, l2,..,ln in H} &\to \texttt{apply l1 in H; sf apply l2 in H; ..;}\\
                              &\phantom{\to\ \ } \texttt{sf apply ln in H}\\
  \texttt{assert x by t} &\to \texttt{assert x; sf solve[t]} \\
  \texttt{enough x by t} &\to \texttt{assert x; sl solve[t]} \\
  \texttt{generalize t1 t2..tn} &\to \texttt{generalize tn; ..; generalize t2;} \\
                              &\phantom{\to\ \ }\texttt{generalize t1}\\
  \texttt{destruct t1, t2,.., tn} &\to \texttt{destruct t1; destruct t2; ..;} \\
                              &\phantom{\to\ \ }\texttt{ destruct tn}\\
  \texttt{destruct t} &\to \texttt{try intros until t; destruct (t)} \\
  \texttt{unfold d1, d2,..,dn} &\to \texttt{unfold d1; unfold d2; ..; unfold dn} \\
  \texttt{rewrite l1, l2,..,ln} &\to \texttt{rewrite l1; sf rewrite l2; ..;} \\
                              &\phantom{\to\ \ } \texttt{sf rewrite ln}\\
  \texttt{rewrite l by t} &\to \texttt{rewrite l; [| solve[t]..]} \\
  \texttt{rewrite n l} &\to \texttt{do n (sf rewrite l)} \\
  \texttt{rewrite n?l} &\to \texttt{do n (try sf rewrite l)} \\
  \texttt{rewrite ?l} &\to \texttt{repeat (try sf rewrite l)}\\
  \texttt{rewrite !l} &\to \texttt{rewrite l; sf rewrite ?l}\\
  \texttt{replace x with y by t} &\to \texttt{replace x with y; sl solve[t]}\\
\end{align*}
In addition to these hard-coded rules, there is also support for expanding
user-defined tactics. For example, Coq's standard library defines the following
notation:
\begin{minted}{coq}
Tactic Notation "now" tactic(t) := t; easy.
\end{minted}
Because this is trivial syntactic sugar, this notation has been registered as
decomposable, adding the rewrite rule $\texttt{now t} \to \texttt{t; easy}$.
Such custom expansions need to be manually registered because it is generally
not possible to automatically distinguish a user-written tactic procedure that
should not be decomposed from a simple syntactic abbreviation.

Finally, we also have dedicated support for unraveling
\textit{intropatterns}, which are a particularly compact way to introduce new
hypotheses into a proof state and perform simple transformations on those
hypotheses like case analysis and rewriting. We expand intropatterns into a
sequence of atomic tactics with the same semantic meaning. To illustrate this,
we provide a radically simplified version of the rewrite rules.
\begin{align*}
  \texttt{intros p1..pn} &\to \texttt{intros p1; ..; intros pn}\\
  \texttt{intros p} &\to \texttt{let v := fresh in}\\
                         &\phantom{\to\ \ }\texttt{intro v; } \text{expand}(\texttt{v}, \texttt{p})\\
  \texttt{apply l in v as p} &\to \texttt{apply l in v; sf } \text{expand}(\texttt{v}, \texttt{p})\\
  \texttt{assert t as p} &\to \texttt{let v := fresh in}\\
                         &\phantom{\to\ \ }\texttt{assert (v:t); } \text{expand}(\texttt{v}, \texttt{p})\\
  \texttt{injection t as p} &\to \texttt{injection t; intros p}\\
  \texttt{case t as [p1..pn]} &\to \texttt{case t; intros p1..pn}\\
  \texttt{case t as [p1|..|pn]} &\to\texttt{case t; [intros p1|..|intros pn]}\\
  \texttt{destruct t as [p1..pn]} &\to \texttt{case t; intros p1..pn}
\end{align*}
\begin{align*}
  \text{expand}(\texttt{v}, \texttt{->}) &= \texttt{intropattern subst -> in v}\\
  \text{expand}(\texttt{v}, \texttt{<-}) &= \texttt{intropattern subst <- in v}\\
  \text{expand}(\texttt{v}, \texttt{\_}) &= \texttt{clear v}\\
  \text{expand}(\texttt{v}, \texttt{[= p1..pn]}) &= \texttt{first [discriminate | injection v as p1..pn]}\\
  \text{expand}(\texttt{v}, \texttt{p \% l1..ln}) &= \texttt{apply l1,..,ln in v as p}\\
  \text{expand}(\texttt{v}, \texttt{[p1..pn]}) &= \texttt{case v as [p1..pn]}\\
  \text{expand}(\texttt{v}, \texttt{[p1|..|pn]}) &= \texttt{case v as [p1|..|pn]}
\end{align*}
The actual rewrite rules employed are substantially more complex. For example,
in practice the rule for \texttt{intros p1..pn} is not always valid. The
semantics of intropatterns specifies that dependent wildcard (\texttt{\_})
hypotheses must be \texttt{clear}ed only after all other actions in the pattern
have been processed. Furthermore, the case analysis rules for \texttt{destruct}
and \texttt{case} must carefully take into account the location of a hypothesis
and any dependent hypotheses. Every other rewrite rule contains similar
subtleties that we have omitted for presentational reasons.

\subsubsection{Dynamic Tactic Decomposition}
\label{sec:dynamic-tactic-decomposition}
Some tactic scripts employ complex loops that dynamically determine which tactic
to execute based on the shape of the proof state in focus. This is often done
for proofs that would otherwise be long and tedious. Such scripts can be seen as
ad-hoc tactic procedures that are specifically tailored for a single proof. As
such, these procedures are not factored into a named tactic and are not expected
to be useful again. For this reason, it would be preferable for machine learning
agents to learn the behavior of such a loop, rather than the learn to synthesize
the loop itself.

To analyze the current proof state, such scripts may use a pattern-matching
construct.
For example, the following tactic looks for a propositional hypothesis
\texttt{H} whose conclusion if \texttt{False} and whose premise is either a
conjunction or a disjunction. Then, depending on the shape of the premise, a
rewrite with the appropriate lemma is performed.
\begin{minted}{coq}
match goal with
| H: ?P \/ ?Q -> False |- _ => rewrite (not_or_iff P Q) in H
| H: ?P /\ ?Q -> False |- _ => rewrite (not_and_iff P Q) in H
end
\end{minted}
During the pattern matching, the variables \texttt{?P} and \texttt{?Q} are bound
to a term. These variables may then be referenced in the tactic expression that
should be executed. To decompose this pattern match, we instrument only
the appropriate rewrite tactic, such that the matched variables have been
appropriately substituted. When this example is executed on a proof
state with hypothesis \verb|R: (X\/Y)/\Z -> False|, we will record the
execution of the fully substituted tactic
\verb|rewrite (not_and_iff (X\/Y) Z) in R|.
We call such a decomposition \textit{dynamic} because it cannot be performed
on the syntax of the proof script alone. Instead, the proof state on which it is
executed needs to be known, such that the correct tactic expression can be
selected and substituted.

A similar substitution procedure can be performed for other constructs that
involve binders. For example, some let-binders can be eliminated:
\begin{multline*}
  \texttt{let t := auto in intros; [t | contradiction | t]} \\
  \to \texttt{intros; [auto | contradiction | auto]}
  \end{multline*}
Note, however, that such
substitutions are performed only in select circumstances, due to the rather complex
behavior of binders and substitutions in the tactic engine.

\subsubsection{Tactic Anonymization}\label{sec:tactic-anonymization}
As a final, optional post-processing step, we anonymize recorded tactics. This
is done to make tactics name-invariant, similar to how the graph representation
of terms is name-invariant.
\begin{align*}
  \texttt{intros H} &\to \texttt{intros ?}\\
  \texttt{intro H} &\to \texttt{intros ?}\\
  \texttt{destruct H1 as [H2 H3 ..]} &\to \texttt{destruct H1}\\
  \texttt{assert T as H} &\to \texttt{assert T}\\
  \texttt{pose H:=T} &\to \texttt{pose T}\\
  \texttt{injection H1 as H2 H3 ..} &\to \texttt{injection H1}\\
  \texttt{apply T in H1 as H2} &\to \texttt{apply T in H1 as ?}\\
\end{align*}
With these transformations, we ask Coq to generate names for new hypotheses
instead of supplying them manually. Such a transformation breaks proof scripts,
because it is not known what the name of a new hypothesis will be ahead of time,
preventing subsequent tactics from referencing it. When combined with a
graph-based representation of proof states, this is not a problem, however.
There, hypotheses are just a node in a graph that does not change when the
hypothesis is renamed.

Note that the transformation presented above is not complete. Especially for
complex tactic expressions that have not been decomposed, names may remain.
These tactics cannot currently be properly represented as a base tactic with
graph-based arguments.

\section{Modes of Interacting with Coq}\label{sec:modes-of-interaction}

Interacting with Coq through the datastructures described in
Section~\ref{sec:cic-representation} and~\ref{sec:tactic-representation} can be
done either by extracting an offline dataset (Section~\ref{sec:datasets}) or
by interfacing directly with Coq. The latter option gives external agents the
ability to prove actual theorems in Coq through a novel interface. An external
program can interface either by acting as a \textit{prediction server}
(Section~\ref{sec:prediction-servers}) or as a \textit{proving client}
(Section~\ref{sec:proving-clients}). Agents can be benchmarked on existing Coq
developments through a massively parallel benchmarking framework
(Section~\ref{sec:benchmarking}).

The crucial difference between a prediction server and a proving client is
where the proof exploration logic is handled: A prediction server is not
responsible for proof exploration. Rather, it is responsible for suggesting a
list of appropriate tactics when given a proof state. This information is then
connected to Tactician's \texttt{Suggest} command and \texttt{synth}
tactic~\cite{DBLP:conf/mkm/BlaauwbroekUG20}, which are responsible for suggesting
these results to Coq users or to synthesize a proof using these suggestions.
In this format, a prediction server has no agency. It simply gives replies to
queries.
On the other hand, a proving client does have agency. Whenever a proving session
is initiated, the client temporarily takes control of the Coq process. It may
arbitrarily explore the search space of the currently open proof by running
tactics on proof states and observing the result. Exploration may continue until
either one or more proofs are completed or the agent aborts its exploration.

\subsection{Communication Protocol}
The communication protocol and data storage format for all modes of interacting
with Coq are based on the Cap'n Proto data serialization and remote procedure calling
protocol~\cite{capnp}. This protocol was determined to be an excellent fit for
these purposes:
\begin{itemize}
\item Cap'n Proto specifies a fast, binary, memory-efficient data serialization
  format that can be used both for storage and communication. When stored on
  disk, the encoded data can be memory-mapped which gives fast random access to
  any piece of information encoded in the data without having to load the entire
  dataset into memory. This allows us to quickly and arbitrarily traverse the entire
  mono-graph of mathematical knowledge extracted from Coq, even when this graph
  is much larger than a machine's working memory. Such fast, random access is
  crucial for training machine learning models using batches of randomly
  selected examples.
\item Cap'n Proto allows for interfacing with our datastructures in a wide variety of
  languages, including Python, C++, Haskell, Rust, Java, Go, C\# and, crucially,
  OCaml (the implementation language of Coq). This enables the implementation of
  agents in any supported language.
\item Beyond communication through simple messages, Cap'n Proto implements a
  remote procedure calling (RPC) protocol based on the distributed
  object-capability model\cite{RobustComposition}. A distributed capability is a
  transferable token that represents the right to execute operations on an
  object that may exist in a different process or on a different machine. This allows
  us to ``export'' proof states to external agents, by viewing them as
  capability objects on which tactics can be executed. Agents can then
  seamlessly explore by executing tactics on arbitrary proof states in the
  search space and observing the results.
\end{itemize}

Because of the popularity of the Python programming language for machine
learning and data analysis, we provide the PyTactician library on top of Cap'n
Proto that further streamlines access to datasets and communication with Coq
from Python.

\subsection{Datasets}\label{sec:datasets}
For the purpose of offline machine learning and data analysis, there is the
ability to extract graph-based and text-based data from Coq during the
compilation of a mathematical theory. Such a dataset contains a single graph, that
efficiently encodes the global context of available definitions (see
Section~\ref{sec:meta-graph}) at every
time point in the compilation process. It also includes the tactic-based proof
tree for each lemma in the theory (see Section~\ref{sec:tactic-representation}).
Every proof state and definition in the graph is additionally represented in a
textual format.

A dataset contains a single data file for each compilation unit (source file) in
the theory. The data file contains an index of all definitions encountered in
the compilation unit and their corresponding graph. When compilation unit $A$
imports the theory from compilation unit $B$, the graph encoded in $A$'s
data file may reference nodes from $B$'s data-file. Memory-mapping all
data files of the theory assembles the partial graphs of each data file into a single
mono-graph that can be arbitrarily indexed and traversed, even if the resulting
graph does not fit in the machine's main memory. The PyTactician Python library
was created to make traversing the graph as painless as possible. It includes a
data sanity checker and visualization webserver that allows for interactive
exploration of the data.

We provide a large
dataset~\cite{blaauwbroek_2023_10028721} of mathematical knowledge extracted from 120
different Coq packages available in the Coq Package
Index~\cite{coqpackageindex}. These packages were automatically selected by a
SAT solver connected to the Opam package manager~\cite{opam} as the largest set
of mutually co-installable packages compatible with Coq v8.11. An online
visualization of this data is
available\footnote{\url{http://grid01.ciirc.cvut.cz:8080}}.
The mathematical domains included in the dataset vary wildly. It includes
analysis, compiler and programming language formalization, separation logic,
homotopy type theory, and much more. The graph extracted from these
formalizations consists of over 250 million nodes, which encode 520k
definitions, of which 266k are theorems, with a total of 4.6 million proof state
transformations.~\footnote{Roughly half of the definitions
  are derived from each other through Coq's module and section
  mechanism.}

The figure on the front page shows a rendering of a small section of the mathematical
universe contained in the dataset. In particular, only the most basic of
mathematical concepts, that are part of the ``Prelude'' in Coq's standard
library, are rendered. The full graph would be over
3000 times larger. The size and color of a node in the rendering is dependent on
how many times it is referenced. As nodes get more popular, they increase in
size. Examples of large nodes are the universes Prop and Type, the inductive
definition for Leibnitz equality, and the inductive definitions for true and
false. Not all popular nodes are definitions, however. Other popular
mathematical entities include hypotheses that posit the existence of natural
numbers and booleans, and even some anonymous subterms that happen to occur very
often.

The placement of each node is determined by the \textit{sfdp} force-directed graph drawing
engine of the Graphviz visualization suite~\cite{hu2005efficient}. As a result,
related nodes will be rendered close together. This is particularly apparent for
inductive definitions and their constructors, where it often appears as if the
constructors are ``planets'' that orbit around a corresponding inductive
that acts as their ``sun''. The color of each edge is determined by its length
as determined by the rendering engine.

\subsection{Prediction Servers}\label{sec:prediction-servers}
Prediction servers connect to the Tactician plugin running inside Coq through a
communication socket. Over this socket, Tactician may send a proof state and
request a list of recommended tactics to execute on that proof state. Different
settings exist to provide either a text-based or graph-based representation (or both) of the
proof state to the server. In response, suggested tactics may be encoded as
text, or as a base tactic with arguments that point to nodes in the graph.
Currently, graph-based tactic arguments can point either to global definitions
or local hypotheses of the proof state. In the future, we plan to extend this to
arbitrary terms encoded within the graph. Synthesis of entirely new graph-based
terms for tactic arguments remains future work.

The output of a prediction server may be accessed from within Coq through
Tactician's \texttt{Suggest} command and \texttt{synth} tactic. \texttt{Suggest}
will simply display the list of tactic suggestions for the current proof state
to the user. The \texttt{synth} tactic will employ a search procedure to
synthesize a proof based on the server's predictions. This tight integration
allows any prediction server to immediately be used by Coq users from the
comfort of their usual development environment.

The graph of a proof state may reference any definition that is available in
Coq's global context. Furthermore, tactics may also reference definitions from
the global context. Therefore, a prediction server requires access to this
graph. Due to the potentially large size of the global context, it is not
feasible to send the entire graph to the server on each request. To remedy this,
the graph of the global context can be cached. In cached mode, Tactician will send a
stack of partial graphs to the server. Each partial graph in the stack may
reference nodes from graphs in the stack below it. Then, whenever Coq's global
context is amended with new definitions, a new graph may be pushed on top of the
stack to synchronize the state with the prediction server. The prediction server
can stitch all partial graphs in the stack together to obtain the final
mono-graph (the PyTactician library can help with such operations). In
interactive mode, when a user navigates back and forth through a Coq document,
the graph is continuously synchronized in real time by pushing and popping
partial graphs on and off the stack. This way, we can keep Coq's entire internal
state synchronized with the prediction server with only a constant
time overhead.

The PyTactician library ships with some simple example prediction servers as
well as oracles. Oracles are prediction servers that interface with a dataset.
When they receive a proof state, they check for the existence of that proof
state in the dataset. If found, the corresponding tactic is suggested.
Otherwise, no tactic is suggested. The oracle can accept proof states both in
text-mode and graph-mode. Oracles are useful for debugging purposes, and to
check how faithful the proofs are encoded in the dataset (see the evaluation in
Section~\ref{sec:evaluation}).

\subsection{Proving Clients}\label{sec:proving-clients}
A proving client is an agent that autonomously explores a proof search space.
Like a prediction server, it connects to the Tactician plugin through a socket.
The graph of Coq's global context is also synchronized with the proving client.
When \texttt{Tactician Explore} command is issued in an open proof, a proving
session is started with the client. The current proof state is exported to the
client as a Cap'n Proto capability object. This allows the client not only to inspect
the graph and text of the proof state but also to execute tactics on it.
The act of executing a tactic on a proof state \textit{unfolds} the proof search
tree. If the executed tactic generates new proof states, they are again exported
to the client as a capability. The client may execute arbitrarily many tactics
on the same proof state in an arbitrary order in order to explore the search
tree. Proof states are automatically garbage collected when the client ceases to
hold a reference to them.

A typical application for a proving client is to implement a reinforcement
learning agent that explores a proof through Monte-Carlo Tree Search
(MCTS)~\cite{DBLP:conf/aiide/ChaslotBSS08}. Instead of learning from an offline
dataset, such an agent learns by experimenting with different actions in its
environment until it understands which actions are effective in what circumstances.

\subsection{Massively Parallel Benchmarking}
\label{sec:benchmarking}

To assess the strength of proving agents, a benchmarking framework was created
that can perform an automatic evaluation of the agent's performance on every
theorem contained in a package on the Coq package index~\cite{coqpackageindex}.
When given a set of Coq packages and a reference to a particular version of the
Tactician plugin and the agent that should be evaluated, the benchmarking
framework will query the Opam package manager to create a software environment
in which an appropriate version of Coq is installed as well as all of the
dependencies of the Coq packages to be benchmarked. Then, before the target Coq
packages are compiled and installed, the benchmark framework will inject
the Tactician plugin into the Coq compilation system. Tactician will report back to
the benchmarking framework which source files are being compiled, the arguments
used to invoke the \texttt{coqc} compiler and a list of all lemmas present in
the source file.

After the benchmark system collects a complete index of lemmas to benchmark, the
benchmarking phase commences. For this, the benchmarking framework must be given
access to computational resources. These resources can vary from a single CPU on
a laptop, a single server with many CPUs, to a cluster of machines accessible over
SSH. It can also request auto-scalable resources on a High-Performance Computing
(HPC) cluster through a workload manager like
SLURM~\cite{DBLP:conf/jsspp/JetteW23} or PBS~\cite{openpbs}. When a
computational resource is acquired, the required software environment is copied
onto that resource as needed. The resource is then assigned a Coq compilation
unit to re-compile in benchmark mode. Whenever a lemma is encountered, a central
work distribution manager decides whether the agent should be benchmarked on that
lemma (or if it has already been benchmarked on a different, parallel resource).
This allows the work to be distributed optimally between massively parallel
resources. Grouping lemmas from the same compilation unit onto the
same resource as much as possible is preferred to reduce the overhead of
re-compiling the same compilation many times. However, large compilation
units containing many lemmas will be automatically split up and distributed over
available resources to improve parallelism. Computational resources can be
dynamically acquired and relinquished based on availability. This framework has
been shown to scale to benchmarking over 120k lemmas in parallel on more than a
1000 CPUs distributed over many machines. Machine learning agents
that require a GPU can also be utilized.

For each lemma, the benchmarking system reports whether the agent successfully
synthesized a proof and the tactic script that was created. Additional
information includes the time required to find the proof and the number of tactic
inferences made. Any logs outputted by Coq and the proving agent are also
collected and saved.

\todo[inline]{Add suggested training tasks section}

\section{Evaluation}\label{sec:evaluation}
We perform several experiments to evaluate how the different data formats,
datasets and interaction protocols described in the previous sections perform.

\paragraph{Effectiveness of Graph Sharing}
Sharing bisimilar graphs allows us to reduce the size of the graph. How much the
graph is reduced in size is a good measure of how much duplication exists in
terms. We measure this by extracting a
dataset from Coq's standard library with different sharing settings. When
sharing is completely disabled, the standard library is extracted into a graph
containing 157 million nodes and 294 million edges. On average, every node has
approximately 1.86 edges. Next, the graph-sharing algorithm was used to share
subterms only within a single definition, but not across them. For
theorems, we include all intermediate proof states and proof terms that form the
final proof. Subterms are shared between all proof states of the proof.
This reduces the number of nodes by 85\% to 22.7 million nodes. Allowing
graph-sharing between all nodes in a single compilation unit eliminates another
2\% of the nodes. Finally, allowing graph-sharing between all compilation units
that depend on each other results in another 1.3\% size reduction for a total
reduction of 88.3\% to 18.3 million nodes. The number of edges has reduced to
40.8 million.

This shows that terms indeed contain a large amount of duplication. Most of that
duplication occurs within individual definitions and proofs. Much of the
duplication is caused by proof states that may be nearly identical and share
common hypotheses. Additionally, proof terms generated by tactics
may be inefficient. It is surprising that relatively little sharing
occurs between definitions. We hypothesize that different Coq packages in fact
contain a significant amount of duplicated effort. However, this is not
detectable by the sharing algorithm because two otherwise identical definitions
that exist in a different global context are not shared by default (see
Section~\ref{sec:graph-sharing}).

\paragraph{Oracle Evaluation}
Both the text-based and the graph-based data formats have limitations to how
many proofs they are able to faithfully represent. For text, this is caused by
limitations and bugs in Coq's printing and parsing facilities. Especially when
complex notations are involved, Coq's printing procedure may not be an exact
inverse of its parsing procedure. For graphs, the main limitation is caused by
the inability to represent some tactics and their arguments. Tactic
arguments are limited to pointers into the graph. Other types of arguments, like
integers, strings, and unresolvable identifiers are not supported. Furthermore,
support for Ssreflect-style tactics is limited.

Using the oracles described in Section~\ref{sec:prediction-servers}, we can
evaluate how many proofs encoded in the dataset can be reconstructed within Coq.
The text-based oracle allows us to reconstruct 70.0\% of the 131 thousand
original lemmas in the dataset, while the graph-based oracle only reaches
28.4\%. For Coq's standard library, the text oracle can prove 96.0\% out of 11
thousand lemmas while the graph oracle reaches 47.8\%. These numbers are higher
because the standard library generally uses less advanced Coq features.

These numbers show that text is currently far more accurate than the graph
representation. Much of this is not a fundamental obstacle. The graph-based
representation of tactics can be improved in the future. Note, however,
that the performance of oracles represents a lower bound on the theorems that can
be proven. It is possible, and even likely, that a lemma with a proof that
cannot be faithfully represented has an alternative proof that can be
represented. As such, even if graph-based representations of tactics are less
accurate, their richer representation of proofs and definitions may still give
them an advantage over text-based representations. This is confirmed by the
performance of neural models discussed in the next paragraph.

\paragraph{Neural Models}
In a separate publication, two neural models were trained based on the
dataset~\cite{graph2tac}.
For a full discussion, we refer the reader there. The first model is a
straightforward language model that is trained to predict tactics based on the
textual representations of the proof states in the dataset. The second, more
advanced model, called Graph2Tac, takes advantage of the graph encoding to learn to extract
feature representations of definitions. This aids the model in gaining an
appropriate understanding of the definitions referenced by proof states, and
which lemmas might be relevant to that proof state. Furthermore, when faced with
a definition that the model has not seen during training, it has the ability to
generate a new feature representation on the fly and incorporate it into its
model without additional training. This allows the model to gain an immediate
understanding of new mathematical concepts, which is shown to be crucial for its
performance.

These models were evaluated on a testing set of 2000 theorems, together with
Tactician's state-of-the-art $k$-nearest neighbor ($k$-NN)
model~\cite{DBLP:conf/mkm/ZhangBPCKU21} and
CoqHammer~\cite{DBLP:journals/jar/CzajkaK18}. The Graph2Tac and $k$-NN solver,
both able to incorporate new data into their model on the fly, perform the best,
both with a 26\% pass rate. The language model proves 15\% and CoqHammer 18\%.

We can conclude two lessons from these results: First, agents that incorporate
as much online data into their model as possible are strongly advantaged over
agents that rely on a static set of background knowledge. We hope that by
providing such data to agents in a convenient format, this can be exploited even
more in the future. Second, it appears that the theoretical limitations of
graph-based tactic representations are not a major bottleneck for real-world
agents. As stated above, agents can often work around these limitations by
finding alternative proofs. Furthermore, in their current state, agents are
likely not capable enough to correctly synthesize complex proofs that are
difficult to represent. This may change in the future, at which point the
communication format needs to be improved.

\section{Threats to Validity}\label{sec:threats-to-validity}
Here, we will discuss potential problems that may occur as a result of the design
decisions and implementation details that were made in this paper.

\paragraph{Kernel vs Non-kernel Objects}
The graph representation presented in this paper is based on Coq's kernel-level
Gallina terms. As such, any information about terms that live outside of the
kernel is lost. This includes implicit arguments, typeclasses and many other
features. On the positive side, this vastly simplifies the format of the data that is
presented to agents. On the other hand, this data is often quite different, and
more larger, than the representation that is usually presented to Coq users.
It is unclear whether an agent may benefit from a data representation where
implicit arguments are omitted.

The inability of our data format to represent anything that is not a first-class
kernel object is somewhat limiting. For Coq's section and module system,
special care had to be taken to record which definitions are derived from which
other definitions (see Section~\ref{sec:meta-graph}). Even then, consumers of
this data must be careful not to use derived definitions inappropriately. This
has been an area where mistakes were shown to be easily made in practice.

Another example of data that we do not represent are hint-databases used by the
\texttt{auto} tactic. The behavior of this tactic is entirely governed by the hint
database that is selected. As such, it may be difficult to properly learn in
what circumstances \texttt{auto} may be appropriate.

It is also unfortunate that tactics themselves are not kernel-level objects.
This forces us to represent tactics on a different level of abstraction than
terms, leading to many implementation problems. It is particularly problematic
to interpret the arguments of tactics as nodes in the graph. This requires us to
convert these arguments to kernel-level objects, which is not always possible.
It would have been preferable if Coq's Gallina language could be used as the
meta-language in which tactics are implemented.

\paragraph{Tactic Decomposition}
The decomposition of complex tactic expressions into atomic pieces should also
be subject to scrutiny. Tactic decomposition is done, because we do not wish to
synthesize complex tactic procedures. We rather see tactics as simple commands that
help us build proof terms. However, such re-imagining of tactics is not without
problems. It can be difficult to know what the optimal level of decomposition
for an expression is, and which tactics should be considered as ``atomic
tactics''. Especially for custom-defined tactics it can be difficult to
automatically determine how to decompose them. The Ssreflect proof language
cannot currently be decomposed at all, leading to poor performance of agents on
developments that use this language.

\paragraph{Choices of Text Representation}
The text-based representations we use are generated by Coq's default printer of
terms, proof states, and tactics. This printer is not without its problems. We have shown
that Coq's parser and printer are not perfect inverses of each other. In
addition, there are many settings that may influence how Coq may print terms.
This could affect the performance of a text-based agent. Printing of implicit
arguments, notations, fully qualified names, universe levels, existential
variable instances, coercions, additional parentheses, and more can all be
independently enabled and disabled. We have not investigated the effect of these
settings on agents.

\todo[inline]{Add future work section}

\section{Related work}
We are not the first to create a platform for interacting with and extracting
data from Coq. SerAPI~\cite{GallegoArias2016SerAPI} is a library for
machine-readable interaction with Coq based on s-expressions. Compared to
SerAPI, the communication model with Coq is inverted in our platform. With
SerAPI, a client drives Coq by submitting commands to and receiving answers from
Coq. This is similar to how a user might interact with Coq. On the other hand,
for our platform, agents are connected to the Tactician plugin, which runs as
part of the Coq executable. Communication with an agent is initiated from Coq,
when a vernacular command like \texttt{synth}, \texttt{Suggest} or
\texttt{Tactician Explore} is executed in a document (see
Sections~\ref{sec:prediction-servers} and~\ref{sec:proving-clients}). At that point,
either a prediction server is queried, or control over the current proof is
given to a proving client. The best model of interaction depends on the
use-case. However, Tactician's setup does allow agents to be easily integrated
into Coq. This is not the case for SerAPI, which has limited the practical adoption of
machine learning agents and platforms based on it, such as
ProverBot9001~\cite{DBLP:conf/pldi/Sanchez-SternAS20},
CoqGym~\cite{DBLP:conf/icml/YangD19},
Passport~\cite{DBLP:journals/corr/abs-2204-10370} and
TacTok~\cite{DBLP:journals/pacmpl/FirstBG20}.

GamePad~\cite{DBLP:conf/iclr/HuangDSS19} is another machine learning platform
for Coq that drives a Coq process from Python. It uses a modified version of
Coq that outputs manually constructed s-expressions similar to SerAPI. Those
expressions are then parsed into Python datastructures. A difference with
SerAPI, is that sub-expressions can be shared through
hash-consing~\cite{DBLP:conf/ml/FilliatreC06}. Note, however, that hash-consing
shared all terms with an equal memory representation and does not respect
$\alpha$-equivalence like we do in Section~\ref{sec:graph-sharing}.

Many machine learning platforms for other proof assistants also exist. For
Isabelle~\cite{DBLP:books/sp/NipkowPW02} there is
IsarStep~\cite{DBLP:conf/iclr/LiYWP21}. For HOL
Light~\cite{DBLP:conf/tphol/Harrison09a} there is
HolStep~\cite{DBLP:conf/iclr/KaliszykCS17} and
HOList~\cite{DBLP:conf/icml/BansalLRSW19}. And for
Lean~\cite{DBLP:conf/cade/MouraKADR15}, there is
LeanStep~\cite{DBLP:conf/iclr/HanRWAP22},
\texttt{lean-gym}~\cite{DBLP:conf/iclr/PoluHZBBS23} and
LeanDojo~\cite{DBLP:journals/corr/abs-2306-15626}.

\todo[inline]{Add conclusion section}

\bmhead{Acknowledgments}
This work was partially supported by the  Amazon Research Awards, EU ICT-48
2020 project TAILOR no. 952215, and the European Regional Development Fund
under the Czech project AI\&Reasoning with identifier
CZ.02.1.01/0.0/0.0/15\_003/0000466.

\bibliography{bibliography}


\begin{thebibliography}{41}
\ifx \bisbn   \undefined \def \bisbn  #1{ISBN #1}\fi
\ifx \binits  \undefined \def \binits#1{#1}\fi
\ifx \bauthor  \undefined \def \bauthor#1{#1}\fi
\ifx \batitle  \undefined \def \batitle#1{#1}\fi
\ifx \bjtitle  \undefined \def \bjtitle#1{#1}\fi
\ifx \bvolume  \undefined \def \bvolume#1{\textbf{#1}}\fi
\ifx \byear  \undefined \def \byear#1{#1}\fi
\ifx \bissue  \undefined \def \bissue#1{#1}\fi
\ifx \bfpage  \undefined \def \bfpage#1{#1}\fi
\ifx \blpage  \undefined \def \blpage #1{#1}\fi
\ifx \burl  \undefined \def \burl#1{\textsf{#1}}\fi
\ifx \doiurl  \undefined \def \doiurl#1{\url{https://doi.org/#1}}\fi
\ifx \betal  \undefined \def \betal{\textit{et al.}}\fi
\ifx \binstitute  \undefined \def \binstitute#1{#1}\fi
\ifx \binstitutionaled  \undefined \def \binstitutionaled#1{#1}\fi
\ifx \bctitle  \undefined \def \bctitle#1{#1}\fi
\ifx \beditor  \undefined \def \beditor#1{#1}\fi
\ifx \bpublisher  \undefined \def \bpublisher#1{#1}\fi
\ifx \bbtitle  \undefined \def \bbtitle#1{#1}\fi
\ifx \bedition  \undefined \def \bedition#1{#1}\fi
\ifx \bseriesno  \undefined \def \bseriesno#1{#1}\fi
\ifx \blocation  \undefined \def \blocation#1{#1}\fi
\ifx \bsertitle  \undefined \def \bsertitle#1{#1}\fi
\ifx \bsnm \undefined \def \bsnm#1{#1}\fi
\ifx \bsuffix \undefined \def \bsuffix#1{#1}\fi
\ifx \bparticle \undefined \def \bparticle#1{#1}\fi
\ifx \barticle \undefined \def \barticle#1{#1}\fi
\bibcommenthead
\ifx \bconfdate \undefined \def \bconfdate #1{#1}\fi
\ifx \botherref \undefined \def \botherref #1{#1}\fi
\ifx \url \undefined \def \url#1{\textsf{#1}}\fi
\ifx \bchapter \undefined \def \bchapter#1{#1}\fi
\ifx \bbook \undefined \def \bbook#1{#1}\fi
\ifx \bcomment \undefined \def \bcomment#1{#1}\fi
\ifx \oauthor \undefined \def \oauthor#1{#1}\fi
\ifx \citeauthoryear \undefined \def \citeauthoryear#1{#1}\fi
\ifx \endbibitem  \undefined \def \endbibitem {}\fi
\ifx \bconflocation  \undefined \def \bconflocation#1{#1}\fi
\ifx \arxivurl  \undefined \def \arxivurl#1{\textsf{#1}}\fi
\csname PreBibitemsHook\endcsname

\bibitem[\protect\citeauthoryear{{The Coq Development
  Team}}{2020}]{the_coq_development_team_2020_4021912}
\begin{botherref}
\oauthor{\bsnm{{The Coq Development Team}}}:
The Coq Proof Assistant.
\doiurl{10.5281/zenodo.4021912} .
\url{https://doi.org/10.5281/zenodo.4021912}
\end{botherref}
\endbibitem

\bibitem[\protect\citeauthoryear{Vaswani
  et~al.}{2017}]{DBLP:conf/nips/VaswaniSPUJGKP17}
\begin{bchapter}
\bauthor{\bsnm{Vaswani}, \binits{A.}},
\bauthor{\bsnm{Shazeer}, \binits{N.}},
\bauthor{\bsnm{Parmar}, \binits{N.}},
\bauthor{\bsnm{Uszkoreit}, \binits{J.}},
\bauthor{\bsnm{Jones}, \binits{L.}},
\bauthor{\bsnm{Gomez}, \binits{A.N.}},
\bauthor{\bsnm{Kaiser}, \binits{L.}},
\bauthor{\bsnm{Polosukhin}, \binits{I.}}:
\bctitle{Attention is all you need}.
In: \beditor{\bsnm{Guyon}, \binits{I.}},
\beditor{\bsnm{Luxburg}, \binits{U.}},
\beditor{\bsnm{Bengio}, \binits{S.}},
\beditor{\bsnm{Wallach}, \binits{H.M.}},
\beditor{\bsnm{Fergus}, \binits{R.}},
\beditor{\bsnm{Vishwanathan}, \binits{S.V.N.}},
\beditor{\bsnm{Garnett}, \binits{R.}} (eds.)
\bbtitle{Advances in Neural Information Processing Systems 30: Annual
  Conference on Neural Information Processing Systems 2017, December 4-9, 2017,
  Long Beach, CA, {USA}},
pp. \bfpage{5998}--\blpage{6008}
(\byear{2017}).
\burl{https://proceedings.neurips.cc/paper/2017/hash/3f5ee243547dee91fbd053c1c4a845aa-Abstract.html}
\end{bchapter}
\endbibitem

\bibitem[\protect\citeauthoryear{Scarselli
  et~al.}{2009}]{DBLP:journals/tnn/ScarselliGTHM09}
\begin{barticle}
\bauthor{\bsnm{Scarselli}, \binits{F.}},
\bauthor{\bsnm{Gori}, \binits{M.}},
\bauthor{\bsnm{Tsoi}, \binits{A.C.}},
\bauthor{\bsnm{Hagenbuchner}, \binits{M.}},
\bauthor{\bsnm{Monfardini}, \binits{G.}}:
\batitle{The graph neural network model}.
\bjtitle{{IEEE} Trans. Neural Networks}
\bvolume{20}(\bissue{1}),
\bfpage{61}--\blpage{80}
(\byear{2009})
\doiurl{10.1109/TNN.2008.2005605}
\end{barticle}
\endbibitem

\bibitem[\protect\citeauthoryear{Gallego~Arias}{2016}]{GallegoArias2016SerAPI}
\begin{botherref}
\oauthor{\bsnm{Gallego~Arias}, \binits{E.J.}}:
{SerAPI: Machine-Friendly, Data-Centric Serialization for Coq}.
Technical report,
MINES ParisTech
(October 2016).
\url{https://hal-mines-paristech.archives-ouvertes.fr/hal-01384408}
\end{botherref}
\endbibitem

\bibitem[\protect\citeauthoryear{Yang and Deng}{2019}]{DBLP:conf/icml/YangD19}
\begin{bchapter}
\bauthor{\bsnm{Yang}, \binits{K.}},
\bauthor{\bsnm{Deng}, \binits{J.}}:
\bctitle{Learning to prove theorems via interacting with proof assistants}.
In: \beditor{\bsnm{Chaudhuri}, \binits{K.}},
\beditor{\bsnm{Salakhutdinov}, \binits{R.}} (eds.)
\bbtitle{Proceedings of the 36th International Conference on Machine Learning,
  {ICML} 2019, 9-15 June 2019, Long Beach, California, {USA}}.
\bsertitle{Proceedings of Machine Learning Research},
vol. \bseriesno{97},
pp. \bfpage{6984}--\blpage{6994}.
\bpublisher{{PMLR}}, \blocation{???}
(\byear{2019}).
\burl{http://proceedings.mlr.press/v97/yang19a.html}
\end{bchapter}
\endbibitem

\bibitem[\protect\citeauthoryear{Blaauwbroek}{2023}]{blaauwbroek_2023_10028721}
\begin{botherref}
\oauthor{\bsnm{Blaauwbroek}, \binits{L.}}:
Tactician's Web of Large-Scale Formal Knowledge.
\doiurl{10.5281/zenodo.10028721} .
\url{https://doi.org/10.5281/zenodo.10028721}
\end{botherref}
\endbibitem

\bibitem[\protect\citeauthoryear{Blaauwbroek
  et~al.}{2023a}]{lasse_blaauwbroek_2023_10445966}
\begin{botherref}
\oauthor{\bsnm{Blaauwbroek}, \binits{L.}},
\oauthor{\bsnm{Pestun}, \binits{V.}},
\oauthor{\bsnm{Olsak}, \binits{M.}}:
{coq-tactician/coq-tactician-api: Tactician's API V15 for Coq 8.11}.
\doiurl{10.5281/zenodo.10445966} .
\url{https://doi.org/10.5281/zenodo.10445966}
\end{botherref}
\endbibitem

\bibitem[\protect\citeauthoryear{Blaauwbroek
  et~al.}{2023b}]{lasse_blaauwbroek_2023_10445964}
\begin{botherref}
\oauthor{\bsnm{Blaauwbroek}, \binits{L.}},
\oauthor{\bsnm{Pestun}, \binits{V.}},
\oauthor{\bsnm{Olsak}, \binits{M.}}:
Coq-tactician/coq-tactician-api: PyTactician V15.1.
\doiurl{10.5281/zenodo.10445964} .
\url{https://doi.org/10.5281/zenodo.10445964}
\end{botherref}
\endbibitem

\bibitem[\protect\citeauthoryear{Rute et~al.}{2024}]{graph2tac}
\begin{barticle}
\bauthor{\bsnm{Rute}, \binits{J.}},
\bauthor{\bsnm{Olšák}, \binits{M.}},
\bauthor{\bsnm{Blaauwbroek}, \binits{L.}},
\bauthor{\bsnm{Massolo}, \binits{F.I.S.}},
\bauthor{\bsnm{Piepenbrock}, \binits{J.}},
\bauthor{\bsnm{Pestun}, \binits{V.}}:
\batitle{Graph2tac: Learning hierarchical representations of math concepts in
  theorem proving}.
\bjtitle{arXiv preprint}
(\byear{2024})
\doiurl{10.48550/arXiv.2401.02949}
{\href{https://arxiv.org/abs/2401.02949}{{arXiv:2401.02949}}}
{[cs.LG]}
\end{barticle}
\endbibitem

\bibitem[\protect\citeauthoryear{Paulin{-}Mohring}{1993}]{DBLP:conf/tlca/Paulin-Mohring93}
\begin{bchapter}
\bauthor{\bsnm{Paulin{-}Mohring}, \binits{C.}}:
\bctitle{Inductive definitions in the system coq - rules and properties}.
In: \beditor{\bsnm{Bezem}, \binits{M.}},
\beditor{\bsnm{Groote}, \binits{J.F.}} (eds.)
\bbtitle{Typed Lambda Calculi and Applications, International Conference on
  Typed Lambda Calculi and Applications, {TLCA} '93, Utrecht, The Netherlands,
  March 16-18, 1993, Proceedings}.
\bsertitle{Lecture Notes in Computer Science},
vol. \bseriesno{664},
pp. \bfpage{328}--\blpage{345}.
\bpublisher{Springer}, \blocation{???}
(\byear{1993}).
\doiurl{10.1007/BFb0037116} .
\burl{https://doi.org/10.1007/BFb0037116}
\end{bchapter}
\endbibitem

\bibitem[\protect\citeauthoryear{Sanchez-Stern
  et~al.}{2023}]{DBLP:journals/corr/abs-2204-10370}
\begin{botherref}
\oauthor{\bsnm{Sanchez-Stern}, \binits{A.}},
\oauthor{\bsnm{First}, \binits{E.}},
\oauthor{\bsnm{Zhou}, \binits{T.}},
\oauthor{\bsnm{Kaufman}, \binits{Z.}},
\oauthor{\bsnm{Brun}, \binits{Y.}},
\oauthor{\bsnm{Ringer}, \binits{T.}}:
Passport: Improving automated formal verification using identifiers.
ACM Trans. Program. Lang. Syst.
\textbf{45}(2)
(2023)
\doiurl{10.1145/3593374}
\end{botherref}
\endbibitem

\bibitem[\protect\citeauthoryear{Blaauwbroek et~al.}{2024}]{sharing-paper}
\begin{barticle}
\bauthor{\bsnm{Blaauwbroek}, \binits{L.}},
\bauthor{\bsnm{Olšák}, \binits{M.}},
\bauthor{\bsnm{Geuvers}, \binits{H.}}:
\batitle{Hashing modulo context-sensitive $\alpha$-equivalence}.
\bjtitle{arXiv preprint}
(\byear{2024})
\doiurl{10.48550/arXiv.2401.02948}
{\href{https://arxiv.org/abs/2401.02948}{{arXiv:2401.02948}}}
{[cs.PL]}
\end{barticle}
\endbibitem

\bibitem[\protect\citeauthoryear{Delahaye}{2000}]{DBLP:conf/lpar/Delahaye00}
\begin{bchapter}
\bauthor{\bsnm{Delahaye}, \binits{D.}}:
\bctitle{A tactic language for the system coq}.
In: \beditor{\bsnm{Parigot}, \binits{M.}},
\beditor{\bsnm{Voronkov}, \binits{A.}} (eds.)
\bbtitle{Logic for Programming and Automated Reasoning, 7th International
  Conference, {LPAR} 2000, Reunion Island, France, November 11-12, 2000,
  Proceedings}.
\bsertitle{Lecture Notes in Computer Science},
vol. \bseriesno{1955},
pp. \bfpage{85}--\blpage{95}.
\bpublisher{Springer}, \blocation{???}
(\byear{2000}).
\doiurl{10.1007/3-540-44404-1\_7} .
\burl{https://doi.org/10.1007/3-540-44404-1\_7}
\end{bchapter}
\endbibitem

\bibitem[\protect\citeauthoryear{P{\'e}drot}{2019}]{pedrot2019ltac2}
\begin{bchapter}
\bauthor{\bsnm{P{\'e}drot}, \binits{P.-M.}}:
\bctitle{Ltac2: tactical warfare}.
In: \bbtitle{The Fifth International Workshop on Coq for Programming Languages,
  CoqPL},
pp. \bfpage{13}--\blpage{19}
(\byear{2019})
\end{bchapter}
\endbibitem

\bibitem[\protect\citeauthoryear{Kaiser
  et~al.}{2018}]{DBLP:journals/pacmpl/KaiserZKRD18}
\begin{barticle}
\bauthor{\bsnm{Kaiser}, \binits{J.}},
\bauthor{\bsnm{Ziliani}, \binits{B.}},
\bauthor{\bsnm{Krebbers}, \binits{R.}},
\bauthor{\bsnm{R{\'{e}}gis{-}Gianas}, \binits{Y.}},
\bauthor{\bsnm{Dreyer}, \binits{D.}}:
\batitle{Mtac2: typed tactics for backward reasoning in coq}.
\bjtitle{Proc. {ACM} Program. Lang.}
\bvolume{2}(\bissue{{ICFP}}),
\bfpage{78}--\blpage{17831}
(\byear{2018})
\doiurl{10.1145/3236773}
\end{barticle}
\endbibitem

\bibitem[\protect\citeauthoryear{Tassi}{2018}]{tassi2018elpi}
\begin{botherref}
\oauthor{\bsnm{Tassi}, \binits{E.}}:
Elpi: an extension language for coq (metaprogramming coq in the elpi
  $\lambda$prolog dialect)
(2018)
\end{botherref}
\endbibitem

\bibitem[\protect\citeauthoryear{Kiselyov
  et~al.}{2005}]{DBLP:conf/icfp/KiselyovSFS05}
\begin{bchapter}
\bauthor{\bsnm{Kiselyov}, \binits{O.}},
\bauthor{\bsnm{Shan}, \binits{C.}},
\bauthor{\bsnm{Friedman}, \binits{D.P.}},
\bauthor{\bsnm{Sabry}, \binits{A.}}:
\bctitle{Backtracking, interleaving, and terminating monad transformers:
  (functional pearl)}.
In: \beditor{\bsnm{Danvy}, \binits{O.}},
\beditor{\bsnm{Pierce}, \binits{B.C.}} (eds.)
\bbtitle{Proceedings of the 10th {ACM} {SIGPLAN} International Conference on
  Functional Programming, {ICFP} 2005, Tallinn, Estonia, September 26-28,
  2005},
pp. \bfpage{192}--\blpage{203}.
\bpublisher{{ACM}}, \blocation{???}
(\byear{2005}).
\doiurl{10.1145/1086365.1086390} .
\burl{https://doi.org/10.1145/1086365.1086390}
\end{bchapter}
\endbibitem

\bibitem[\protect\citeauthoryear{Blaauwbroek
  et~al.}{2020}]{DBLP:conf/mkm/BlaauwbroekUG20}
\begin{bchapter}
\bauthor{\bsnm{Blaauwbroek}, \binits{L.}},
\bauthor{\bsnm{Urban}, \binits{J.}},
\bauthor{\bsnm{Geuvers}, \binits{H.}}:
\bctitle{The tactician - {A} seamless, interactive tactic learner and prover
  for coq}.
In: \beditor{\bsnm{Benzm{\"{u}}ller}, \binits{C.}},
\beditor{\bsnm{Miller}, \binits{B.R.}} (eds.)
\bbtitle{Intelligent Computer Mathematics - 13th International Conference,
  {CICM} 2020, Bertinoro, Italy, July 26-31, 2020, Proceedings}.
\bsertitle{Lecture Notes in Computer Science},
vol. \bseriesno{12236},
pp. \bfpage{271}--\blpage{277}.
\bpublisher{Springer}, \blocation{???}
(\byear{2020}).
\doiurl{10.1007/978-3-030-53518-6\_17} .
\burl{https://doi.org/10.1007/978-3-030-53518-6\_17}
\end{bchapter}
\endbibitem

\bibitem[\protect\citeauthoryear{Varda}{}]{capnp}
\begin{botherref}
\oauthor{\bsnm{Varda}, \binits{K.}}:
Cap’n Proto Serialization and Capability based Remote Procedure Calling
  Framework.
\url{https://capnproto.org/}
\end{botherref}
\endbibitem

\bibitem[\protect\citeauthoryear{Miller}{2006}]{RobustComposition}
\begin{botherref}
\oauthor{\bsnm{Miller}, \binits{M.S.}}:
Robust composition: Towards a unified approach to access control and
  concurrency control.
PhD thesis,
Johns Hopkins University,
Baltimore, Maryland, USA
(May 2006)
\end{botherref}
\endbibitem

\bibitem[\protect\citeauthoryear{{he Coq Development Team}}{}]{coqpackageindex}
\begin{botherref}
\oauthor{\bsnm{{he Coq Development Team}}}:
Coq Package Index.
\url{https://coq.inria.fr/coq-package-index}
\end{botherref}
\endbibitem

\bibitem[\protect\citeauthoryear{{The Opam Development Team}}{}]{opam}
\begin{botherref}
\oauthor{\bsnm{{The Opam Development Team}}}:
Opam Package Manager.
\url{https://opam.ocaml.org/}
\end{botherref}
\endbibitem

\bibitem[\protect\citeauthoryear{Hu}{2005}]{hu2005efficient}
\begin{barticle}
\bauthor{\bsnm{Hu}, \binits{Y.F.}}:
\batitle{Efficient and high quality force-directed graph drawing}.
\bjtitle{The Mathematica Journal}
\bvolume{10},
\bfpage{37}--\blpage{71}
(\byear{2005})
\end{barticle}
\endbibitem

\bibitem[\protect\citeauthoryear{Chaslot
  et~al.}{2008}]{DBLP:conf/aiide/ChaslotBSS08}
\begin{bchapter}
\bauthor{\bsnm{Chaslot}, \binits{G.}},
\bauthor{\bsnm{Bakkes}, \binits{S.}},
\bauthor{\bsnm{Szita}, \binits{I.}},
\bauthor{\bsnm{Spronck}, \binits{P.}}:
\bctitle{Monte-carlo tree search: {A} new framework for game {AI}}.
In: \beditor{\bsnm{Darken}, \binits{C.}},
\beditor{\bsnm{Mateas}, \binits{M.}} (eds.)
\bbtitle{Proceedings of the Fourth Artificial Intelligence and Interactive
  Digital Entertainment Conference, October 22-24, 2008, Stanford, California,
  {USA}}.
\bpublisher{The {AAAI} Press}, \blocation{???}
(\byear{2008}).
\burl{http://www.aaai.org/Library/AIIDE/2008/aiide08-036.php}
\end{bchapter}
\endbibitem

\bibitem[\protect\citeauthoryear{Jette and
  Wickberg}{2023}]{DBLP:conf/jsspp/JetteW23}
\begin{bchapter}
\bauthor{\bsnm{Jette}, \binits{M.A.}},
\bauthor{\bsnm{Wickberg}, \binits{T.}}:
\bctitle{Architecture of the slurm workload manager}.
In: \beditor{\bsnm{Klus{\'{a}}cek}, \binits{D.}},
\beditor{\bsnm{Corbal{\'{a}}n}, \binits{J.}},
\beditor{\bsnm{Rodrigo}, \binits{G.P.}} (eds.)
\bbtitle{Job Scheduling Strategies for Parallel Processing - 26th Workshop,
  {JSSPP} 2023, St. Petersburg, FL, USA, May 19, 2023, Revised Selected
  Papers}.
\bsertitle{Lecture Notes in Computer Science},
vol. \bseriesno{14283},
pp. \bfpage{3}--\blpage{23}.
\bpublisher{Springer}, \blocation{???}
(\byear{2023}).
\doiurl{10.1007/978-3-031-43943-8\_1} .
\burl{https://doi.org/10.1007/978-3-031-43943-8\_1}
\end{bchapter}
\endbibitem

\bibitem[\protect\citeauthoryear{{The OpenPBS Development Team}}{}]{openpbs}
\begin{botherref}
\oauthor{\bsnm{{The OpenPBS Development Team}}}:
OpenPBS Open Source Project.
\url{https://www.openpbs.org/}
\end{botherref}
\endbibitem

\bibitem[\protect\citeauthoryear{Zhang
  et~al.}{2021}]{DBLP:conf/mkm/ZhangBPCKU21}
\begin{bchapter}
\bauthor{\bsnm{Zhang}, \binits{L.}},
\bauthor{\bsnm{Blaauwbroek}, \binits{L.}},
\bauthor{\bsnm{Piotrowski}, \binits{B.}},
\bauthor{\bsnm{Cern{\'{y}}}, \binits{P.}},
\bauthor{\bsnm{Kaliszyk}, \binits{C.}},
\bauthor{\bsnm{Urban}, \binits{J.}}:
\bctitle{Online machine learning techniques for coq: {A} comparison}.
In: \beditor{\bsnm{Kamareddine}, \binits{F.}},
\beditor{\bsnm{Coen}, \binits{C.S.}} (eds.)
\bbtitle{Intelligent Computer Mathematics - 14th International Conference,
  {CICM} 2021, Timisoara, Romania, July 26-31, 2021, Proceedings}.
\bsertitle{Lecture Notes in Computer Science},
vol. \bseriesno{12833},
pp. \bfpage{67}--\blpage{83}.
\bpublisher{Springer}, \blocation{???}
(\byear{2021}).
\doiurl{10.1007/978-3-030-81097-9\_5} .
\burl{https://doi.org/10.1007/978-3-030-81097-9\_5}
\end{bchapter}
\endbibitem

\bibitem[\protect\citeauthoryear{Czajka and
  Kaliszyk}{2018}]{DBLP:journals/jar/CzajkaK18}
\begin{barticle}
\bauthor{\bsnm{Czajka}, \binits{L.}},
\bauthor{\bsnm{Kaliszyk}, \binits{C.}}:
\batitle{Hammer for coq: Automation for dependent type theory}.
\bjtitle{J. Autom. Reason.}
\bvolume{61}(\bissue{1-4}),
\bfpage{423}--\blpage{453}
(\byear{2018})
\doiurl{10.1007/s10817-018-9458-4}
\end{barticle}
\endbibitem

\bibitem[\protect\citeauthoryear{Sanchez{-}Stern
  et~al.}{2020}]{DBLP:conf/pldi/Sanchez-SternAS20}
\begin{bchapter}
\bauthor{\bsnm{Sanchez{-}Stern}, \binits{A.}},
\bauthor{\bsnm{Alhessi}, \binits{Y.}},
\bauthor{\bsnm{Saul}, \binits{L.K.}},
\bauthor{\bsnm{Lerner}, \binits{S.}}:
\bctitle{Generating correctness proofs with neural networks}.
In: \beditor{\bsnm{Sen}, \binits{K.}},
\beditor{\bsnm{Naik}, \binits{M.}} (eds.)
\bbtitle{Proceedings of the 4th {ACM} {SIGPLAN} International Workshop on
  Machine Learning and Programming Languages, MAPL@PLDI 2020, London, UK, June
  15, 2020},
pp. \bfpage{1}--\blpage{10}.
\bpublisher{{ACM}}, \blocation{???}
(\byear{2020}).
\doiurl{10.1145/3394450.3397466} .
\burl{https://doi.org/10.1145/3394450.3397466}
\end{bchapter}
\endbibitem

\bibitem[\protect\citeauthoryear{First
  et~al.}{2020}]{DBLP:journals/pacmpl/FirstBG20}
\begin{barticle}
\bauthor{\bsnm{First}, \binits{E.}},
\bauthor{\bsnm{Brun}, \binits{Y.}},
\bauthor{\bsnm{Guha}, \binits{A.}}:
\batitle{Tactok: semantics-aware proof synthesis}.
\bjtitle{Proc. {ACM} Program. Lang.}
\bvolume{4}(\bissue{{OOPSLA}}),
\bfpage{231}--\blpage{123131}
(\byear{2020})
\doiurl{10.1145/3428299}
\end{barticle}
\endbibitem

\bibitem[\protect\citeauthoryear{Huang
  et~al.}{2019}]{DBLP:conf/iclr/HuangDSS19}
\begin{bchapter}
\bauthor{\bsnm{Huang}, \binits{D.}},
\bauthor{\bsnm{Dhariwal}, \binits{P.}},
\bauthor{\bsnm{Song}, \binits{D.}},
\bauthor{\bsnm{Sutskever}, \binits{I.}}:
\bctitle{Gamepad: {A} learning environment for theorem proving}.
In: \bbtitle{7th International Conference on Learning Representations, {ICLR}
  2019, New Orleans, LA, USA, May 6-9, 2019}.
\bpublisher{OpenReview.net}, \blocation{???}
(\byear{2019}).
\burl{https://openreview.net/forum?id=r1xwKoR9Y7}
\end{bchapter}
\endbibitem

\bibitem[\protect\citeauthoryear{Filli{\^{a}}tre and
  Conchon}{2006}]{DBLP:conf/ml/FilliatreC06}
\begin{bchapter}
\bauthor{\bsnm{Filli{\^{a}}tre}, \binits{J.}},
\bauthor{\bsnm{Conchon}, \binits{S.}}:
\bctitle{Type-safe modular hash-consing}.
In: \beditor{\bsnm{Kennedy}, \binits{A.}},
\beditor{\bsnm{Pottier}, \binits{F.}} (eds.)
\bbtitle{Proceedings of the {ACM} Workshop on ML, 2006, Portland, Oregon, USA,
  September 16, 2006},
pp. \bfpage{12}--\blpage{19}.
\bpublisher{{ACM}}, \blocation{???}
(\byear{2006}).
\doiurl{10.1145/1159876.1159880} .
\burl{https://doi.org/10.1145/1159876.1159880}
\end{bchapter}
\endbibitem

\bibitem[\protect\citeauthoryear{Nipkow
  et~al.}{2002}]{DBLP:books/sp/NipkowPW02}
\begin{bbook}
\bauthor{\bsnm{Nipkow}, \binits{T.}},
\bauthor{\bsnm{Paulson}, \binits{L.C.}},
\bauthor{\bsnm{Wenzel}, \binits{M.}}:
\bbtitle{Isabelle/HOL - {A} Proof Assistant for Higher-Order Logic}.
\bsertitle{Lecture Notes in Computer Science},
vol. \bseriesno{2283}.
\bpublisher{Springer}, \blocation{???}
(\byear{2002}).
\doiurl{10.1007/3-540-45949-9} .
\burl{https://doi.org/10.1007/3-540-45949-9}
\end{bbook}
\endbibitem

\bibitem[\protect\citeauthoryear{Li et~al.}{2021}]{DBLP:conf/iclr/LiYWP21}
\begin{bchapter}
\bauthor{\bsnm{Li}, \binits{W.}},
\bauthor{\bsnm{Yu}, \binits{L.}},
\bauthor{\bsnm{Wu}, \binits{Y.}},
\bauthor{\bsnm{Paulson}, \binits{L.C.}}:
\bctitle{Isarstep: a benchmark for high-level mathematical reasoning}.
In: \bbtitle{9th International Conference on Learning Representations, {ICLR}
  2021, Virtual Event, Austria, May 3-7, 2021}.
\bpublisher{OpenReview.net}, \blocation{???}
(\byear{2021}).
\burl{https://openreview.net/forum?id=Pzj6fzU6wkj}
\end{bchapter}
\endbibitem

\bibitem[\protect\citeauthoryear{Harrison}{2009}]{DBLP:conf/tphol/Harrison09a}
\begin{bchapter}
\bauthor{\bsnm{Harrison}, \binits{J.}}:
\bctitle{{HOL} light: An overview}.
In: \beditor{\bsnm{Berghofer}, \binits{S.}},
\beditor{\bsnm{Nipkow}, \binits{T.}},
\beditor{\bsnm{Urban}, \binits{C.}},
\beditor{\bsnm{Wenzel}, \binits{M.}} (eds.)
\bbtitle{Theorem Proving in Higher Order Logics, 22nd International Conference,
  TPHOLs 2009, Munich, Germany, August 17-20, 2009. Proceedings}.
\bsertitle{Lecture Notes in Computer Science},
vol. \bseriesno{5674},
pp. \bfpage{60}--\blpage{66}.
\bpublisher{Springer}, \blocation{???}
(\byear{2009}).
\doiurl{10.1007/978-3-642-03359-9\_4} .
\burl{https://doi.org/10.1007/978-3-642-03359-9\_4}
\end{bchapter}
\endbibitem

\bibitem[\protect\citeauthoryear{Kaliszyk
  et~al.}{2017}]{DBLP:conf/iclr/KaliszykCS17}
\begin{bchapter}
\bauthor{\bsnm{Kaliszyk}, \binits{C.}},
\bauthor{\bsnm{Chollet}, \binits{F.}},
\bauthor{\bsnm{Szegedy}, \binits{C.}}:
\bctitle{Holstep: {A} machine learning dataset for higher-order logic theorem
  proving}.
In: \bbtitle{5th International Conference on Learning Representations, {ICLR}
  2017, Toulon, France, April 24-26, 2017, Conference Track Proceedings}.
\bpublisher{OpenReview.net}, \blocation{???}
(\byear{2017}).
\burl{https://openreview.net/forum?id=ryuxYmvel}
\end{bchapter}
\endbibitem

\bibitem[\protect\citeauthoryear{Bansal
  et~al.}{2019}]{DBLP:conf/icml/BansalLRSW19}
\begin{bchapter}
\bauthor{\bsnm{Bansal}, \binits{K.}},
\bauthor{\bsnm{Loos}, \binits{S.M.}},
\bauthor{\bsnm{Rabe}, \binits{M.N.}},
\bauthor{\bsnm{Szegedy}, \binits{C.}},
\bauthor{\bsnm{Wilcox}, \binits{S.}}:
\bctitle{Holist: An environment for machine learning of higher order logic
  theorem proving}.
In: \beditor{\bsnm{Chaudhuri}, \binits{K.}},
\beditor{\bsnm{Salakhutdinov}, \binits{R.}} (eds.)
\bbtitle{Proceedings of the 36th International Conference on Machine Learning,
  {ICML} 2019, 9-15 June 2019, Long Beach, California, {USA}}.
\bsertitle{Proceedings of Machine Learning Research},
vol. \bseriesno{97},
pp. \bfpage{454}--\blpage{463}.
\bpublisher{{PMLR}}, \blocation{???}
(\byear{2019}).
\burl{http://proceedings.mlr.press/v97/bansal19a.html}
\end{bchapter}
\endbibitem

\bibitem[\protect\citeauthoryear{de~Moura
  et~al.}{2015}]{DBLP:conf/cade/MouraKADR15}
\begin{bchapter}
\bauthor{\bsnm{Moura}, \binits{L.M.}},
\bauthor{\bsnm{Kong}, \binits{S.}},
\bauthor{\bsnm{Avigad}, \binits{J.}},
\bauthor{\bsnm{Doorn}, \binits{F.}},
\bauthor{\bsnm{Raumer}, \binits{J.}}:
\bctitle{The lean theorem prover (system description)}.
In: \beditor{\bsnm{Felty}, \binits{A.P.}},
\beditor{\bsnm{Middeldorp}, \binits{A.}} (eds.)
\bbtitle{Automated Deduction - {CADE-25} - 25th International Conference on
  Automated Deduction, Berlin, Germany, August 1-7, 2015, Proceedings}.
\bsertitle{Lecture Notes in Computer Science},
vol. \bseriesno{9195},
pp. \bfpage{378}--\blpage{388}.
\bpublisher{Springer}, \blocation{???}
(\byear{2015}).
\doiurl{10.1007/978-3-319-21401-6\_26} .
\burl{https://doi.org/10.1007/978-3-319-21401-6\_26}
\end{bchapter}
\endbibitem

\bibitem[\protect\citeauthoryear{Han et~al.}{2022}]{DBLP:conf/iclr/HanRWAP22}
\begin{bchapter}
\bauthor{\bsnm{Han}, \binits{J.M.}},
\bauthor{\bsnm{Rute}, \binits{J.}},
\bauthor{\bsnm{Wu}, \binits{Y.}},
\bauthor{\bsnm{Ayers}, \binits{E.W.}},
\bauthor{\bsnm{Polu}, \binits{S.}}:
\bctitle{Proof artifact co-training for theorem proving with language models}.
In: \bbtitle{The Tenth International Conference on Learning Representations,
  {ICLR} 2022, Virtual Event, April 25-29, 2022}.
\bpublisher{OpenReview.net}, \blocation{???}
(\byear{2022}).
\burl{https://openreview.net/forum?id=rpxJc9j04U}
\end{bchapter}
\endbibitem

\bibitem[\protect\citeauthoryear{Polu
  et~al.}{2023}]{DBLP:conf/iclr/PoluHZBBS23}
\begin{bchapter}
\bauthor{\bsnm{Polu}, \binits{S.}},
\bauthor{\bsnm{Han}, \binits{J.M.}},
\bauthor{\bsnm{Zheng}, \binits{K.}},
\bauthor{\bsnm{Baksys}, \binits{M.}},
\bauthor{\bsnm{Babuschkin}, \binits{I.}},
\bauthor{\bsnm{Sutskever}, \binits{I.}}:
\bctitle{Formal mathematics statement curriculum learning}.
In: \bbtitle{The Eleventh International Conference on Learning Representations,
  {ICLR} 2023, Kigali, Rwanda, May 1-5, 2023}.
\bpublisher{OpenReview.net}, \blocation{???}
(\byear{2023}).
\burl{https://openreview.net/pdf?id=-P7G-8dmSh4}
\end{bchapter}
\endbibitem

\bibitem[\protect\citeauthoryear{Yang
  et~al.}{2023}]{DBLP:journals/corr/abs-2306-15626}
\begin{botherref}
\oauthor{\bsnm{Yang}, \binits{K.}},
\oauthor{\bsnm{Swope}, \binits{A.M.}},
\oauthor{\bsnm{Gu}, \binits{A.}},
\oauthor{\bsnm{Chalamala}, \binits{R.}},
\oauthor{\bsnm{Song}, \binits{P.}},
\oauthor{\bsnm{Yu}, \binits{S.}},
\oauthor{\bsnm{Godil}, \binits{S.}},
\oauthor{\bsnm{Prenger}, \binits{R.}},
\oauthor{\bsnm{Anandkumar}, \binits{A.}}:
Leandojo: Theorem proving with retrieval-augmented language models.
CoRR
\textbf{abs/2306.15626}
(2023)
\doiurl{10.48550/arXiv.2306.15626}
{\href{https://arxiv.org/abs/2306.15626}{{2306.15626}}}
\end{botherref}
\endbibitem

\end{thebibliography}

\end{document}